\documentclass[draft]{agujournal2019}
\usepackage{url} 
\usepackage{lineno}
\usepackage[inline]{trackchanges} 
\usepackage{soul}
\draftfalse

\journalname{JGR: Atmospheres}

\begin{document}

%
%


\title{Observations of the Origin of Downward Terrestrial Gamma-Ray Flashes}

%
%




\makeatletter
\newcommand{\ssymbol}[1]{^{\@fnsymbol{#1}}}
\makeatother

\authors{J.W.~Belz\affil{1}, P.R.~Krehbiel\affil{2}, J.~Remington\affil{1}, M.A.~Stanley\affil{2}, R.U.~Abbasi\affil{3}, R.~LeVon\affil{1}, W.~Rison\affil{2}, D.~Rodeheffer\affil{2}
\\ {\em and the Telescope Array Scientific Collaboration} \\
T.~Abu-Zayyad\affil{1}, M.~Allen\affil{1}, E.~Barcikowski\affil{1}, D.R.~Bergman\affil{1}, S.A.~Blake\affil{1}, M.~Byrne\affil{1}, R.~Cady\affil{1}, B.G.~Cheon\affil{6}, M.~Chikawa\affil{8}, A.~di~Matteo\affil{9\ssymbol{1}}, T.~Fujii\affil{10}, K.~Fujita\affil{11}, R.~Fujiwara\affil{11}, M.~Fukushima\affil{12,13}, G.~Furlich\affil{1}, W.~Hanlon\affil{1}, M.~Hayashi\affil{14}, Y.~Hayashi\affil{11}, N.~Hayashida\affil{15}, K.~Hibino\affil{15}, K.~Honda\affil{16}, D.~Ikeda\affil{17}, T.~Inadomi\affil{18}, N.~Inoue\affil{4}, T.~Ishii\affil{16}, H.~Ito\affil{19}, D.~Ivanov\affil{1}, H.~Iwakura\affil{18}, H.M.~Jeong\affil{20}, S.~Jeong\affil{20}, C.C.H.~Jui\affil{1}, K.~Kadota\affil{21}, F.~Kakimoto\affil{5}, O.~Kalashev\affil{22}, K.Kasahara\affil{23}, S.~Kasami\affil{24}, H.~Kawai\affil{25}, S.~Kawakami\affil{11}, K.~Kawata\affil{12}, E.~Kido\affil{12}, H.B.~Kim\affil{6}, J.H.~Kim\affil{1}, J.H.~Kim\affil{11}, V.~Kuzmin\affil{22\ssymbol{2}}, M.~Kuznetsov\affil{9,22}, Y.J.~Kwon\affil{26}, K.H.~Lee\affil{20}, B.~Lubsandorzhiev\affil{22}, J.P.~Lundquist\affil{1}, K.~Machida\affil{16}, H.~Matsumiya\affil{11}, J.N.~Matthews\affil{1}, T.~Matuyama\affil{11}, R.~Mayta\affil{11}, M.~Minamino\affil{11}, K.~Mukai\affil{16}, I.~Myers\affil{1}, S.~Nagataki\affil{19}, K.~Nakai\affil{11}, R.~Nakamura\affil{18}, T.~Nakamura\affil{27}, Y.~Nakamura\affil{18}, T.~Nonaka\affil{12}, H.~Oda\affil{11}, S.~Ogio\affil{11,28}, M.Ohnishi\affil{12}, H.~Ohoka\affil{12}, Y.~Oku\affil{24}, T.~Okuda\affil{29}, Y.~Omura\affil{11}, M.~Ono\affil{19}, A.~Oshima\affil{36}, S.~Ozawa\affil{23}, I.H.~Park\affil{20},M.~Potts\affil{1}, M.S.~Pshirkov\affil{22,30}, D.C.~Rodriguez\affil{1}, G.~Rubtsov\affil{22}, D.~Ryu\affil{31}, H.~Sagawa\affil{12}, R.~Sahara\affil{11}, K.~Saito\affil{12}, Y.~Saito\affil{18}, N.~Sakaki\affil{12}, T.~Sako\affil{12}, N.~Sakurai\affil{11}, K.~Sano\affil{18}, T.~Seki\affil{18}, K.~Sekino\affil{12}, F.~Shibata\affil{16}, T.~Shibata\affil{12}, H.~Shimodaira\affil{12}, B.K.~Shin\affil{11}, H.S.~Shin\affil{12}, J.D.~Smith\affil{1}, P.~Sokolsky\affil{1}, N.~Sone\affil{18}, B.T.~Stokes\affil{1}, T.A.~Stroman\affil{1}, Y.~Takagi\affil{11}, Y.~Takahashi\affil{11}, M.~Takeda\affil{12}, R.~Takeishi\affil{20}, A.~Taketa\affil{17}, M.~Takita\affil{12}, Y.~Tameda\affil{24}, K.~Tanaka\affil{32}, M.~Tanaka\affil{33}, Y.~Tanoue\affil{11}, S.B.~Thomas\affil{1}, G.B.~Thomson\affil{1}, P.~Tinyakov\affil{9,22}, I.~Tkachev\affil{22}, H.~Tokuno\affil{5}, T.~Tomida\affil{18}, S.~Troitsky\affil{22}, Y.~Tsunesada\affil{11,28}, Y.~Uchihori\affil{34}, S.~Udo\affil{15}, T.~Uehama\affil{18}, F.~Urban\affil{35}, M.~Wallace\affil{1}, T.~Wong\affil{1}, M.~Yamamoto\affil{18}, H.~Yamaoka\affil{33}, K.~Yamazaki\affil{36}, K.~Yashiro\affil{7}, M.~Yosei\affil{24}, H.~Yoshii\affil{37}, Y.~Zhezher\affil{12,22}, Z.~Zundel\affil{1}}


\affiliation{1}{Department of Physics and Astronomy, University of Utah, Salt Lake City, Utah, USA}
\affiliation{2}{Langmuir Laboratory for Atmospheric Research, New Mexico Institute of Mining and Technology, Socorro, NM, USA}
\affiliation{3}{Department of Physics, Loyola University Chicago, Chicago, Illinois, USA}
\affiliation{4}{The Graduate School of Science and Engineering, Saitama University, Saitama, Saitama, Japan}
\affiliation{5}{Graduate School of Science and Engineering, Tokyo Institute of Technology, Meguro, Tokyo, Japan}
\affiliation{6}{Department of Physics and The Research Institute of Natural Science, Hanyang University, Seongdong-gu, Seoul, Korea}
\affiliation{7}{Department of Physics, Tokyo University of Science, Noda, Chiba, Japan}
\affiliation{8}{Department of Physics, Kindai University, Higashi Osaka, Osaka, Japan}
\affiliation{9}{Service de Physique Th\'{e}orique, Universit\'{e} Libre de Bruxelles, Brussels, Belgium}
\affiliation{10}{The Hakubi Center for Advanced Research, Kyoto University, Kitashirakawa-Oiwakecho, Sakyo-ku, Kyoto, Japan}
\affiliation{11}{Graduate School of Science, Osaka City University, Osaka, Osaka, Japan}
\affiliation{12}{Institute for Cosmic Ray Research, University of Tokyo, Kashiwa, Chiba, Japan}
\affiliation{13}{Kavli Institute for the Physics and Mathematics of the Universe (WPI), University of Tokyo, Kashiwa, Chiba, Japan}
\affiliation{14}{Information Engineering Graduate School of Science and Technology, Shinshu University, Nagano, Nagano, Japan}
\affiliation{15}{Faculty of Engineering, Kanagawa University, Yokohama, Kanagawa, Japan}
\affiliation{16}{Interdisciplinary Graduate School of Medicine and Engineering, University of Yamanashi, Kofu, Yamanashi, Japan}
\affiliation{17}{Earthquake Research Institute, University of Tokyo, Bunkyo-ku, Tokyo, Japan}
\affiliation{18}{Academic Assembly School of Science and Technology Institute of Engineering, Shinshu University, Nagano, Nagano, Japan}
\affiliation{19}{Astrophysical Big Bang Laboratory, RIKEN, Wako, Saitama, Japan}
\affiliation{20}{Department of Physics, Sungkyunkwan University, Jang-an-gu, Suwon, Korea}
\affiliation{21}{Department of Physics, Tokyo City University, Setagaya-ku, Tokyo, Japan}
\affiliation{22}{Institute for Nuclear Research of the Russian Academy of Sciences, Moscow, Russia}
\affiliation{23}{Advanced Research Institute for Science and Engineering, Waseda University, Shinjuku-ku, Tokyo, Japan}
\affiliation{24}{Department of Engineering Science, Faculty of Engineering, Osaka Electro-Communication University, Neyagawa-shi, Osaka, Japan}
\affiliation{25}{Department of Physics, Chiba University, Chiba, Chiba, Japan}
\affiliation{26}{Department of Physics, Yonsei University, Seodaemun-gu, Seoul, Korea}
\affiliation{27}{Faculty of Science, Kochi University, Kochi, Kochi, Japan}
\affiliation{28}{Nambu Yoichiro Institute of Theoretical and Experimental Physics, Osaka City University, Osaka, Osaka, Japan}
\affiliation{29}{Department of Physical Sciences, Ritsumeikan University, Kusatsu, Shiga, Japan}
\affiliation{30}{Sternberg Astronomical Institute, Moscow M.V. Lomonosov State University, Moscow, Russia}
\affiliation{31}{Department of Physics, Ulsan National Institute of Science and Technology, UNIST-gil, Ulsan, Korea}
\affiliation{32}{Graduate School of Information Sciences, Hiroshima City University, Hiroshima, Hiroshima, Japan}
\affiliation{33}{Institute of Particle and Nuclear Studies, KEK, Tsukuba, Ibaraki, Japan}
\affiliation{34}{National Institute of Radiological Science, Chiba, Chiba, Japan}
\affiliation{35}{CEICO, Institute of Physics, Czech Academy of Sciences, Prague, Czech Republic}
\affiliation{36}{Engineering Science Laboratory, Chubu University, Kasugai, Japan}
\affiliation{37}{Department of Physics, Ehime University, Matsuyama, Ehime, Japan}

\let\thefootnote\relax\footnote{$\ssymbol{1}$ Currently at INFN, sezione di Torino, Turin, Italy}
\let\thefootnote\relax\footnote{$\ssymbol{2}$ Deceased}
\addtocounter{footnote}{-1}\let\thefootnote\svthefootnote





\correspondingauthor{Jackson Remington}{jremington@cosmic.utah.edu}




\begin{keypoints}

\item Downward Terrestrial Gamma-ray Flashes occur during strong initial breakdown pulses of negative cloud-to-ground and cloud lightning.

\item The initial breakdown pulses consist of streamer-based fast negative breakdown having transient sub-pulse conducting events, or `sparks'.

\item The streamer to leader transition of negative stepping occurs during strong currents in the final stage of initial breakdown pulses.

\end{keypoints}

%
%

%
%


\begin{abstract}
In this paper we report the first close, high-resolution observations
of downward-directed terrestrial gamma-ray flashes (TGFs) detected
by the large-area Telescope Array cosmic ray observatory, obtained in
conjunction with broadband VHF interferometer and fast electric field
change measurements of the parent discharge.  The results show that
the TGFs occur during strong initial breakdown pulses (IBPs) in the first few milliseconds of negative cloud-to-ground and low-altitude intracloud flashes, and that the IBPs are
produced by a newly-identified streamer-based discharge process called
fast negative breakdown.  The observations indicate the relativistic
runaway electron avalanches (RREAs) responsible for producing the
TGFs are initiated by embedded spark-like transient conducting events (TCEs) within
the fast streamer system, and potentially also by individual fast
streamers themselves.  The TCEs are inferred to be the
cause of impulsive sub-pulses that are characteristic features of classic IBP sferics. Additional development of the avalanches would be
facilitated by the enhanced electric field ahead of the advancing front
of the fast negative breakdown.  In addition to showing the nature of
IBPs and their enigmatic sub-pulses, the observations also provide
a possible explanation for the unsolved question of how the streamer
to leader transition occurs during the initial negative breakdown,
namely as a result of strong currents flowing in the final stage of successive IBPs, extending backward through both the IBP itself and the negative streamer breakdown preceding the IBP.
\end{abstract}


%
%

\section{Introduction}
%


%
%
%
%

The interplay between lightning and high-energy particle physics was realized
over two decades ago with the serendipitous observation of gamma radiation
emanating from the Earth. The BATSE (Burst and Transient Source Experiment)
instrument aboard NASA's Compton Gamma-Ray Observatory was designed to detect
radiation from Gamma Ray Bursts (GRBs), deep-space events which are considered
the most intense sources of electromagnetic radiation in the Universe. In
1994, BATSE unexpectedly recorded a series of brief, intense flashes of gamma
rays, which appeared to originate at high altitudes ($\geq$15~km above ground
level) above thunderstorm regions~\cite{carlson2007,fishman1994}.  The 
terrestrial gamma-ray flashes (TGFs) lasted from hundreds of microseconds up
to a millisecond or more, and their energy spectrum was consistent with
bremsstrahlung emission from electrons with energies of several million
electron volts (MeV) or greater.

Subsequent observations, now numbering in the thousands of events, aboard the
Ramaty High Energy Solar Spectroscopic Imager (RHESSI)
satellite~\cite{gjesteland2012,grefenstette2009}, NASA's Fermi Gamma-ray Space
Telescope~\cite{briggs2013,foley2014,roberts2017}, and the Astrorivelatore Gamma a Immagini Leggero 
(AGILE) satellite~\cite{marisaldi2014} have shown that, instead of being produced at high
altitude above storms, the TGFs originate at lower altitudes commensurate with
being inside storms.  In particular, it has been shown that the TGFs are
produced at the altitudes of intracloud (IC) lightning flashes, during upward negative breakdown at the beginning of the flashes~\cite{cummer2011,cummer2015,lu2010, lyu2016,mailyan2016,shao2010,Stanley2006}. The early RHESSI
observations were found to be associated with millisecond-duration initial
breakdown activity that occurs in the beginning stages of IC flashes.
However, a direct connection with the initial breakdown events was uncertain
due to a 1-3~ms timing uncertainty in the RHESSI data~\cite{lu2011}.

In recent years, a small subset of TGFs has been associated with high-peak current
(few hundred kiloampere) IC discharge events, called energetic in-cloud pulses
(EIPs)~\cite{lyu2015}.  EIPs are energetic versions of what are called
preliminary or initial breakdown pulses~\cite{marshall2013}, that are
characteristic features of the beginning stages of IC and negative
cloud-to-ground ($-$CG) flashes. The EIP studies have utilized data from the
Gamma-ray Burst Monitor (GBM) on Fermi~\cite{briggs2010}, which detects individual photons with microsecond timing accuracy, allowing more accurate correlation with
ground-based low frequency (LF) radio atmospheric or ``sferic'' observations.
Although EIPs are infrequent and the number of documented cases is
small (a dozen or so), TGFs have been detected for 100\% of EIPs that occurred
within view of the Fermi satellite and within range of ground-based sferic
sensors.  As a result of this predictability, EIPs are considered to be
high-probability producers of at least a class of TGF-generating lightning
events~\cite{cummer2017,lyu2016,lyu2018a}. However, the detailed discharge
processes that produce EIPs has not been understood, due to the lack of
measurements of the parent flashes with ground-based instrumentation (such observations of a close EIP by \citeA{tilles2020}, reported while this paper was in review, provides the first detailed information on the discharge processes and storm environment that led to its occurrence, as discussed later).

As satellite-based observations of upward TGFs have accumulated, the question
has been whether lightning produces downward TGFs that could be detected on the
ground below or near thunderstorms. In particular, negative-polarity
cloud-to-ground ($-$CG) discharges begin with downward negative breakdown that
would be expected to produce TGFs directed earthward.  Until recently, only a
few TGFs had been detected at ground level in association with
overhead lightning. Instead of being produced in the early stages of natural
lightning, however, the gamma rays occurred either during the upward
ascent of artificial trailing-wire, rocket-triggered lightning
discharges~\cite{dwyer2004,hare2016}, or at a later time in natural flashes,
following high-current return strokes of $-$CG
discharges~\cite{dwyer2012b,ringuette2013,tran2015}. Also, a particularly strong downward TGF was recently reported during a winter thunderstorm by \citeA{wada2019} at the time of lightning discharge in the storm that appeared to be produced at low altitude ($\simeq$400~m) above ground. Otherwise, significant impediments to
detecting downward TGFs have been a) the increasingly strong attenuation of
gamma radiation at low altitudes in the atmosphere, and b) the ground-based
detectors being either too far below and/or not widespread enough to detect
the forward-beamed radiation. Both issues have been addressed with
observations from the large-area (700~km$^2$) Telescope Array Surface Detector
(TASD) cosmic ray facility in central Utah.

In data collected between 2008 and 2013 there were ten occasions in which the
TASD was triggered by multiple bursts of energetic particles --- not arising
from cosmic rays. The events occurred within a millisecond of being detected
by the U.S.\ National Lightning Detection Network (NLDN)~\cite{abbasi2017},
which identified them as being produced during $-$CG flashes.  Follow-up
observations with the TASD by the authors of the present study, obtained
between 2014 and 2016 in coordination with a 3-D lightning mapping array (LMA)
and a lightning electric field change sensor, detected ten additional events,
each consisting of three to five lightning-initiated bursts~\cite{abbasi2018}.
The bursts were typically $\simeq$10 $\mu$s or less in duration, and
occurred over several hundred $\mu$s time intervals during the first
millisecond of downward negative breakdown at the beginning of $-$CG flashes.
Scintillator responses and simulation studies showed that the bursts
primarily resulted from gamma radiation and collectively comprised low-fluence
TGFs.  The LMA observations showed the bursts coincided with impulsive
in-cloud VHF radiation events during energetic downward negative breakdown,
\hbox{3-4~km} above ground level.  Although the TASD and LMA observations had
sub-microsecond time resolution, the electric field change measurements
recorded only the relatively slow electrostatic field change, with
insufficient bandwidth to detect the faster electric field changes of the
initial breakdown activity.


Here we report observations of downward TGFs produced by four additional
flashes (three --CGs and one low-altitude IC flash) obtained in 2018 during
continued studies with the Telescope Array.  For this study, the TASD and LMA
observations were augmented with crucially important, high-resolution VHF
interferometric and fast electric field change measurements of the parent
lightning discharges, obtained in relatively close proximity (16--24~km) to the
TGFs.  Coupled with sub-microsecond TGF measurements at TASD stations
immediately below and near the flashes, the observations document the TGF
occurrence with a high degree of temporal and spatial resolution not available
before now.  In each of the four flashes, the TGFs show a clear correspondence
with downward negative breakdown during strong initial breakdown pulse
(IBP) events in the first millisecond or so of the flashes.  The negative breakdown
progresses at a fast average speed ($\simeq$1--3~$\times 10^7$~m/s), indicative of a newly-recognized type of discharge process called fast negative
breakdown (FNB)~\cite{tilles2019a}.  Such breakdown is the negative analog of
fast positive breakdown found in an earlier study to be the cause of
high-power discharges called narrow bipolar events (NBEs)~\cite{rison2016}.

For both polarities, the breakdown is produced by a propagating system
of streamers that substantially enhance (up to 50\% or more) the electric field ahead of the streamers' advancing front~\cite{attanasio2019}. For the negative polarity version, electron avalanches produced within the streamer system would propagate through and ahead of the advancing front, producing downward-directed
gamma radiation. Detailed analysis of the observations indicate that the TGFs are
often initiated at the time of characteristic ``sub-pulses'' that occur during large-amplitude, `classic' sferics. From this, we infer that the sub-pulses are
produced by transient spark-like discharges embedded within the
negative streamer system, the conducting tips of which would initiate
relativistic electron avalanches, whose further development is
facilitated by the enhanced E field ahead of and beyond the streamer front. In other
instances, TGFs appear to be initiated during brief episodes of accelerated-speed FNB.

Although obtained for downward negative breakdown of --CG flashes, the results are expected to apply equally well
to negative breakdown at the beginning of upward IC flashes, for which the initial breakdown pulse activity is fundamentally the same as for downward CG flashes. Together, the results establish that downward TGFs of --CG flashes and
satellite-detected upward TGFs of IC flashes are variants of the same
phenomenon, and are produced during fast negative breakdown early in the
developing negative leader stage of CG and IC flashes.

\section{Results}


\subsection{Observations}

\begin{figure}
\includegraphics[width=0.98\linewidth]{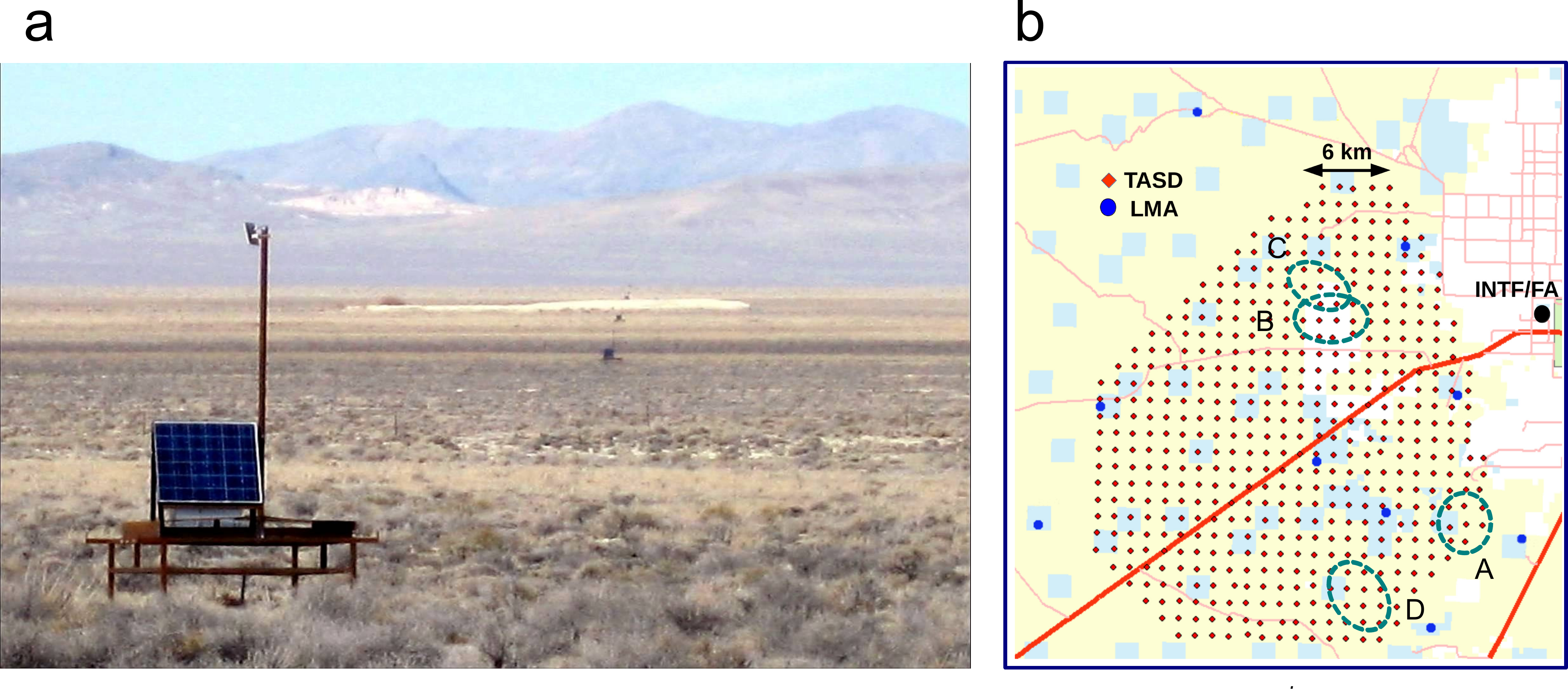}
\caption{\label{fig:detectors} Telescope Array Surface Detector. {\normalfont
({\bf a}) View of a close and distant surface detector stations on the desert
plain west of Delta, Utah.  Each detector unit consists of two 3~m$^2$ by
1.2~cm thick scintillator planes separated by a 0.1~cm steel
sheet~\cite{abuzayyad2012}.  Photo by M.~Fukushima.  ({\bf b}) Map of the 
TASD stations, showing the locations of TGFs A--D (dashed ellipses).  A total
of 512 surface detectors have been deployed over a 700~km$^2$ area on a 1.2~km
grid since 2008.  A nine-station 3-D lightning mapping array (LMA) has been
operated at the TASD since 2013 (blue dots). In July 2018, a VHF
interferometer (INTF) and fast electric field sferic sensor (FA) were deployed
6~km east of the TASD, only a few days prior to observing the TGFs reported here. }}
\end{figure}

Figure~\ref{fig:detectors} shows the layout of the
Telescope Array Surface Detector (TASD) and the Lightning Mapping Array (LMA) used
in both the earlier and present studies. The VHF interferometer (INTF) and
fast electric field change antenna (FA) were located 6~km east of the TASD,
and utilized three receiving antennas with 106--121~m baselines oriented to maximize
angular resolution over the TASD (see Methods Appendix~A1).

On August 2, 2018, two small, localized storms occurred over the TASD that
produced three TGFs relatively close (17~km) to the INTF.  The first
TGF-producing discharge occurred at 14:17:20~UT and was a $-$CG flash that
generated two TASD triggers $\simeq$1~ms after it began.  The flash was
initiated at $\simeq$5.5~km MSL altitude by a moderately high-power
($+$28~dBW, 630~W) upward fast positive narrow bipolar event  (Supporting
Figure~S6). The ensuing downward negative breakdown went to ground in
$\simeq$8~ms, corresponding to a stepped leader speed of
$\simeq$~$5\times$~$10^5$~m/s, somewhat faster than the normal stepped leader
speeds of 1--2$\times 10^5$~m/s.  The two triggers recorded three gamma-ray
bursts, jointly called TGF~A, when the breakdown was at $\simeq$~4.5~km MSL
altitude (3.1~km above ground level).

\newpage
\clearpage

\newpage
\begin{figure}[b]
\includegraphics[width=1.0\linewidth]{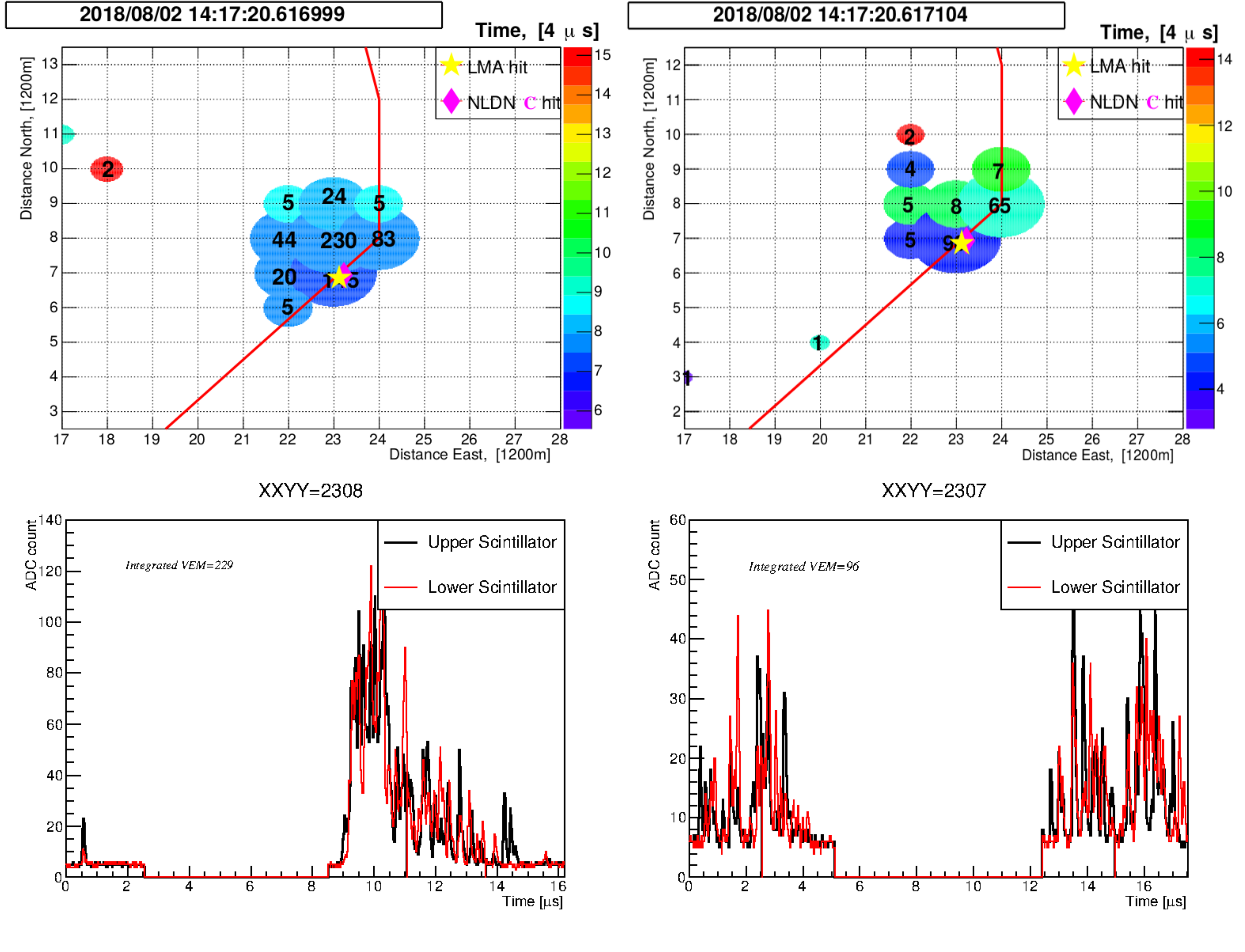}
\caption{\label{fig:sddisplay} TASD observations of TGF~A. {\normalfont {\em
Top left and right}: Surface scintillator ``footprints'' for the three gamma-ray
showers of TGF~A.  The grid spacing is in units of 1.2~km. The area of each
circle is proportional to the logarithm of the energy deposit, and color
indicates timing in 4~$\mu$s steps relative to the event trigger, corresponding to the approximate onset time of the gamma events at the ground.  The yellow star shows the
LMA-estimated plan location of the TGF, and is in close agreement with the
location of its sferic by the National Lightning Detector Network (NLDN,
underlying magenta diamond) making it difficult to distinguish between the two. The red
lines denote the boundary of the TASD array, showing that a portion of both showers
was likely undetected. {\em Bottom left and right}: Scintillator responses of
the surface detector stations having the largest energy deposit during each of the gamma-ray showers. 
The upper scintillator is represented by black traces and the
lower scintillator by red traces. A single Vertical Equivalent Muon (VEM),
or about 2~MeV of energy deposit, corresponds roughly to a pulse 30~ADC counts
above background with 100~ns FWHM on these plots. The horizontal time axes are relative to the detectors' individual triggers (different from the overall `event' trigger, see Appendix~A1).}} 
\end{figure} 

\clearpage

\newpage
\begin{figure}[t]
\begin{center}
\includegraphics[width=1\textwidth]{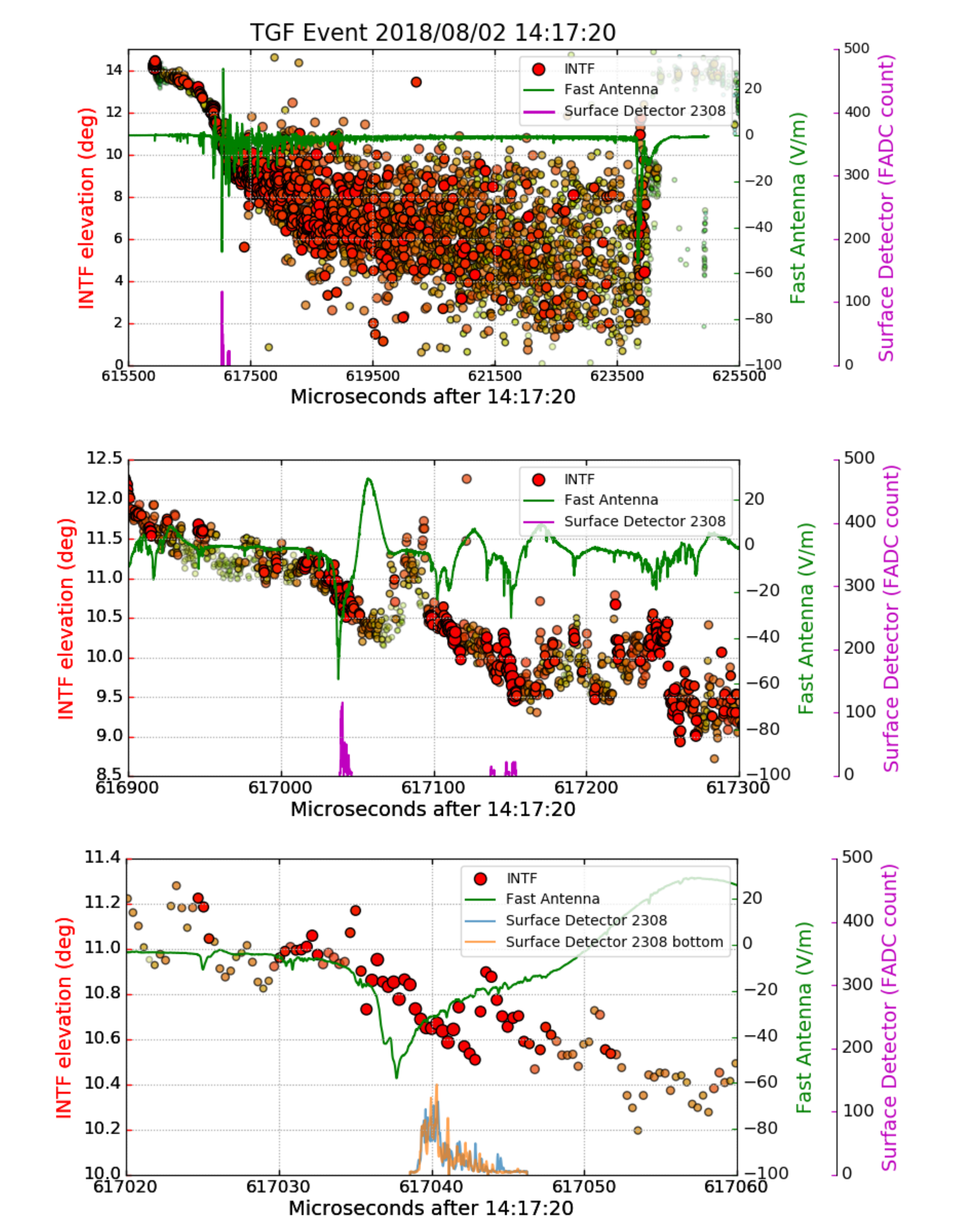}
\caption{\label{fig:overlay3} INTF and FA observations of TGF~A. {\normalfont
Panels show interferometer elevation versus time (circled dots,
sized and colored by power), fast electric field sferic waveform (green
waveform) and TASD particle surface detections (vertical purple bars). {\em Top:} Observations from
initial breakdown through time of --38.3~kA initial cloud-to-ground stroke.
Initial TGF detection occurred in coincidence with the strongest (--36.7~kA)
sferic pulse, 326~$\mu$s after flash start (Supporting Table S1).  
{\em Middle:} 400~$\mu$s of observations around the time of the three gamma-ray
showers of the flash, showing their correlation with the two largest amplitude
initial breakdown pulses (IBPs) and episodes of fast downward negative
breakdown (FNB).  TASD footprints for the showers are shown in
Figure~\ref{fig:sddisplay}.  {\em Bottom:} Detailed 40~$\mu$s view of the
upper and lower scintillator responses (blue and orange traces) relative to the
IBP sferic and the downward FNB.  }}
\end{center}
\end{figure}
\clearpage

Figure~\ref{fig:sddisplay} shows ``footprints'' of the TA surface detections
for each of the two triggered events, along with the corresponding set of
scintillator observations at a central SD station. The triggers occurred within
$\simeq$100~$\mu$s of each other, in the southeastern corner of the TASD.  The
observations are similar to those reported in our previous study~\cite{abbasi2018}, in that they consisted of gamma bursts typically 10~$\mu$s or less
in duration and were detected at 9--12 adjacent SDs, over areas $\simeq$3--4
km in diameter. The initial burst was the most energetic, depositing an
integrated total of 230~Vertical Equivalent Muons (VEM) (471~MeV) in the
nearby TASD station, and a total of 561~VEM (1,150 MeV) over all nine adjacent
stations (see Supporting Table S1).

INTF and FA observations for the flash are presented in
Figure~\ref{fig:overlay3}, which shows how the bursts were related to the
discharge processes.  The top panel provides an overview of the first 10~ms of
the flash, from the start of the downward negative leader through the initial
stroke to ground.  The gamma bursts (vertical purple bars) occurred early in
the flash, $\simeq$1.0 and 1.1~ms after the flash's initiation.  Around this
time, the FA data show a sequence of initial breakdown pulses (IBPs) of
rapidly increasing and then decaying amplitude --- typical of the beginning of
--CG flashes.

The first 1,150~MeV burst was associated with a particularly strong ($-38$~kA)
IBP sferic, comparable in magnitude to the sferic of the ensuing return
stroke, which had an NLDN-detected peak current of $-37$~kA.  The second TGF
was less strong (192 total VEM, or 393 MeV) and was associated with the
next-strongest IBP sferic (middle panel). Both gamma bursts were associated with
episodes of accelerated downward negative breakdown.

The bottom panel of Figure~\ref{fig:overlay3} shows in detail how the initial
gamma burst was related to the VHF radiation and sferic waveform, during a
40~$\mu$s window around the time of the burst. From the INTF elevation angles
and the LMA-indicated 17~km plan distance to the source location, the VHF
radiation sources descended $\simeq$150~m in 10~$\mu$s, corresponding to an average
propagation speed $v\simeq 1.5 \times 10^7$~m/s. By coincidence, this is the same as the
extent and speed of the upward fast positive NBE breakdown at the beginning of
the flash (also $\simeq$150~m in 10~$\mu$s), and is
indicative of the downward activity being caused by analogous fast negative
breakdown (FNB)~\cite{tilles2019a}.  The gamma burst occurred partway
through the fast downward breakdown, \hbox{$\simeq$1--2~$\mu$s} after the peak
of the negative sferic, and continued for about 5~$\mu$s before dying out
shortly after the end of the FNB.

\subsection{Source determination and time shifting}

\begin{figure}[t]
\begin{center}
\includegraphics[width=1.0\textwidth]{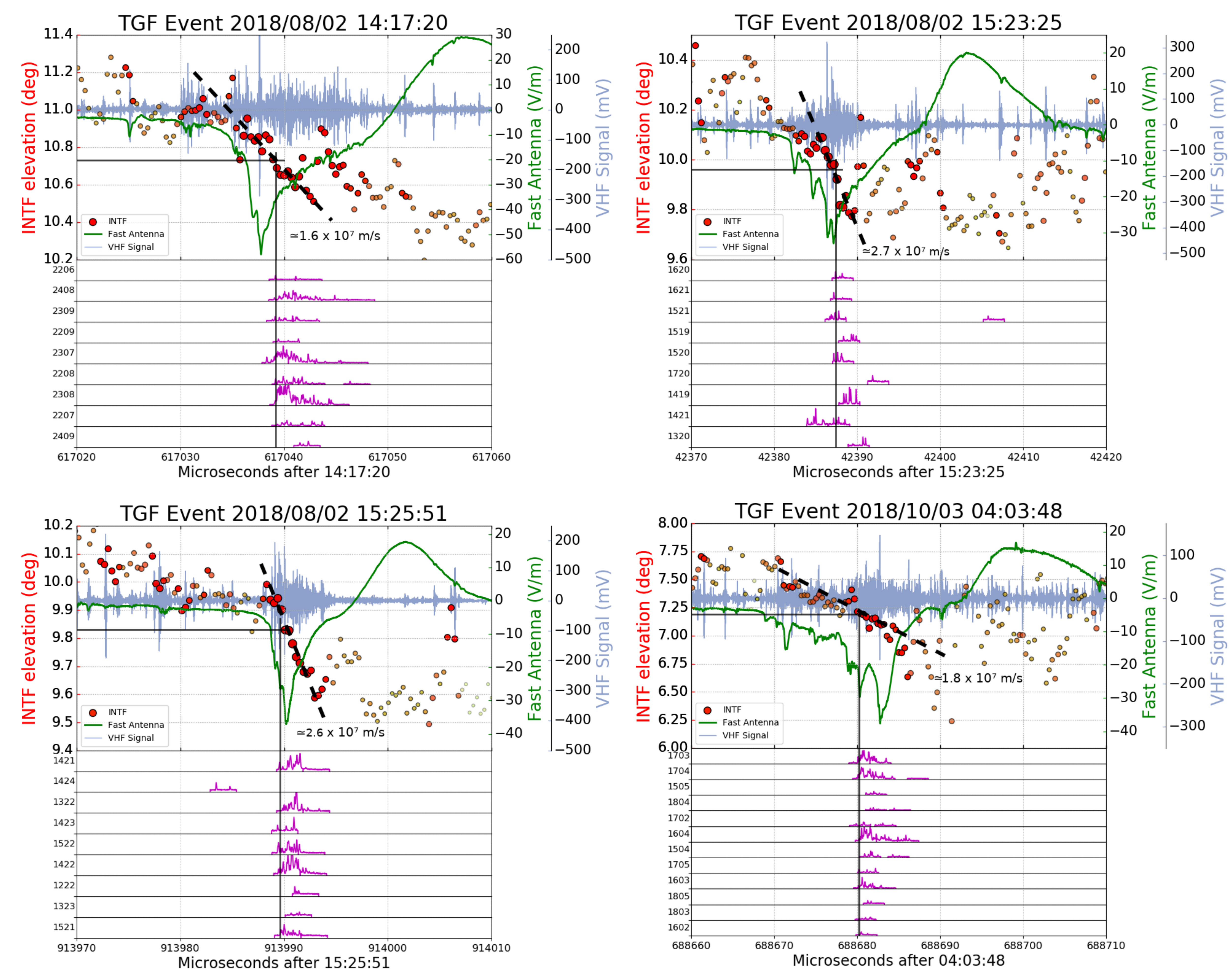}
\caption{\label{fig:all_tgfs} Detailed comparative observations.  {\normalfont
Time-shifted surface detector data for the primary gamma-ray
event during each of the four TGF-producing flashes, showing how the
TASD detections (lower axes) compare to each other, and their
relation to the VHF radiation sources and fast electric field sferics (upper axis) of the
developing discharges.  Black vertical and horizontal lines in each panel show
the median onset time of the gamma burst(s) during the downward FNB, obtained from
analysis of the collective onset times $t_b$ at the different TASD stations
and the observed INTF elevation angle vs.\ time (see Section~2.2).  Light
blue traces show the VHF time series waveform observed by the INTF.
Station numbers XXYY in the lower axes identify each TASD's easterly (XX) and northerly (YY) location within the array in 1.2~km grid spacing units. FNB propagation speeds are indicated by the dashed lines and associated values. Full-page versions of these plots are given as Figs.\ S15--S18. in the Supporting Information}}
\end{center}
\end{figure}

Figure~\ref{fig:all_tgfs} shows observations of the strongest gamma-ray event
for each of the TGF-producing flashes, along with time-shifted scintillator
detections for each participating TASD station. The vertical line for each
flash serves as a reference time for comparing the different SD waveforms with
each other and with the INTF/FA. As described below, it corresponds to the
median onset time at the different SD stations.  Similarly, the horizontal
line indicates the elevation angle corresponding to the median source
altitude immediately around that time.

\clearpage

The coordinate system for comparing the TASD observations with the INTF and FA
data is shown in Figure~\ref{fig:coord_system}a of the Appendix.  It is a
source-centric system in which the plan position on the ground beneath the TGF
serves as the coordinate origin.  To shift the scintillator detection times,
we need to know the slant ranges $r$ and $R$ from the TGF source to the SD and
from the source to the INTF. The $x,y$ plan location of the source is obtained
from the LMA observations within $\pm$1~ms of the TGF, which determines the
plan distances $D$ and $d$ to the INTF and to each TASD station. The TGF is
therefore at point $a$ = [0,0,$z_a$] in the coordinate system, where $z_a$ is
the altitude of the source above a reference plane of 1400~m MSL.  A generic
TASD station is at point~$b$, typically within $\simeq$1--3~km plan distance
of the TGF. The INTF/FA is at point~$c$, typically 15--25 km plan distance
from the TGFs. The net time shift $\Delta t$ between the surface detector data
at a given TASD station and the INTF is given by the difference in propagation
delays. In particular, $\Delta t = (R/c) - (r/c) = (R - r)/c$. Because
the plan distances are known, the slant ranges and hence time shifts $\Delta
t$ are functions only of $z_a$.  Once $z_a$ is determined, the time shifts are
calculated for each TASD individually and used to compare the different
TASD waveforms a) with each other, and b) with the FA sferic and the VHF
source activity and centroid observations, as seen in
Figure~\ref{fig:all_tgfs}. For each TASD station, the onset time at the INTF
is given by $t_c = t_b + \Delta_t$, where $t_b$ is the onset time at the TASD
in question.  As mentioned above, the vertical line in Figure~\ref{fig:all_tgfs}
corresponds to the median of the onset times at the different stations. At the
same time it also serves as a reference point for identifying stations having
onset times that differ from the median value.

Because the LMA typically mislocates non-impulsive, VHF-noisy sources, the
TGF's altitude is determined from the INTF elevation angles $\theta_c$.  The
difficulty with doing this is that the angle changes with time during the IBP, 
namely $\theta_c = \theta_c(t_c)$, making it unclear which time to pick.  Even
though the elevation change corresponds only to a $\simeq$100--200 m spread in
the source altitude, it corresponds to the full 10--20~$\mu$s duration of
the VHF and FA sferic observations.  The ambiguity is resolved by recognizing
that two independent measurements are necessary to determine the two unknowns,
namely the source altitude $z_a$ and time $t_a$. In addition to the INTF
elevation angle $\theta_c$, the second measurement comes from onset time $t_b$
at the particular TASD in question. Although this provides enough
information to obtain the solution, the different variables of the problem,
namely [$\theta_c$, $t_c$, $z_a$, $t_a$], wind up depending upon each
other, requiring an iterative approach to obtain the solution.

Figure~S14 shows a block diagram of the iteration process.  For each
TASD the onset time $t_b$ is used along with an initial value of
$z_a$ to determine the corresponding onset time $t_c$ at the INTF.
The INTF data relating $t_c$ and $\theta_c$ is then used to determine
the corresponding source altitude $z_a$ and time $t_a$. If the
resulting $z_a$ is different from the initially assumed value, the new
value is used as the starting altitude for the next step.  The
iteration is stable and convergence is reached within a couple of
steps.  The process is repeated for each of the participating TASDs to
obtain a set of $z_a$, $t_a$, $t_c$, and $\theta_c$ values, from which
the median is determined. Table~S2 lists the full set of solutions for
each TASD of the different TGFs. The median $t_c$ and $\theta_c$
values are shown in bold and correspond to the vertical and horizontal
lines in Figure~\ref{fig:all_tgfs}. For TGFs~A, C and D, the participating
TASDs all have similar onset times.  The exception is TGF~B, which has two
or more onset times, as discussed in the next section.  An analogous but
somewhat different method of time-shifting and comparing the TASD and INTF/FA
observations, developed independently during the course of the study, is
described in Appendix A2 and shown in the Supporting Figures.  The approach
utilized measurements at two TASD stations having the strongest detections to
determine the time shifts for the other TASDs and alignment with the INTF/FA
observations, and provided an alternative way of investigating the
observations.


\subsection{Temporal comparisons}

The above analyses provide accurately-determined estimates of i) each TGF's
plan location $x_a$,~$y_a$, altitude $z_a$, and time $t_a$, ii) the onset
times $t_c$ of the gamma events during the IBP, and iii) the INTF elevation
angle $\theta_c$ corresponding to $t_c$ and $z_a$. The $t_c$ and $\theta_c$
values are shown by the vertical and horizontal lines in each of the panels of
Figure~\ref{fig:all_tgfs}. We re-emphasize the fact that the $t_c$ values serve
as reference times for comparing the different TASD detections with each
other.  For TGFs~A, C, and D, most or all of the stations detected the onset
at the same time. The onset times are well-identified by the analysis technique
and are indicative of the TGFs in question all having a single onset. An
important exception is TGF~B, for which TASD 1421 had a noticeably earlier
onset time. Three other stations (1519, 1419, and 1320) appeared to have
slightly delayed onsets. As discussed below, the different apparent onsets are
notable because the footprint of the stations involved were systematically
displaced in a fully 360 degree circular pattern around a central hole.
The observations are also illustrative of the comparisons being able to
identify multiple onset times.


For each of the four flashes, the gamma bursts were associated with
well-defined episodes of downward-propagating fast negative breakdown.  The
average propagation speeds during the episodes ranged from $\simeq$1.6 to 2.7
$\times 10^7$~m/s (slanted dashed lines in each panel of
Figure~\ref{fig:all_tgfs}). This is compared to average speeds of $\simeq$1.0
to 2.5$ \times 10^6$~m/s for the breakdown immediately preceding the IBPs and
TGFs (Figure~\ref{fig:overlay3} and Supporting Figures~S7--S9).  The sferics
associated with the TGFs constituted the strongest initial breakdown pulses of
the flashes.  Whereas the onset time of the gamma burst of TGF~A
(Figure~\ref{fig:all_tgfs}a) occurred slightly after the main peak of the IBP
sferic, the bursts during other flashes occurred during or at various times
prior to the peak.  For TGF~C, the onset was at or shortly after the beginning
of the IBP and FNB, while for TGF~B, the primary onset was closely correlated
with the main IBP peak. For TGF~D, the onset appeared to be exclusively
correlated with a strong, leading-edge sub-pulse during the IBP's FNB.  IBPs
having such sub-pulses are called ``classic'' IBPs~\cite{karunarathne2014,marshall2013,nag2009,shi2019}.  The sub-pulse feature of the preliminary
breakdown has long been recognized, beginning with \citeA{weidman1979}, but
the cause both of IBPs and their sub-pulses has remained unknown. The present
results show that the IBPs are produced by fast negative breakdown, and that
the sub-pulses are capable of initiating gamma bursts.

\begin{figure}[t]
\begin{center}
\includegraphics[width=1.0\textwidth]{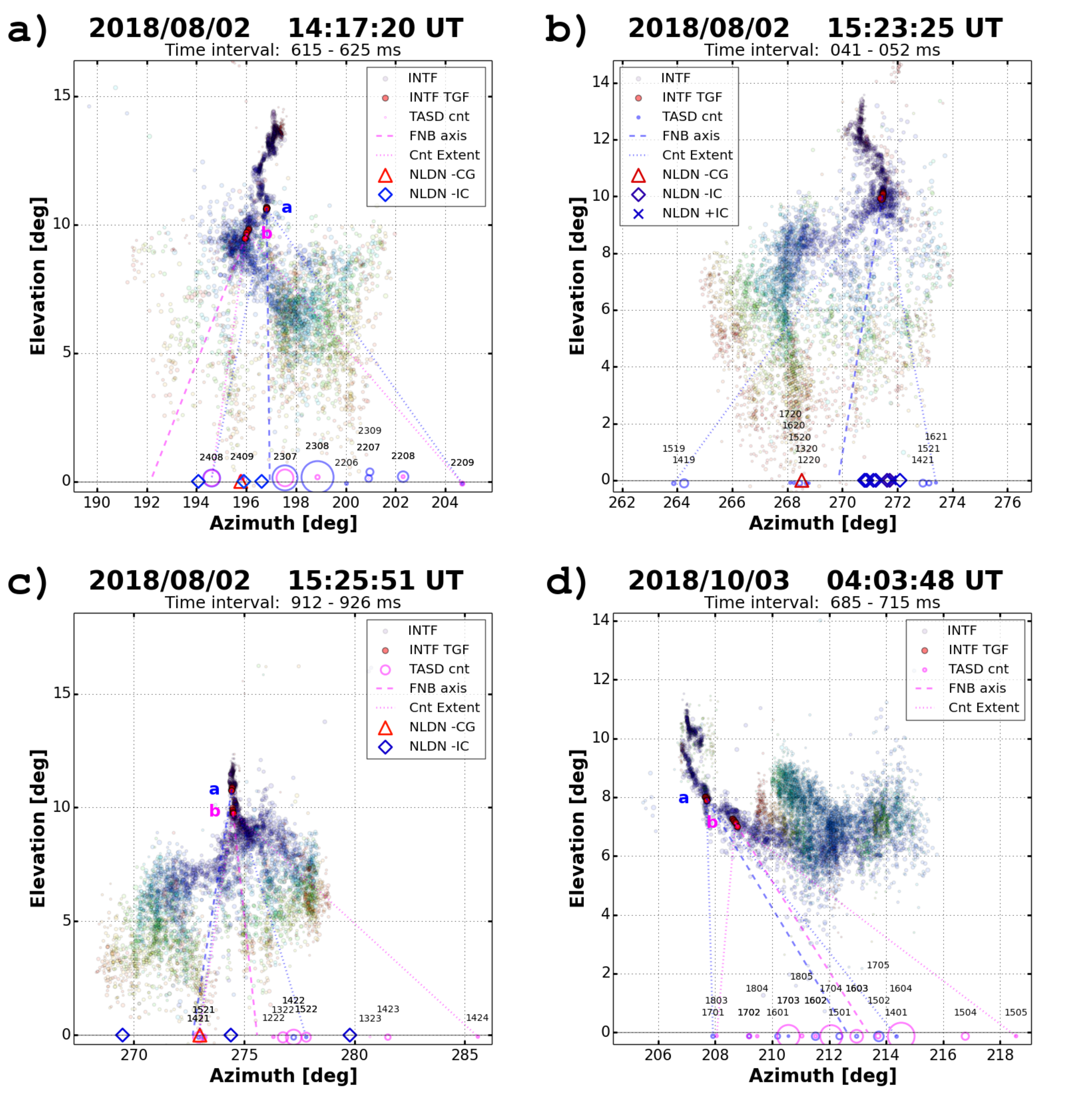}
\caption{\label{fig:newfig} INTF observations of the TGF-producing flashes.
{\normalfont  Azimuth-elevation plots of INTF observations for the parent
flashes of TGFs~A--D, showing the initial downward development leading to the
TGF occurrences (dark red sources and a,b labels, indicating the TGF
altitudes).  Continuation of VHF activity is shown up to the time of the
initial stroke to ground for --CG flashes A,B,C, and for a comparable time
during the low-altitude intracloud flash of TGF~D.  Dashed lines indicate the
directions of the FNB associated with each TGF, and the inferred possible
beaming direction.  Baseline circles indicate detected TGF strength (VEM
counts) and azimuthal directions of participating TASDs. Dotted line pairs
indicate maximum angular spread (labelled `Cnt Extent') of the SD detections,
as viewed in the transverse plane from the INTF site. Vertical/horizontal
aspect ratios are adjusted to show true angular extent.  TGF~B appeared to
have multiple onset times at the different TASDs, and therefore narrower
beaming than indicated by the overall angular extent.  Baseline symbols show
NLDN locations of CG and IC events. Full page versions of each panel are given
in Figures~S19-S22 of the Supporting Information}}
\end{center} 
\end{figure}

For TGF~A at 14:17:20 (Figures~\ref{fig:all_tgfs}a and \ref{fig:newfig}a), the
scintillator detections in Figure~\ref{fig:overlay3} are from TASD~2308,
corresponding to the station having the most energetic footprint.  However,
the estimated plan location of the burst from the LMA observations, as well as
the NLDN location for the sferic associated with the burst, indicate the
breakdown was almost directly above TASD~2307, 1.2~km to the south and 17~km
southwest of the INTF (Supp.\ Figs.\ S1 and S10e). The energy deposit in
TASD~2307 was slightly weaker than that in 2308 (145 vs.\ 230 VEM), indicating
that the gamma burst was tilted slightly northward from vertical.  A
significant feature of the observations in Figure~\ref{fig:all_tgfs}a is that the apparent onset time of the
burst coincided with a step discontinuity in the VHF elevation centroid
values.  We later show (Figure~\ref{fig:coord_system}b) that the discontinuity
was due to a brief interval of enhanced propagation speed, in which the FNB
descended $\simeq$50~m in 1.5~$\mu$s, corresponding to a speed $v \simeq
3\times 10^7$~m/s, two times faster than the average speed of the IBP's FNB.  Observations of the second set of gamma bursts during the
flash shows them to be similarly associated with brief episodes of enhanced
fast breakdown speeds ($\simeq 2.3 \times 10^7$ and $4.6 \times 10^7$~m/s;
Supp.\ Fig.\ S10d,g).

\clearpage

TGFs~B and C (Figures~\ref{fig:all_tgfs}b,c and \ref{fig:newfig}b,c) occurred
in a later storm over the north-central part of the TASD, but at the same plan
distances (16--17 km) from the INTF. Both were relatively weak in comparison
with TGF~A, with total surface detections of 112 and 212 VEM, respectively
(Supp.\ Table~S1 and Figs.\ S2 and S3).  The parent flash of TGF~B was
similar to that of TGF~A in terms of its initiation altitude ($\simeq$3.9~km
AGL, 5.3~km MSL) and average leader speed ($1.5 \times 10^6$~m/s).  The
gamma bursts began 0.65~ms after flash start, again during the strongest
initial breakdown pulse of the flash, whose peak current was as strong as that
of TGF~A (--30~kA).  However, instead of the SD waveforms having a common
onset time, as for TGF~A, the onset times varied noticeably at different sets
of TASDs.  In addition, the overall footprint of the TGF was annular-shaped
around a central hole (Supp.\ Fig.\ S2). The LMA and NLDN observations
indicate the burst's source was over the western side of the footprint,
adjacent to the hole. The initial burst was detected only at a single
station, SD~1421 immediately northeast of the source.  The primary onset
occurred 2--3~$\mu$s later, and was detected at four adjacent stations
2--3~km to the east on the opposite side of the hole (SDs~1521, 1520, and
1621, 1620).  This was followed by the two southern stations having an
additionally delayed onset (SDs 1519 and 1419), and finally a fourth onset
back at the western-most station, almost directly below the source (SD~1720).

Concerning the correlation with the INTF and FA data for TGF~B, the early
gamma-ray detection at TASD 1421 coincided with a prominent sub-pulse of the
IBP, and represents a separate onset time.  The sub-pulse occurred during an
apparently brief interlude of upward rather than downward development of the
VHF radiation sources.  Subsequently, the gamma-ray activity occurred during
downward fast negative breakdown having a propagation speed of $2.7\times
10^7$~m/s, with the primary onset time coinciding with the main sferic peak.
Less than a microsecond after the peak, the elevation centroids exhibited a
20--30~m step discontinuity similar to that seen during TGF~A, which appeared
to initiate the bursts detected at the southern TASDs.

The parent flash of TGF~C occurred 2.5 minutes later in essentially the same
location as TGF~B, and produced two gamma bursts 117~$\mu$s apart in time,
similar to TGF~A. In contrast with TGF~B, both bursts were relatively simple
and provide canonical examples of the basic processes of TGF production.  For
each event the gamma radiation was downward-directed and detected immediately
below and north of the source (Supp.\ Figs.\ S3 and S12).  The first
event was weaker and produced a total of 35~VEM (72~MeV) at four adjacent
TASDs below the source.  Figure~\ref{fig:all_tgfs}c focuses on the second
event, which was stronger and produced a total of 212~VEM (434~MeV) at nine
adjacent stations below the source.  As seen in Figures~\ref{fig:all_tgfs}c and
S12d, the parent IBP was temporally isolated from preceding and subsequent activity, and a sudden
increase of the VHF radiation signaled the onset of downward negative
breakdown and the IBP sferic.  The breakdown descended $\simeq$120~m in
4.7~$\mu$s at a steady rate $2.6\times 10^7$~m/s, indicative of FNB.  In this
simple case, the gamma radiation began immediately after the start of the
FNB and continued with varying but generally increasing intensity
through the entire descent until the breakdown ceased. In the process, several
unresolved sub-pulses occurred, similar to the sub-pulses of TGF~A. Also seen in other IBPs but more clearly shown in this flash, onset
of the FNB was immediately preceded by brief upward-developing
VHF sources, indicative of characteristic FPB breakdown that appeared to trigger the downward FNB.

TGF~D (Figures~\ref{fig:all_tgfs}d and \ref{fig:newfig}d) occurred during a
nocturnal storm on October~3 in a similar southward direction as TGF~A, but
further to the south at 24~km plan distance over the southeastern corner of
the TASD (Supp.\ Fig.\ S4 and S13). Again, the flash produced two triggers,
the first of which contained three weak gamma bursts that were partially
outside the southern boundary.  The second trigger and burst  occurred
140~$\mu$s later, $\simeq$800~$\mu$s after the flash start.  Its footprint was
shifted about 2~km northward from that of the first burst, placing it entirely
inside the TASD.  The apparent source of the bursts was on the eastern part of
the overlapping region between the two footprints (Supp.\ Fig.\ S13e). The
first burst was therefore beamed southwestward from its source and the second
burst was beamed northwestward.  The westward component of the beaming is clearly evident in the INTF observations of Figure~\ref{fig:newfig}, which showed
an increasingly strong WNW-ward tilt of the azimuthal locations as the
breakdown descended, with the tilt angle becoming as large as $45^\circ$ from
vertical by the time of the gamma burst.  A total of 440~VEM (962~MeV) was detected
at 12 stations during the second burst, compared to a partial total of 100~VEM
(205~MeV) at 9~stations during the first burst.

Concerning the second trigger and main burst of TGF~D, the IBP of the burst
had a complex, relatively long-duration (15~$\mu$s) sferic waveform that was
accompanied by steady downward development of the VHF radiation sources.
Overall, the breakdown descended $\simeq$240~m in 13.4~$\mu$s at an average
rate of $1.8 \times 10^7$~m/s.  The gamma burst was initiated partway through
the descent, coincident with a major sub-pulse and the onset of increased VHF
radiation.  The sequence of events is similar to that of TGF~C in that the
radiation increase and corresponding sub-pulse was preceded by a brief
interval of fast upward positive breakdown. The ensuing fast downward activity
exhibited a small step discontinuity in the VHF centroids that coincided with
the onset of the gamma burst and sub-pulse. As in each of the other TGF
flashes, the gamma radiation continued up until the approximate end of the
FNB, shortly after the main negative peak of the IBP sferic.

\section{Discussion}

\subsection{Observational Results}

The results of this study demonstrate that TGFs are produced
during strong initial breakdown pulses (IBPs) in the beginning stages of
negative-polarity breakdown.  This is shown with a high degree of temporal and
spatial resolution provided by a unique combination of a state-of-the-art
cosmic-ray facility, coupled with high-quality VHF and LF sferic observations
of the parent lightning discharges.  In addition to showing how TGFs are
related to IBPs, the observations reveal how the initial breakdown pulses
themselves are produced, which has remained unknown for over 50 years. In
particular, IBPs are produced by a recently-identified type of discharge
process called fast negative breakdown (FNB)~\cite{tilles2019a}.  FNB is the
negative-polarity analog of fast positive breakdown that has been identified
as the cause of high-power narrow bipolar events (NBEs), and which is
instrumental in initiating lightning~\cite{rison2016}.  Both polarities of
fast breakdown propagate at speeds around 1/10 the speed of light, with FPB
sometimes reaching $(1/3)\,\!c$. FPB is understood to be produced by a system
of propagating positive streamers that, when occurring at the beginning of a
flash, is initiated by corona from ice hydrometeors in a locally strong
electric field region inside storms~(\citeA{rison2016}; \citeA{attanasio2019}).


Although the nature of fast negative breakdown is uncertain~\cite{tilles2019a},
its similarities with FPB strongly suggest that FNB is also streamer-based,
except for being of negative polarity.  Independent of polarity or
direction, both positive and negative fast streamer systems would
significantly enhance the ambient electric field ahead of their advancing
front~\cite{attanasio2019}, facilitating the development of high energy
electron avalanches necessary for gamma-ray production. 

Owing to its simplicity, TGF~C provides a canonical example of the basic
processes involved during an IBP.  In particular, the IBP of TGF~C was
initiated by a brief (1--2~$\mu$s) interval of fast upward positive breakdown,
immediately followed by a sudden increase in the VHF radiation and the onset
of oppositely-directed downward FNB (Figures~4c and S8).  The positive breakdown
began slightly beyond the lowest extent of the preceding negative breakdown
and propagated weakly but rapidly back into preceding activity, whereupon it
initiated oppositely-directed and VHF-strong FNB back down and beyond the path
of the upward FPB, extending the negative breakdown to lower altitude (see
also Fig.\ S12d,g). Similar sequences of upward positive/downward negative
breakdown were associated with TGF-producing IBPs of the other flashes,
including a preceding, weaker gamma-ray event of TGF~C (Fig.\ S12c,f).

The TGF observations show that the onset of the electron avalanching and
gamma-ray production occurred at various stages during the IBPs. For TGF~A, the
onset occurred after the sferic peak, but during still-continuing FNB. TGF~C
occurred at or shortly after the beginning of its IBP and FNB onset. For the
more complex discharges of TGFs~B and D, the onset was often associated with
leading-edge sub-pulses that are a characteristic feature of classic
IBPs~\cite{weidman1979,nag2009,karunarathne2014}.  Like IBPs, the nature and
cause of sub-pulses has continued to be a mystery (e.g.,
\citeA{dasilva2015,stolzenburg2016}).  The results of the present study show
that the main driving force of the IBPs is fast negative breakdown, which has
the sub-pulses as embedded components.  Basically, the sub-pulses are
indicative of repeated breakdown events within the developing IBP discharge.
The observation that TGFs are often associated with sub-pulses,
and that this occurs during fast negative streamer breakdown, provides a possible
explanation for the sub-pulses' occurrence. Namely, that they are produced by
spark-like transient conducting events (TCEs) embedded within the negative
streamer system. That the events are spark-like is indicated by the pointed,
cusp-like nature of their sferics, evidence of a sudden current onset and
rapid turnoff, and also by the sub-pulses repeating several times
as the IBP progresses. It should be noted that the final peak of the overall IBP
sferic is also cusp-like, indicating that it too is produced by a spark-like
sub-pulse. 

Once initiated, the gamma radiation typically lasts $\simeq$3 to 5~$\mu$s for
the flashes of this study. GEANT4 simulations presented in Figure~S24 of the
Supporting Information show that multipath Compton scattering does not
artificially extend the duration, as 95\% of detectable particles produced by 10~MeV (100~MeV) photons at 3~km AGL will arrive within 20~ns (60~ns). The total energy available for deposit after the first 100~ns is small enough to be indistinguishable from background levels, thus the observed durations reflect the intrinsic duration of the sources. An important implication of this result is that relativistic avalanching lasting 3--5~$\mu$s would propagate a distance of $\simeq$1--1.5~km, substantially beyond the 100--200~m extent of the FNB and IBP. This would provide the electron avalanches with additional amounts of electric potential energy until the ambient electric field drops below the threshold for avalanche propagation ($\simeq 2 \times 10^5$~ V/m)~\cite{dwyer2003}.

Before proceeding, we emphasize the fact that the TASD is detecting multi-MeV
gamma radiation from the lightning discharges, and not lower energy
$x$-radiation.  We repeat here the simple arguments for this, presented by Abbasi~et~al. (2018) and based on the well-understood physics of Compton electron production and the well-calibrated TASD response to minimum-ionizing charged particles.  In particular, TASD responses for the events of the present and earlier studies (e.g.
Supplemental Figure~S3) can clearly be resolved into individual
minimum-ionizing Compton electrons that result in the deposit of approximately 2.4~MeV into
either the upper or lower scintillator plane, or in correlated deposits into both planes. A property of particles above the minimum-ionization threshold is that higher-energy particles would still deposit only 2.4~MeV per plane~\cite{pdg2020}. Thus, the TASD cannot determine the maximum energy of Compton electrons, but it can place a lower limit on the energy values. Compton
electrons that deposit 2.4~MeV into one plane are produced by a photon with
no less than 2.6~MeV (Supplemental Figure~S9 of Abbasi 2018). 
Electrons that deposit 2.4~MeV into both planes, and also traverse the
1~mm steel separating sheet, have a total energy loss of 6.2~MeV and must be
produced by photons with a minimum energy of 6.4~MeV.

The above inferred photon energies should be interpreted as minimal values, as they assume that the Compton electrons are produced by head-on collisions in which the gamma ray is backscattered and transfers the maximum amount of energy to the electron. The likely contributions of grazing incidence collisions to our signal would imply the actual photon energies are several times higher, depending on the grazing angle (Supplemental Figure~S10 of \citeA{abbasi2018}).  Even for single-scintillator layer detections, these are comparable to the average 7--8~MeV energy of relativistic runaway spectra detected by satellites.  In any case, there is no question that the TASDs are detecting multi-MeV gamma-rays.




\subsection{Extension to intracloud flashes}

\begin{figure}[t]
\begin{center}
\includegraphics[width=0.85\linewidth]{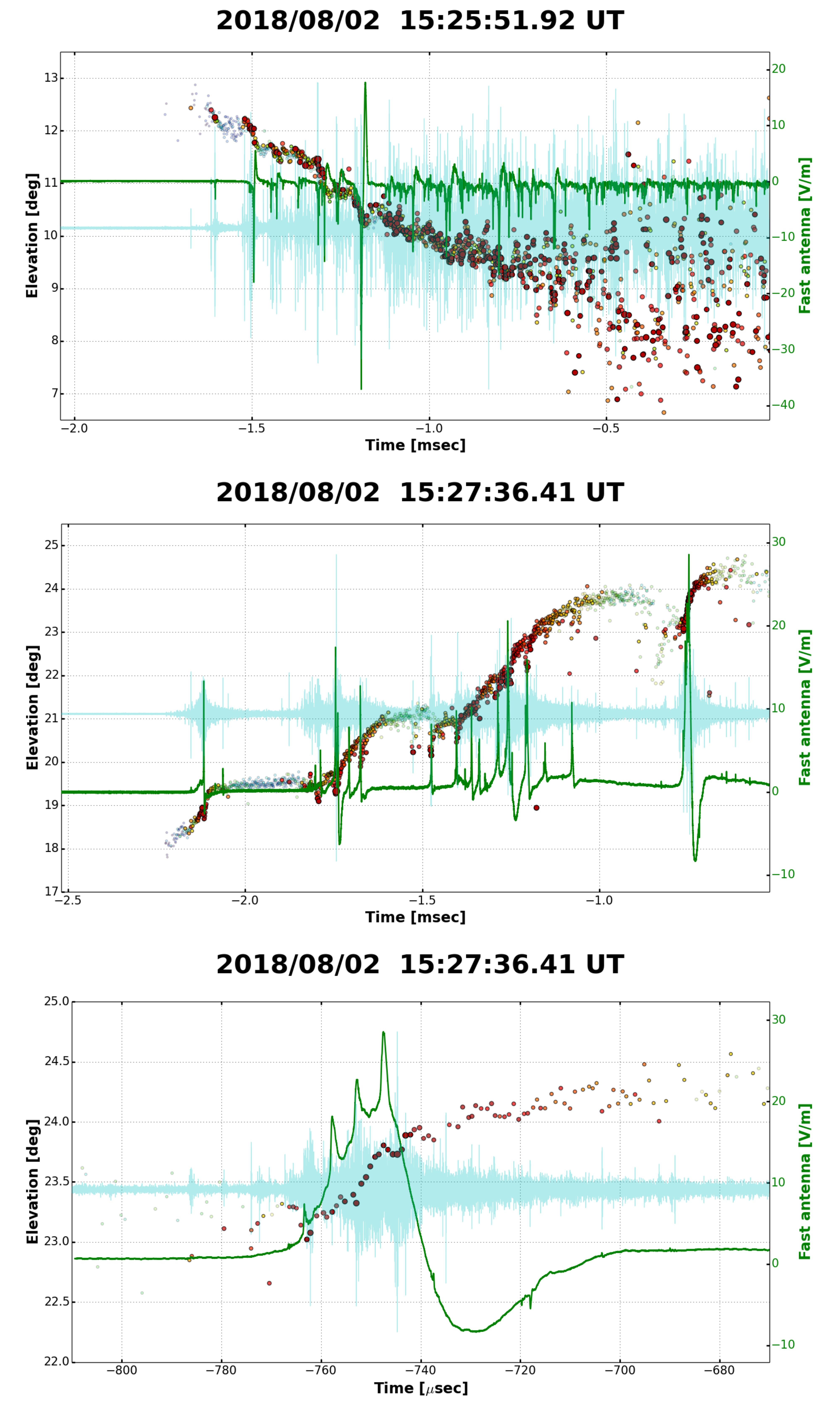}
\caption{\label{fig:IC_compare}
{\normalfont Comparison of the  --CG flash that produced TGF~C with the
IC flash that was the next flash in the storm, illustrating the differences
and similarities between the two types of flashes. Top two panels show 2~ms
of observations for the downward  --CG and upward IC.  Bottom panel shows an expanded view of the large IBP near the end of
the IC interval which, except for polarity and overall duration, is
basically identical to the IBP that produced TGF~B three flashes earlier
in the storm. The propagation speed of the upward FNB is also similar,
being $\simeq1.5\times 10^7$~m/s.}}
\end{center} 
\end{figure}

Although obtained for downward negative breakdown at the beginning of --CG and
low-altitude IC flashes, the results apply equally well to upward negative
breakdown at the beginning of normal-polarity IC flashes at higher altitudes in storms.
Figure~\ref{fig:IC_compare} compares INTF and FA observations of the --CG flash of TGF C with those of an IC 
flash that was the next lightning discharge in the
storm (see Figs.\ S27--S29 for additional observations of the flashes).  The top
two panels show 2~ms of data for the two flashes with time scales of
500~$\mu$s/division. The bottom panel shows an expanded view of the
large-amplitude classic IBP near the end of the IC interval.  Taken together,
the plots illustrate the differences and similarities of the initial breakdown
processes of IC and --CG flashes.  In particular, and as has long been known
(e.g., Kitagawa and Brook, 1960; Weidman and Krider, 1979), the downward
negative breakdown of --CG flashes intensifies more rapidly and continuously
than the negative breakdown of upward IC flashes.  The difference is clearly
seen in the top two panels and is due to a combination of effects: first, the IC flashes needing to propagate through a
relatively large vertical extent of quasi-neutral charge before reaching upper
positive storm charge, compared with little or no spacing of the lower positive
charge during --CG flashes (e.g., Fig.\ 1 of \citeA{krehbiel2008}, and
Fig.\ 3 of \citeA{dasilva2015}), and secondly the IC discharges occurring
at reduced pressure.  The overall result is that IC flashes develop more
intermittently and with longer stepping lengths than --CG flashes (e.g., Edens, 2014).

Despite the intensification differences, individual initial breakdown pulses
of IC flashes exhibit the same features as those of --CG flashes.  In
both instances, classic IBP sferics consist of an initial strong electric
field change having embedded sub-pulses, followed by a characteristically
large and relatively slow opposite-polarity field change.  The
similarity is illustrated by comparing an expanded plot (bottom panel of Figure~\ref{fig:IC_compare}) of the large-amplitude IBP at the end of the middle panel with that of TGF~B seen in Figures~\ref{fig:all_tgfs}b and S16, which
occurred in the same storm $\simeq$4~min earlier, three flashes before the
IC flash. Except for polarity, the sferics are virtually identical. More
importantly, the INTF data shows both are produced in the same manner, namely
by fast negative breakdown.  Owing to the increased stepping distance, 
IC IBPs tend to have longer durations than those of --CGs; lasting
$\simeq$70~$\mu$s for the IC IBP vs.\ $\simeq$35~$\mu$s for the IBP of TGF~B.
The fast negative breakdown component of the IC IBPs is also similarly longer,
being $\simeq$20~$\mu$s for the IC vs.\ $\simeq$10~$\mu$s for TGF~B.  The
factor of two overall duration difference agrees with the study by Smith et
al. (2018) of median durations of large IBP sferics in Florida storms. 
Another example of a similar classic IC IBP sferic is seen in Fig.\ 4 of
the study of Florida IBPs by \citeA{marshall2013}, which had a duration
of $\simeq$100~$\mu$s and was considered to be a `candidate' TGF flash.
At this point it should be noted that in many instances the durations of IC
and CG IBPs are the same for both types of flashes.  This is seen in the
scatter diagram of Figure~5 of \citeA{smith2018}, and is shown in detail by
the comprehensive observations of \citeA{tilles2020}.  Figures~9.3 and 9.4 of
the latter study, conducted in Florida with the same INTF and FA
instrumentation as in the present Utah study, show that (except for polarity)
the IC and --CG IBPs were essentially indistinguishable both in terms of their
sferics and durations.

\clearpage

\begin{figure}[t]
\begin{center}
\includegraphics[width=0.9\linewidth]{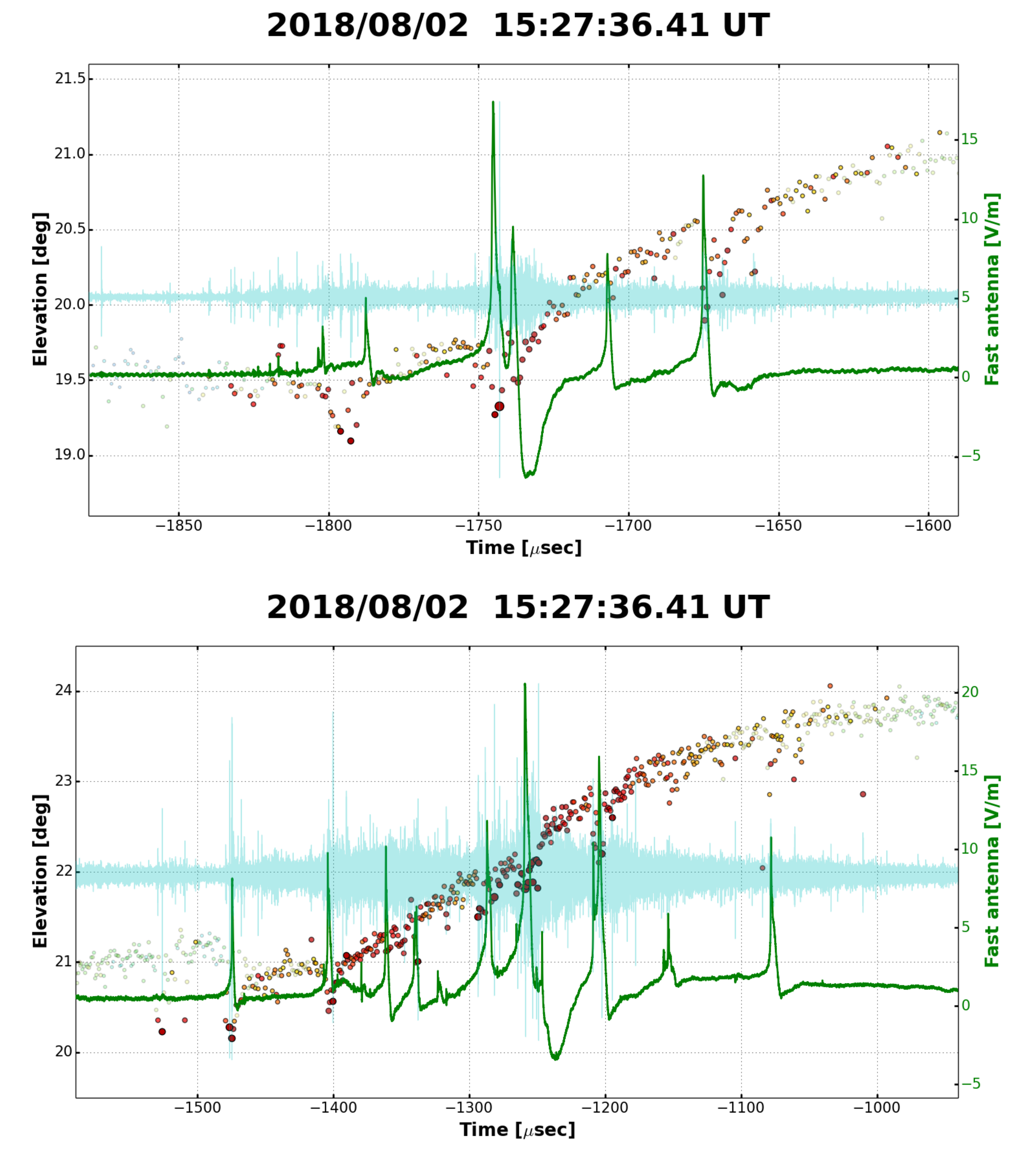}
\caption{\label{fig:IC_IBPs}
{\normalfont Expanded views of the complex IBP clusters of the IC flash
of Figure 6, showing the increased number and highly-impulsive nature of
the sub-pulses.  The FNB breakdown of the IBPs and the sub-pulses are
each embedded in continuous upward negative streamer breakdown having a
propagation speed of $\simeq$2--3$\times 10^6$~m/s, showing that negative
streamer breakdown doesn't have to travel at speeds of $10^7$~m/s to
produce the sub-pulse sparks.  The durations of the two clusters were
$\simeq$130 and 400~$\mu$s, respectively, with the sferic of the first
cluster resembling that of the TGF-producing IBP of Figs.\ 2 of ~\citeA{lyu2018a,pu2019}, and the second cluster resembling the sferic
of another complex TGF-producing sferic of Pu et al. }} 
\end{center} 
\end{figure} 
\clearpage

Due to the TGF-producing storms having low flashing rates (typically 1--2~min
between flashes in the present study), the electrification is allowed to build up to large values,
causing both the --CG and IC flashes to be highly energetic when they
finally occur.  For the IC flash of Figure~\ref{fig:IC_compare}, this is
reflected not only in the amplitude and duration of the classic IBP, but
also by the preceding activity being produced by two complex sequences (clusters) of IBPs and sub-pulses, seen in the middle panel.  Each of the 
clusters is linked together by continuous, upward-developing high power negative breakdown,
producing long-duration complex steps.  The overall durations of the two clusters were
$\simeq$130 and 400~$\mu$s, respectively.  Expanded views of the complex
IBPs are seen in Figure~\ref{fig:IC_IBPs}, which show the sferics were
dominated by increasing numbers of sub-pulses that assisted in
continuing the negative breakdown and extending the cluster durations.  In
addition to their increased numbers, the sub-pulses are dramatically more impulsive
and stronger in amplitude than those of the --CG flashes.  The
IC sub-pulses had amplitudes of $\simeq$10--20~V/m, compared to
$\simeq$5--10~V/m for the sub-pulses (at essentially the same distances) of the TGF-producing IBPs of Figure~\ref{fig:all_tgfs} (seen in larger scale in Supporting Figs.\ S15--S18).
Given that the simpler IBPs of the --CG flashes produced TGFs, the IC flash would likely have been
equally or more capable of generating upward TGFs.  Due to relativistic avalanching being a
strong function of the potential difference being shorted out by the
spark-like sub-pulses~\cite{celestin2015}, as well as the sub-pulses being
more dynamic~\cite{celestin2012b} and repetitively impulsive, the
resulting avalanching and TGFs would be more energetic, as well as longer lasting.
Similar observations were obtained for an IC flash that occurred between
TGFs~B and C, which are compared with TGF~B in Figs.\ S24--S26.

\subsection{Implications for TGF production mechanisms}

As summarized in the recent modeling study of TGFs by \citeA{mailyan2019},
there are two classes of models for TGF production: First, what is
termed the relativistic runaway electron avalanche (RREA) or relativistic
feedback (RFD) model, in which electron avalanches develop in km-scale regions of
strong electric fields in storms~\cite{dwyer2003}. In this model, the
avalanching is enhanced by relativistic feedback that increases the avalanche
currents by several orders of magnitude~\cite{dwyer2012a}.  The second class
is broadly termed the `leader' model, in which the relativistic avalanches are
initiated in the highly concentrated electric field produced at the negative
tip of a conducting leader channel.  The electric field at the tip is
extremely strong as a result of the leader having kilometer-scale extents and
shorting out tens to a few hundred MV of potential difference in the storm.
Whereas the RREA process by itself requires cosmic ray-produced or other seed
relativistic electrons to get started, the leader process begins with low
energy thermal electrons, and requires exceedingly large electric fields
($\simeq3\times 10^7$~V/m --- an order of magnitude larger than the breakdown
strength of air) to be accelerated into the runaway electron regime, where
their number and energy increases exponentially with time and distance~(e.g.,
\citeA{dwyer2004}).  Electric fields of this strength are produced only 
at the tips of conducting leader-type channels, and then only transiently during
rapid channel development.  Thermal electrons are accelerated into the
relativistic regime as a result of transient negative streamers within the
strong E region (the so-called `negative corona flash'), as described by \citeA{moss2006}, \citeA{celestin2011}, and \citeA{celestin2015}.  Once the leader/streamer-initiated avalanches are started they would be able to initiate the relativistic
feedback process.

While relativistic feedback can explain the large currents and fluxes of highly
energetic satellite-detected events, it does not appear to be playing a role
in initiating the smaller-scale observations of the present study. Instead,
the inference that IBP sub-pulses are caused by spark-like transient
discharges embedded within the fast negative streamer system points to the
leader/streamer model as playing an important and possibly dominant role in generating
runaway avalanches and TGFs.  Once initiated, the runaway electrons would additionally increase in energy while propagating through the enhanced field region ahead of and beyond
the relatively broad streamer front~\cite{attanasio2019}.

An important question is whether the conducting channels of the sub-pulses
(which we refer to as transient conducting events, or TCEs) are isolated
within the negative streamer system and from each other, or if they are
connected back into, or originate from, the conducting channel of the incoming
negative leader.  If so connected, the potential drop beyond the negative
tip of the sub-pulse channel would be comparable to the amount shorted out by
the km-long or longer leader, envisioned to be as large as 60 to 200~MV or
more (e.g., \citeA{celestin2015,mailyan2019}).  Such a leader is termed a
`high potential' leader, which by itself can produce the large
($\simeq10^{16}$--$10^{18}$) gamma photon fluxes inferred by satellite
observations \cite{celestin2015}.

To address the question of the sub-pulse connectivity, we note that the
sub-pulses continue to occur until one suddenly causes the IBP sferic to begin
transitioning to an opposite-polarity field change during the final part of
the IBP.  Although the flash current does not change direction, the electric
field waveform becomes dominated by the electrostatic and induction
components, which are inverted in polarity from the radiation component due to
the flash being beyond the reversal distance $d$ for vertical dipolar
discharges, where $d = \sqrt{2}\,h$ and $h$ is the discharge height above
ground level (e.g., \citeA{macgorman1998}).  At the same time, the fast negative breakdown continues to propagate
for several microseconds before finally dying out.  From the large amplitude and
relatively long duration of the opposite-polarity field change, one can infer
that the current is not constrained to the IBP itself but develops
retrogressively back through the negative breakdown leading to the IBP,
converting a potentially weak streamer-leader channel to a hot conducting
leader and completing the step.  That the current during a negative leader
step develops in a retrograde manner back along the incoming breakdown channel
has been shown by in-situ balloon-borne observations of negative leader
stepping during an IC flash by \citeA{winn2011}, and by high speed video
observations around the time of IBPs of --CG flashes by \citeA{stolzenburg2013}, as discussed later.  Because sub-pulses previous to the final
sub-pulse do not initiate the opposite polarity field change, one can
infer they are not connected to the incoming leader breakdown, but instead are
isolated from the leader and from each other.  The question then becomes
whether the sub-pulse sparks short out enough potential difference to account
for the observed TGFs.


In terms of the space stem/space leader model of negative leader stepping
(e.g., \citeA{petersen2008,biagi2010}), the sub-pulse sparks would
correspond to conducting space leaders that occur in the negative streamer
region ahead of the developing leader.  Continuing the space leader
interpretation, the final sub-pulse develops back into the incoming
leader, at which point the leader's potential rapidly advances to the opposite
end of the space leader, producing the negative corona flash that launches the
relativistic electrons.  This scenario could explain TGF~A, which was
initiated a few microseconds after the final, sharply-pointed negative peak of the
sferic (Figs.~4a and S15). TGF~A also produced the most surface-detected
energy of the different TGFs (561~VEM total, or 1150 MeV; Table~S1).
Because the TGF occurred just above the TASD boundary (Figs.~2 and S1),
the detected energy could have been up to 50\% larger had it been entirely captured. Similarly, the scenario could also explain the main onset of
TGF~B, which occurred at the same time as the final sub-pulse peak
(solid vertical line in Figs.~4b and S16).

For TGFs~C and D, however, and for the early initial detection of TGF~B, the
TGF onsets were associated with sub-pulses that did not initiate a
retrograde current (Figs.~4, S17, S18, and the left-most vertical dotted line in Fig.~S16).
These and the other early sub-pulses of the IBPs would be characterized as
attempted space leaders, and may have somehow paved the way for the final
sub-pulse, but otherwise appeared to be independent of each other and not
connected back to an incoming leader.  The gamma events of TGFs~C and D had
total surface detections of 212 and 440 VEM (434 and 902 MeV), respectively,
with TGF~D being the second strongest TGF after TGF~A.  At the same time, the
total activity of TGF~B, which was most closely associated with the IBP's
final sub-pulse and presumably the best candidate for being connected to the
incoming leader, had the weakest total surface detection of all, 112 VEM (229 MeV).

Storm-to-storm variability, as well as that from flash to flash in the same storm, coupled with the
small sample size makes it difficult to compare the different observations.
However, the fact that three TGF events (C, D, and the initial lone detection of TGF~B) were initiated by sub-pulses that
did not connect back into the incoming breakdown of the IBP, and the subsequent activity of TGF~B producing a weak TGF despite its sub-pulse eventually connecting back into the incoming breakdown, indicates that the
occurrence and strength of the gamma bursts are determined more by the
amplitude and impulsiveness of the initiating sub-pulse rather than by the
incoming breakdown consisting of a hot conducting leader.


From the above results, as well as the IBPs being produced by fast
negative streamer breakdown, the sub-pulses are analogous to the space leader
in negative leader stepping in that they occur within negative streamers
ahead of the leader. Instead of being produced by a relatively slow-developing
thermal space stem, the sub-pulses are impulsive sparks caused by sudden
instabilities in extended-length streamer channels associated with fast
propagation speed of streamers.  And instead of the impulsivity of the step
being produced by the space leader suddenly contacting a conducting leader
channel and rapidly propagating the leader potential forward to the head of
the space leader, the impulsiveness and negative corona burst is produced by
the spark itself.  The succession of sub-pulse sparks eventually causes one to
develop back into a somewhat diffuse leader, giving rise to the
backward-developing current that further establishes and converts the
incoming breakdown into a well-defined hot conducting channel. This scenario
agrees with high-speed video observations by \citeA{stolzenburg2013,stolzenburg2014},
indicating that the `unusual' steps of IBPs occur ahead of a weakly-conducting
nascent leader rather than a continuously hot, conducting channel (see later discussion).



If the space stem/space leader process is what initially advances the
conducting leader channel, a legitimate question concerns how such a hot
leader is produced in propagating from the end of the preceding IBP (or from the
flash start) to the beginning of the IBP in question, in the absence of
discernible space stem/space leader activity.  At some point the leader
becomes self-propagating (e.g., \citeA{dasilva2019}), but apparently this does not occur in the early
stages of the breakdown, as evidenced by the increasing need for and strength
of IBPs in the initial few milliseconds of negative breakdown.  Up until then, the
advancing negative breakdown between IBPs appears to be a system
of relatively weakly conducting negative streamers, which can self-propagate
more readily.

From the INTF observations, the average speed of the downward negative
breakdown at the beginning of the TGF-producing flashes is
$\simeq$1.0--2.5~$\times 10^6$~m/s (e.g., Figure~\ref{fig:overlay3}a and
Supporting Figures~S7--S13), an order of magnitude or so faster than other
estimates of developing leader speeds (e.g., \citeA{behnke2005}).  Similarly
fast progression speeds were reported during the upward development of
TGF-producing IC discharges by~\citeA{cummer2015}, who used ionospheric
reflections to determine the altitude and hence the upward progression speed
of successive radio pulses of TGF-producing IC flashes.  For three different
flashes, the speeds were noted to be remarkably similar and fast, ranging from
0.8--1.0$\times 10^6$~m/s.  As in the present study, the TGFs were produced
partway along the vertical development (in their case upward), when the leader
was $\simeq$1--2~km in extent.  The fact that TGFs were not also produced by
subsequent pulses at higher altitude during the vertical development led them
to ask why this did not happen, in view of the leader lengths being
proportionally longer.  A similar question would apply to the present,
downward-directed observations at the beginning of the --CGs.



Taken together, the results suggest a scenario in which a `step' consists of
a) intermediate-speed negative streamer breakdown being launched at the end of
the previous step's IBP, which progresses in a forward direction until b)
initiating accelerated-speed FNB and an IBP having embedded sub-pulses, one of
which c) initiates a strong current that develops retrogressively backward
through the IBP and its preceding negative breakdown, thermalizing and
extending the negative leader.  The IBP then reverts back to intermediate or
slower-speed negative streamer breakdown, beginning the next step.  Whether a
TGF is produced during the IBP is largely decoupled from the preceding negative
breakdown, explaining the independence of TGF production on the extent of the
negative breakdown up to that point. Where the preceding extent plays a role
is in enhancing the electric field ahead of its developing front, to the point
that the FNB is initiated. The field enhancement is due to the cumulative
dipolar charge transfer of the negative breakdown during each step (e.g.,
\citeA{krehbiel2018,attanasio2019,cummer2020}), causing successive IBPs to
become stronger with time. The TGFs of this study were produced
by the strongest IBP of the flash, but in 3 of the 4 flashes one or two
additional bursts occurred that were associated with separate episodes of FNB
and sub-pulse activity (see Figs.\ S10d,g, S12c,f and S13c,f).  The additional
gamma events occurred during less strong IBPs within $\simeq$100--150~$\mu$s
either before or after the main gamma events, and represent sparsified
examples of the TGF activity that would be expected during the kind of complex
IC IBP events seen in Figures~\ref{fig:IC_compare} and \ref{fig:IC_IBPs}.


The above scenario for the stepping provides an explanation for the optical
observations of \citeA{stolzenburg2013}, in which partially-obscured
luminosity in the first 1--2~ms of a --CG flash advanced downward with a
series of surges associated with bright optical emissions at the times of
successive IBPs. The observations were obtained from high speed video
recordings having 20~$\mu$s time resolution. Each bright surge lasted about
80--100~$\mu$s and was preceded by dim, linearly downward extension of the
channel, with the brightest frame ``immediately followed by backward lighting
of the entire tail'' that preceded the bright surge. The sequence then started
over again with renewed dim downward extension of the channel to a lower
elevation angle, with the process repeating for up to five surges.  In terms
of the above scenario, a) the linear downward channel extensions would
correspond to the intermediate-speed, inter-IBP negative streamer activity, b)
the succeeding bright optical emissions would have been produced by the
spark-like sub-pulses of the IBP, and c) the immediately following upward
propagating light would be produced by the retrograde current traveling back up
along the path of the pre-IBP activity, converting it into a hot conducting
leader.  As noted earlier, \citeA{winn2011} observed similar backward
propagating current events following individual steps of an already-developed
negative leader toward the end an IC flash, using close balloon-borne electric
field change observations of the flash. The correlation of bright optical
pulses with --CG IBPs was extended by \citeA{stolzenburg2016} to be produced
by IC-type IBPs at the beginning of hybrid --CG flashes.  Similar to
\citeA{marshall2013}, the IBPs were considered to be candidate producers of
TGFs, on the basis of the IBPs being complex and having strong sub-pulses.

The mechanism for producing the spark-like sub-pulses and TCEs within the fast
negative breakdown would be essentially the same as that which causes the FPB
and FNB to be the producer of high-power VHF radiation, described as being the
strongest natural source of VHF radiation on Earth~\cite{levine1980}.  Due to
their fast propagation speed, both polarities of streamers would have extended
partially-conducting tails that would become unstable in the strong ambient
fields~\cite{shi2016,malagon2019}. The resulting rapid current cutoff, coupled
with meters-long extents and large numbers, make both polarities of
streamer systems potent radiators at VHF~\cite{rison2016}.  The negative
polarity streamers of FNB would have more robust and extensive tails than
positive streamers, that could occasionally extend over longer distances, with
the resulting instabilities and currents producing hot, spark-like conducting
channels of the sub-pulse TCEs.  In addition to explaining the optical
emissions associated with IBPs, the sudden occurrence of a dynamically
impulsive conducting channel would provide the means for initiating
relativistic electron avalanches~\cite{moss2006,celestin2012,celestin2015}.


It is interesting to note that, in addition to being produced by sub-pulses,
it may also be possible for relativistic electron avalanches to be initiated
by individual negative streamers themselves. This is suggested by the modeling
study of \citeA{moss2006}, who showed that the extremely strong electric
fields sufficient to accelerate electrons into the runaway regime will occur
briefly immediately prior to branching of the streamers.  Electrons produced
in association with branching can reach kinetic energies as large as 2--8~keV
or larger, well into the runaway electron regime. Although determined to occur
in the corona flash and streamer zone at the tip of a conducting leader, the
process might also occur at the tips of streamers having relatively long
conducting tails. The branching process was noted to strongly favor negative
streamers over positive, due to positive streamers requiring photoionization to
sustain their propagation.  If it occurs, the branching mechanism would be a
powerful adjunct to TCEs, since large numbers of individual streamers exist
within a propagating system that are spread over a much larger cross-sectional
area than an individual conducting leader or TCE channel, and are continually
branching.

Other issues of note concerning the observations are a) that the TGFs are
broadly rather than narrowly beamed, favoring a tip-based conducting channel
model~\cite{mailyan2019}, and b) are commonly tilted at substantial angles
from vertical.  From the TASD footprints and source altitudes, the half
angular width of the beaming is on the order of 35$^\circ$ or so ($\simeq$2.4
km radial plan spread for a 3.3~km source altitude).  From the INTF observations of
Figure~\ref{fig:newfig} (repeated in larger scale in Supporting Figs.\
S19--S23), the tilting can be 45$^\circ$ or more, depending on the
3-dimensional development of the discharge. Finally, successive sub-pulses can
be oriented in different directions, as indicated by successive onsets
occurring in different directions for TGF~B (Figs.\ S16 and S20).





We note that the simulations of our previous study~\cite{abbasi2018} implied
TGF fluences on the order of $10^{12}$--$10^{14}$ relativistically-generated
gamma photons, several orders of magnitude less than satellite-inferred
fluences of $\simeq\!10^{16}$--$10^{18}$ photons.  From~\citeA{celestin2015}
(Table~1), total fluences of $10^{12}$--$10^{14}$ photons correspond to
potential drops of $\simeq$10 to 50~MV or so at the conducting channel tips,
while fluences of $\simeq\!10^{16}$--$10^{18}$ photons correspond to larger
potential drops of 160--300~MV.  That the observed fluences are relatively weak
would be consistent with the inference that the TGFs are produced by isolated
conducting sparks that short out lesser amounts of potential difference.
However, if km-long conducting leaders are not involved, the question is
whether sufficient potential difference is available for producing the
relativistic electrons and the observed gamma radiation.  For example, from
\citeA{celestin2015} (Fig.~3), 5--10~MV potential drops would not produce
relativistic electrons greater than $\simeq$1--2~MeV. On the other hand, 60~MV
(160~MV) of potential drop would produce relativistic
electrons up to 9~MeV (20~MeV). From the modeling, then, at least 50~MV of potential drop would be required
to produce the expected gamma energies observed in
this study. The predicted fluences corresponding to 60~MV (160~MV) potential
drop, however, is $\simeq 6\times 10^{14}$ ($\simeq 4\times 10^{16}$) photons, 
two orders of magnitude greater than the inferred fluences of these
TGFs.  Thus the observations are inconsistent with the leader-streamer
modeling, in that the fluences corresponding even to the minimum likely
detected photon energy produced by 60~MV potential drop would be at the
upper end of the implied fluence values of \citeA{abbasi2018}. 






The question of available potential energy can be addressed by considering the
electric field required for streamer propagation, called the stability field
$E_{st}$.  From \citeA{dasilva2013}, at one atmosphere of
pressure $E_{st} \simeq$~$5\times 10^5$~V/m for positive streamers, but
$\simeq12.5\times 10^5$~V/m for negative streamers in virgin air.  The fields
scale according to pressure, so at 5~km altitude (0.5~atm) $E^-_{st} \simeq
6\times 10^5$~V/m. Thus FNB propagating over the 100--240~m long extents of
the TGF IBPs (Table S3) would experience total potential differences of
$\simeq$60 to 150~MV, with 60~MV being consistent with observed photon
energies up to $\simeq$9~MeV.  Some or all of the potential difference that
is not shorted out by the sparking would be available for additional avalanche
growth down to the propagation threshold of $2 \times 10^5$~V/m, which is not
accounted for in the \citeA{celestin2015} calculations.  Also not accounted
for are dynamical effects in initiating the relativistic electrons that are
associated with the sparking being impulsive, which are significant for pulsed
discharges (Section~5.4.3 of \citeA{nijdam2020}). Finally, using the stability
field values doesn't account for the field intensification ahead of the
advancing streamer front, which can be as much as 50\% above the ambient
$E_{st}$ value~(e.g., \citeA{attanasio2019,dasilva2019}).  For IC flashes
at higher altitudes, $E_{st}$ would be reduced by about another factor of two,
but this would be offset by the IC events typically being longer by a factor
of two or more, leaving the total potential differences about the
same.  Finally, we note that vertical profiles of the electric potential in
electrified storms similar to those being studied show the total potential
differences available for IC and --CG flashes are both on the order of 200~MV
(e.g., Fig.\ 1 of \citeA{krehbiel2008}; Fig.\ 3 of \citeA{dasilva2015}).

In short, while the details remain to be understood, taken together,
sufficient potential difference is available to produce gamma radiation into
the 10--20~MeV range or potentially higher, consistent with the observations
and the physics of the Surface Detector responses. The main issue is the
fluence values.  A possible explanation for the fluence inconsistency that
allows both the observational data and the modeling to be correct would be
that the gamma photons are produced by $\simeq$10 to 50~MV of potential drop,
which from Fig.~3 and Table~1 of \citeA{celestin2015} would produce
relativistic electron energies in the range of $\simeq$2--9~MeV and fluences
in the observed range of $10^{12}$--$10^{14}$ photons. Once initiated, the
electron energies would be further accelerated up to $\simeq$10--20~MeV by the
enhanced field ahead of the streamer front and any ambient field beyond
greater than the threshold field of $2\times 10^5$~V/m.  Because the extent of
the field ahead of the streamer system would be less than an e-folding
avalanche length, the fluences would not change significantly while the
electron energies increase.

To the extent that satellite-detected TGFs from IC flashes have substantially
larger fluences, the implication is either a) that the satellite detected
events emanate from the tips of fully-formed, kilometer-length or longer
conducting leaders, in which case fluences of $10^{16}$--$10^{18}$ photons
are achieved directly from the negative corona flash produced by potential
drops as large as several hundred MV, or b) that the fluences of lesser
potential drops are enhanced by the relativistic feedback process.  The
above-mentioned observations by~\citeA{cummer2015} raise the important
question about the leader hypothesis of why TGFs are not produced later in the
development of upward, kilometer or multi-km conducting leaders.  Instead, and
as additionally discussed below, the observational data supports the idea that
the much greater satellite-detected fluences are due to the relativistic
feedback mechanism, which was initially developed to explain this very issue
\cite{dwyer2012a}.

Another substantial difference between the present observations and those
obtained by satellites concerns the durations of the TGFs, being  5--10~$\mu$s
for the downward --CG TGFs, versus $\simeq$20--200 $\mu$s for the upward,
IC-generated TGFs ~(e.g., \citeA{mailyan2016,mailyan2018,ostgaard2019}).  The
difference can be at least partially explained by observations that IC flashes
can often have long-duration, complex sferics, consisting of multiple
sub-pulses and IBPs, each of which would be capable of producing TGFs.
Examples of such sferics are seen in Figures~\ref{fig:IC_compare} and
\ref{fig:IC_IBPs}. Of particular note are the observations of three TGF events
by \citeA{lyu2018a}, in which complex dB/dt events produced Fermi-detected
TGFs having continuous durations of $\simeq$50, 100, and 120~$\mu$s. In the
latter two cases, gamma detections occurred intermittently for an additional
60 and 100~$\mu$s both before and/or after the main activity, extending their
overall durations to $\simeq$160 and 220~$\mu$s, respectively.  For each of
the three events, the TGFs were produced during the occurrence of a slow,
smooth component of the sferic, indicative of being caused by electron
avalanching that produced the TGFs.  Complex, lengthy sferics were also
produced by the other two events of the same Lyu et al.\ study.

Of particular interest, and the best-studied example, was the first event of
4~September 2015 (Fig.~2 of \citeA{lyu2018a}), which occurred over west-central
Florida.  Its sferic closely resembled that of the first complex IBP of the
Utah IC, seen in the top panel of Figure~\ref{fig:IC_IBPs}. In both cases, the
sferic lasted for $\simeq$250~$\mu$s and consisted of several highly impulsive
sub-pulses before and after a central event.  For the Utah IC the central
event was itself a large-amplitude IBP, while for the Florida IC it was the
large-amplitude slow field change of the electron avalanche. The comparison,
along with the other Lyu et al.\ examples illustrates the fact that a)
long-duration TGFs can be produced by IC flashes having complex sferics, and
b) that the only difference between the Utah and Florida ICs is that the
latter initiated strong runaway avalanching, while the former did not, but based on the sferic similarities,
could well have done so. The second complex IBP of the Utah IC, seen in the
bottom panel of Figure~\ref{fig:IC_IBPs}, would have been even more capable of
generating a long-duration TGF based on its greater duration and VHF signal strength.

\citeA{pu2019} extended Lyu et al.'s study to include five additional examples
of continuously and intermittently long-duration TGFs being produced by other
IC flashes having complex IBP sferics. Finally, we call attention to the study
by \citeA{tilles2020b} of a high peak current (247~kA) energetic in-cloud pulse
(EIP) that was observed in Florida with the same physical INTF and FA
instrumentation of the present study. The EIP was produced by a complex
sequence of repeated IBP-type fast breakdown activity, but its sferic was
completely dominated by a sequence of three successive slow, smooth
relativistic avalanches indicative of being produced by relativistic feedback.
No gamma-detecting satellite happened to be in view of the EIP, but the flash
undoubtedly produced an upward TGF~\cite{lyu2016,cummer2017} and is an example
of how IC flashes are capable of producing extremely strong avalanching as a
result of complex IBP-type activity.

\subsection{Summary}

The results can be summarized as follows:

\vspace{-0.5\baselineskip}
\begin{enumerate}

\item Downward TGFs occur during strong, ``classic'' initial breakdown pulses
(IBPs) of downward negative CG and IC flashes.  In turn, the IBPs are
produced by streamer-based fast negative breakdown (FNB).

\item The TGFs consist of short, $\simeq$5--10 $\mu$s~duration bursts of gamma
rays initiated by sub-pulses during the IBPs, and apparently also by brief
episodes of enhanced speed FNB.

\item The correspondence of TGFs with sub-pulses is indicative of the
sub-pulses being produced by spark-like transient conducting events (TCEs),
consistent with their sferics being impulsive or cusp-like and explaining
the bright optical activity observed during IBPs of --CG and IC flashes.

\item In turn, the TCEs are considered to result from instabilities in
occasionally long streamer tails or partially conducting channels embedded
within the FNB of the IBP, and to be isolated from each other and from the
incoming breakdown preceding the IBP.

\item Based solely on the well-understood physics of surface detector
responses and Compton electron production, individual electrons detected by
the TASD surface stations correspond to photon energies no less than 2.6 MeV
if detected in a single scintillator layer and 6.2 MeV if detected in both
layers.

\item  From the electric field required to propagate negative streamers in virgin air
at --CG altitudes, the electric potential difference experienced by the FNB
over the 100-m to 240~m extents of TGF-producing IBPs is $\simeq$60
to 150 MV.  


\item Instead of the breakdown leading up to an IBP being a long conducting
leader, it appears to be due to weakly-conducting negative streamer breakdown that gets
accelerated to produce the IBP.

\item The observational data indicate that the streamer to leader transition of successive
steps is caused by current generated during the characteristic opposite-polarity field change
in the final stage of the step's IBP.

\item The initial upward negative breakdown of IC flashes is shown to be
produced in the same basic manner as the initial downward breakdown of --CG
discharges, but generally lasting longer and having longer step sizes.  

\item The long durations of satellite-detected TGFs can be explained by IC flashes producing complex clusters of sub-pulses and
IBPs, which enable the development of continuous and intermittent electron
avalanching. Sparse versions of this are seen during successive IBPs of --CG flashes.


\end{enumerate}

While the present study has been underway, the TASD has been in the process of expanding by a
factor of four in its coverage area, and the TGF and lightning observations
are continuing.  The LMA network is being similarly expanded, and an
additional VHF interferometer instrument is to be added in the current year.
Detailed analyses of additional observations are the subject of continued
study.
\appendix
\section{Methods}

\subsection{Instruments}

{\bf Telescope Array Surface Detector.} The TASD consists of 507 scintillator
detectors arranged on a 1.2~km square grid.  The array is situated on a
relatively high, 1400~m altitude desert plain in west-central Utah, and covers
an area of $\simeq$700~km$^2$. Each detector has two scintillator planes, each
3~m$^2 \times 1.2$~cm thick, separated by a 1~mm thick steel sheet and housed
inside an RF-sealed and light-tight stainless steel enclosure.  The TASD is
designed to detect the charged components --- primarily electrons, positrons,
and muons --- of the cosmic ray-induced Extensive Air Shower (EAS). An event
trigger is recorded when three adjacent SDs observe a signal greater than that
of 3 Minimum Ionizing Particles (MIPs) ($\simeq$150~FADC counts) within 8~$\mu$s. When an event trigger occurs,
the signals from all individually-triggered SDs within $\pm$~32~$\mu$s are recorded~\cite{abuzayyad2012}. An individual SD trigger occurs upon observing a signal of amplitude greater than 0.3~MIP ($\simeq$15~FADC counts) within 8~$\mu$s.

The TASD is an inefficient detector of gamma radiation, relying on the
production of high-energy electrons through the Compton scattering mechanism
in either the thin scintillator, steel housing, or air above the detector
units. Detailed simulations of this process have been described in
the authors' previous study~\cite{abbasi2018}. Incident gamma-ray photons with
energy above 10~MeV will on average deposit about 20\% (30\%) of the energy of
a MIP in the upper (lower) scintillator. The majority of photons will not
interact in the detector at all; those that do will primarily create Compton
recoil electrons with kinetic energies at or below the photon energy level.
The Compton electrons can then deposit energy up to a MIP (2.4~MeV) in each
plane of the scintillator, though the amount deposited in each plane will
depend on where the Compton scatter occurs.

{\bf Lightning Mapping Array.} As shown in Figure~\ref{fig:detectors}, the LMA consisted of nine
stations located within and around the TASD, and determines accurate 3-D
observations of peak VHF radiation events above threshold in 80~$\mu$s time
intervals.~\cite{rison1999,thomas2004}  In addition to showing the large scale
structure and development of flashes and the lightning flashing rate, its
observations were used to determine the plan distance to the TGF events and
also to finely calibrate the INTF azimuth and elevation values. The angular
calibration was done separately on a flash-by-flash basis for each TGF event.

{\bf VHF lightning interferometer (INTF) and fast electric field change
antenna (FA).} The INTF records broadband (20--80~MHz) waveforms at 180~MHz
from three flat-plate receiving antennas, and determines the
two-dimensional azimuth and elevation arrival directions of the VHF radiation
with sub-microsecond resolution~\cite{stock2014a}.  This is done on a
post-processed basis, and determines the radiation centroid in overlapping 0.7
or 1.4 $\mu$s windows.  Triangular baselines of 106--121~m were used to
maximize the angular resolution over the TASD.  The elevation angles were used
to determine the source altitude of the TGFs, based on the LMA-determined plan
distance to the source, and the amplitude of the received signals was used to
determine the VHF power of the centroids.  The fast electric field change
antenna (FA) provided high resolution (180 MHz) measurements of the low
frequency (LF/ELF) discharge sferics that are key to interpreting the INTF and
LMA observations.

\begin{figure}
\begin{center}
\includegraphics[width=1.0\linewidth]{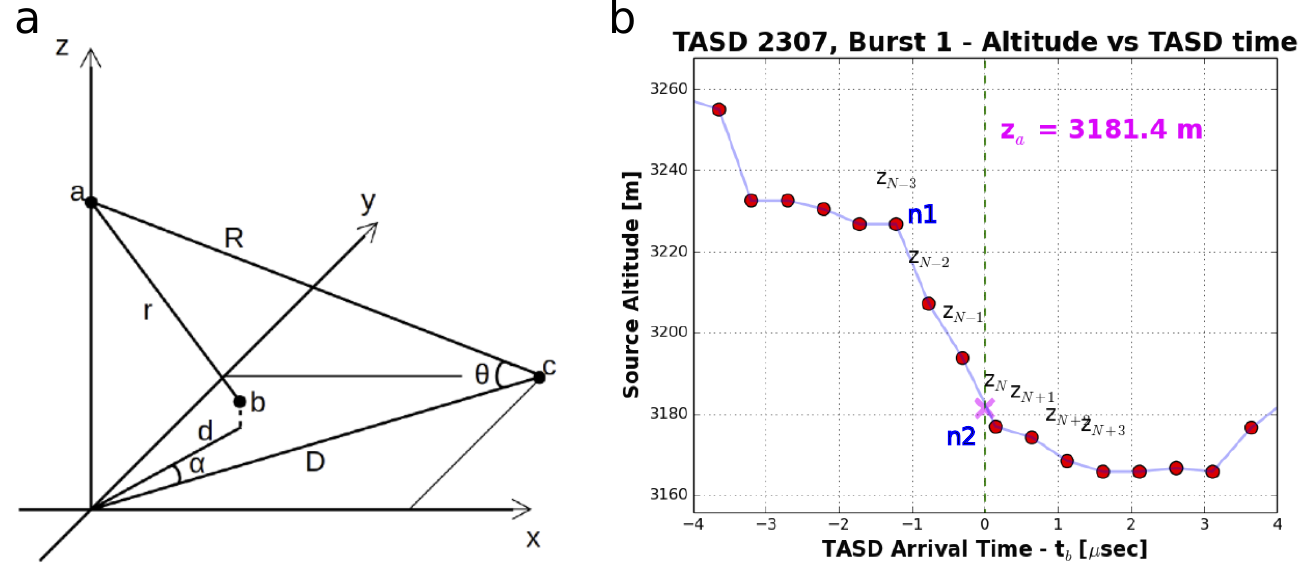}
\caption{\label{fig:coord_system} Methods information. {\normalfont ({\bf a})
Source-centric coordinate system for temporal correlations.  The TGF source
is at ($x_a,y_a,z_a$), with the plan $x$,~$y$ location serving as the
coordinate origin. The TASD station is at location $b$ relative to the origin
and the reference altitude, and the INTF/FA is at the more distant location
$c$. ({\bf b}) Iteration at 0.5~$\mu$s time steps used in the alternative
approach for determining the source altitude (TGF~A in this case), showing the
occurrence of enhanced-speed downward FNB immediately before the TGF onset
(red `x'). }}
\end{center}
\end{figure}
\clearpage

\subsection{Analysis procedures}

Figure~\ref{fig:coord_system}a shows the coordinate system used for
analyzing the INTF and TASD observations.  For simplicity, this is done in a
Cartesian coordinate system centered at the $x_a,y_a$ plan location of the
TGF's source.  The plan location is determined from the mean values of the
latitude and longitude of LMA sources within $\pm 1$~ms of the TGF's
occurrence, seen in Supporting Figs.\ 10e--13e. The altitude values are
determined relative to a 1400~m reference plane, which is within 2~m of the
GPS altitude of the VHF receiving antenna used as the INTF's GPS time base.
The plan locations and altitudes of the TASD stations are precisely known and
fully accounted for in the calculations, with trigger times of each TASD's data
accurate to 40 ns. Similarly, the INTF source directions were carefully
calibrated to within 0.08 degrees in azimuth and 0.26 degrees in elevation,
obtained by comparing accurately-located LMA sources with corresponding INTF
source directions separately for each flash.

Given the LMA-estimated values of $x_a$ and $y_a$, two additional measurements
are needed to determine the TGF's onset altitude $z_a$ and time $t_a$.  The
source altitude can be estimated from the LMA observations, but has
insufficient accuracy and temporal resolution to resolve the fast downward
breakdown that occurs during the parent IBP (typically 100--150 m in
5--10~$\mu$s).  Instead, the altitude is more accurately determined from the
INTF elevation angle $\theta_c$ vs.\ time, which is obtained with
sub-microsecond resolution. In particular, $z_a=D\tan\theta(t_c)=z_a(t_c)$,
where $D=\sqrt{x_c^2 + y_c^2}$ is the plan distance between the INTF and TGF.
For an event at altitude $z_a$ and time $t_a$, the arrival times at TASD $i$
and the INTF are given by

\begin{eqnarray}
t_b &=& t_a + r_{b}/c  \label{eq:t_b} \\
t_c &=& t_a + R/c  \label{eq:t_c} \;, 
\end{eqnarray} 
where $r_b = [x_i^2 + y_i^2 + (z_a-z_i)^2]$ and $R = [x_c^2 + y_c^2 + z_a^2]$ 
are the slant ranges from the TGF source.  Because the plan locations
are considered to be known, $r_b = r_{b}(z_a)$ and $R = R(z_a)$, so the
time-of-arrival equations represent two equations and two unknowns, $t_a$ and
$z_a$.  The unknowns are determined from two measurements, in particular 
the arrival time $t_b$ at a given TASD station, and the INTF elevation
measurements, $\theta_c(t_c)$.  Since $\theta_c$ varies with time during the
IBPs, it is not known in advance which time value $t_c$ to use for determining
$z_a$.  This results from $z_a$ depending on itself in a manner that is not
amenable to analytical inversion. But the equations are readily solved by
iterating over the range of values for $z_a$, or equivalently over the
possible $\theta_c$ or $t_c$ values.


Two semi-independent approaches were used to determine the solutions.  Both
used an alternative form of (\ref{eq:t_c}) obtained by eliminating $t_a$ to
obtain
\begin{equation} 
t_c = t_b + \frac{(R-r_b)}{c} = t_b + \Delta t_b \; ,  \label{t_c2}
\end{equation} 
where $\Delta t_b = (R/c) - (r_{b}/c)$ corresponds to the time shift for comparing
a given TASD's observations with the INTF/FA observations.  For an assumed source
altitude $z_a$, the time shift between the onset time $t_b$ at a given TASD
station and its arrival time $t_c$ at the INTF is readily calculated from
the difference of the slant ranges $R$ and $r_b$ of the source relative to the
INTF and the TASD in question.  In turn, the $t_c$ value can be used to
determine $\theta_c(t_c)$ and hence $z_a$.  Comparing the assumed and inferred
$z_a$ values forms the basis for a closed loop iteration procedure, in which
the assumed $z_a$ is simply replaced by the new $z_a$ value (Supp.\
Fig.\ S14). Consistency is reached in just a few steps. At the same time, the
corresponding INTF elevation angle $\theta_c$ and arrival time $t_c$ at the
INTF is also determined.

The above is the method used by the first approach, as described in Section
2.2.  For each of the primary TGFs shown in Figure~\ref{fig:all_tgfs}, the
source altitudes inferred from the onset times at the different TASDs were in
good agreement, having uncertainties of 30~m, 16~m, 10~m, and 40~m for
TGFs~A, B, C and D, respectively (see Supporting Table~S2).  To guard against
outliers, median values were used for determining the final $z_a$ and $t_a$
values at onset, as well as $\theta_c$ and $t_c$.  The final $t_c$ values
provide a reference time for evaluating the onset times of each
gamma-ray event. As can be seen from the TASD plots in Figure~\ref{fig:all_tgfs},
in most cases the waveforms begin within a microsecond or so of the indicated
$t_c$ onset time.  Detections that begin in advance of or after the indicated
onset, as for TGF~B, are indicative of different onset times.

Instead of using a closed-loop iteration process, as above, the second
approach worked backward from the INTF observations of the elevation angle
$\theta_c$ vs.\ $t_c$ to determine $z_a$ and $\Delta t$ in reverse.  This
was used to predict the arrival times at two of the TASD stations that
detected the TGF most strongly, and involved stepping through the $t_c$
times and corresponding $\theta_c$ values in 0.5~$\mu$s increments and
determining the time when the difference between the predicted and observed
$t_b$ values passed through zero.  The common reference time $t_b$ was defined
to be when the TASD signal first ascended to half of its eventual peak
amplitude on the 2 stations with the strongest signals (short vertical dotted lines in the TASD waveforms of Supp.\ Figs.\ S10-S13c,d), which were averaged to obtain the final estimate of the time
alignment.

Figure~\ref{fig:coord_system}b shows the results of the stepping procedure for
TASD 2307 of TGF~A. The plot shows the difference between the observed and
trial $t_b$ times of the main gamma-ray event, with the interpolated step value
where $\Delta t_b$ goes through zero determining the value of $t_c$ (red `x' in the figure). For this (and the
iterative) procedure to work, the INTF data was processed with higher time
resolution and increased overlap to make $\theta_c(t_c)$ more continuous. This
is a standard procedure for analyzing INTF observations~\cite{stock2014a}, and
allows more detail to be seen in $\theta_c$ vs.\ time. For these analyses, the
higher resolution data was downsampled to 0.5~$\mu$s intervals by using the
median of the higher frequency processing over a $\pm$4~$\mu$s interval around
each 0.5~$\mu$s point (unfilled gray circles in panels c and d of Supp.\
Figs.\ S10-S13).

What is informative and notable about the example of
Figure~\ref{fig:coord_system}b is that the onset time of the strong gamma burst
of TGF~A coincided with the end of a brief interlude of rapid descent in the
source altitude, denoted by the vertical dashed line in the figure.  The speed
of the descent is determined from the spacing between the dots, which occur at
0.5~$\mu$s intervals. In 1.5~$\mu$s (three step intervals), the source
descended about 50~m, corresponding to a downward speed of $3.3\times
10^7$~m/s. This enhanced-speed interlude was unresolved by the normal
processing, and instead caused the step discontinuity seen in
Figure~\ref{fig:all_tgfs}a during the fast negative breakdown. The stepping
method of determining the onset time agreed well with the result of the
iterative approach, which showed the gamma-ray onset to be at the end of the
discontinuity (bold vertical line in Figure~\ref{fig:all_tgfs}a).  The
agreement is not surprising, given that the same basic data was used in the
two analysis approaches.  But the correspondence with different approaches
indicates good precision in the procedures, and reinforces the observation
that the gamma bursts occur in association with intervals of enhanced speed
breakdown.

\subsection{Measurement uncertainties}

Whereas the INTF and FA data are
well-synchronized timewise by being simultaneously digitized at a high rate,
the main question is how accurately the TASD waveforms from the different
TASDs are synchronized with the INTF/FA data.  As discussed
above, this can be qualitatively determined by examining the waveforms from
the different SDs relative to the inferred onset time (vertical line) for each
of the TGF events in Figure~\ref{fig:all_tgfs}. In most cases, the observed onsets are within a microsecond or less of the inferred time, with important exceptions in TGFs B and C. 

A quantitative result can only be obtained from propagating the measurements' standard errors through calculations in the previous section, using the general form of
\begin{equation} 
\delta f= \sqrt{(\frac{\partial f}{\partial x_1} \delta x_1)^2 + ... + (\frac{\partial f}{\partial x_n} \delta x_n)^2}   \label{df}
\end{equation} 
where $f = f(x_1,...,x_n)$. Detector locations are known to centimeter accuracy and have negligible contributions. Similarly, gamma-ray detection trigger times are known on the order of sampling rate (10s~of~ns). Both are taken into account, but have very little effect on final uncertainties. Primary error sources, then, come from the two instances of taking averages described the previous section; TGF source plan locations are taken as the mean GPS location of LMA sources within 1~ms of particle detections, and its uncertainty is the standard error. TGF source elevations are done the same way --- a mean is taken of all INTF sources within 4~$\mu$s of the TGFs inferred arrival at the interferometer (from Equation~A3), and its uncertainty is the standard error.

All subsequent calculations can then be shadowed by their error counterparts using Equation~\ref{df} and are presented in Tables~S2 and S3. Typically, altitude measurements are much less precise for this type of study, but here altitude determination comes from the higher-sampled INTF data whereas plan location data is supplied by only a few LMA points. As a result, altitude uncertainties are 30, 20, 10, and 40 meters for TGFs A, B, C, and D respectively, compared to horizontal location errors of 150, 80, 40, and 300 meters. Timing uncertainties follow the same trend, with 0.7, 0.4, 0.2, and 1.4 $\mu$s for each respective TGF. Standard errors for all other calculations are shown in Tables~S2 and S3. 

Notice that elevation errors are nearly equal (Table~S2), but poor grouping of LMA data at the time of TGF~D means a larger error in the plan location. As the error is propagated through each calculation, quantities for TGF~D continue to be the least reliable among the four, showing that the low LMA sampling rate and possible mislocations during fast breakdown are the main contributors to all further uncertainty.

\acknowledgments
The lightning instrumentation, operation and analyses of this study have been
supported by NSF grants AGS-1205727, AGS-1613260, AGS-1720600 and AGS-1844306.
The Telescope Array experiment is supported by the Japan Society for the
Promotion of Science through Grants-in-Aids for Scientific Research on Specially
Promoted Research (15H05693) and for Scientific Research (S) (15H05741), and
the Inter-University Research Program of the Institute for Cosmic Ray
Research; by the U.S. National Science Foundation awards PHY-0307098,
PHY-0601915, PHY-0649681, PHY-0703893, PHY-0758342, PHY-0848320, PHY-1069280,
PHY-1069286,PHY-1404495, PHY-1404502 and PHY-1607727; by the National Research
Foundation of Korea (2015R1A2A1A01006870, 2015R1A2A1A15055344,
2016R1A5A1013277, 2007-0093860, 2016R1A2B4014967, 2017K1A4A3015188); by the
Russian Academy of Sciences, RFBR grant 16-02-00962a (INR), IISN project No.
4.4502.13, and Belgian Science Policy under IUAP VII/37 (ULB). The foundations
of Dr. Ezekiel R. and Edna Wattis Dumke, Willard L. Eccles, and George S. and
Dolores Dor\'e Eccles all helped with generous donations. The State of Utah
supported the project through its Economic Development Board, and the
University of Utah through the Office of the Vice President for Research. The
experimental site became available through the cooperation of the Utah School
and Institutional Trust Lands Administration (SITLA), U.S. Bureau of Land
Management (BLM), and the U.S. Air Force. We appreciate the assistance of the
State of Utah and Fillmore offices of the BLM in crafting the Plan of
Development for the site.  We also wish to thank the people and the officials
of Millard County, Utah for their steadfast and warm support. We gratefully
acknowledge the contributions from the technical staffs of our home
institutions. An allocation of computer time from the Center for High
Performance Computing at the University of Utah is gratefully acknowledged.
We thank Ryan Said and W. A. Brooks of Vaisala Inc.\ for providing
high-quality NLDN data lightning discharges over and around the TASD under
their academic research use policy as well as several anonymous reviewers for their requests and comments, the responses to which significantly improved the paper. Data that support the conclusions
presented in the manuscript are provided in the figures of the paper.
Additional information can be found in the supporting material and is available on the Open Science Framework (DOI: 10.17605/OSF.IO/Z3XDA).


%
%
\bibliography{ms.bib}

\begin{thebibliography}{}

\bibitem [\protect \citeauthoryear {%
Abbasi%
\ \protect \BOthers {.}}{%
Abbasi%
\ \protect \BOthers {.}}{%
{\protect \APACyear {2017}}%
}]{%
abbasi2017}
\APACinsertmetastar {%
abbasi2017}%
\begin{APACrefauthors}%
Abbasi, R.%
\BCBT {}\ \BOthersPeriod {.}
\end{APACrefauthors}%
\unskip\
\newblock
\APACrefYearMonthDay{2017}{}{}.
\newblock
{\BBOQ}\APACrefatitle {The bursts of high energy events observed by the
  telescope array surface detector} {The bursts of high energy events observed
  by the telescope array surface detector}.{\BBCQ}
\newblock
\APACjournalVolNumPages{Physics Letters A}{381}{32}{2565 - 2572}.
\newblock
\begin{APACrefDOI} \doi{10.1016/j.physleta.2017.06.022} \end{APACrefDOI}
\PrintBackRefs{\CurrentBib}

\bibitem [\protect \citeauthoryear {%
Abbasi%
\ \protect \BOthers {.}}{%
Abbasi%
\ \protect \BOthers {.}}{%
{\protect \APACyear {2018}}%
}]{%
abbasi2018}
\APACinsertmetastar {%
abbasi2018}%
\begin{APACrefauthors}%
Abbasi, R.%
\BCBT {}\ \BOthersPeriod {.}
\end{APACrefauthors}%
\unskip\
\newblock
\APACrefYearMonthDay{2018}{}{}.
\newblock
{\BBOQ}\APACrefatitle {Gamma Ray Showers Observed at Ground Level in
  Coincidence With Downward Lightning Leaders} {Gamma ray showers observed at
  ground level in coincidence with downward lightning leaders}.{\BBCQ}
\newblock
\APACjournalVolNumPages{Journal of Geophysical Research:
  Atmospheres}{123}{13}{6864-6879}.
\newblock
\begin{APACrefDOI} \doi{10.1029/2017JD027931} \end{APACrefDOI}
\PrintBackRefs{\CurrentBib}

\bibitem [\protect \citeauthoryear {%
Abu-Zayyad%
\ \protect \BOthers {.}}{%
Abu-Zayyad%
\ \protect \BOthers {.}}{%
{\protect \APACyear {2013}}%
}]{%
abuzayyad2012}
\APACinsertmetastar {%
abuzayyad2012}%
\begin{APACrefauthors}%
Abu-Zayyad, T.%
\BCBT {}\ \BOthersPeriod {.}
\end{APACrefauthors}%
\unskip\
\newblock
\APACrefYearMonthDay{2013}{}{}.
\newblock
{\BBOQ}\APACrefatitle {{The surface detector array of the Telescope Array
  experiment}} {{The surface detector array of the Telescope Array
  experiment}}.{\BBCQ}
\newblock
\APACjournalVolNumPages{Nuclear Instrumentation and Methods in Physics
  Research}{A689}{}{87-97}.
\newblock
\begin{APACrefDOI} \doi{10.1016/j.nima.2012.05.079} \end{APACrefDOI}
\PrintBackRefs{\CurrentBib}

\bibitem [\protect \citeauthoryear {%
Attanasio%
, Krehbiel%
\BCBL {}\ \BBA {} da Silva%
}{%
Attanasio%
\ \protect \BOthers {.}}{%
{\protect \APACyear {2019}}%
}]{%
attanasio2019}
\APACinsertmetastar {%
attanasio2019}%
\begin{APACrefauthors}%
Attanasio, A.%
, Krehbiel, P.%
\BCBL {}\ \BBA {} da Silva, C.%
\end{APACrefauthors}%
\unskip\
\newblock
\APACrefYearMonthDay{2019}{}{}.
\newblock
{\BBOQ}\APACrefatitle {Griffiths and Phelps lightning initiation model,
  revisited} {Griffiths and phelps lightning initiation model,
  revisited}.{\BBCQ}
\newblock
\APACjournalVolNumPages{Journal of Geophysical Research:
  Atmospheres}{124}{14}{}.
\newblock
\begin{APACrefDOI} \doi{10.1029/2019JD030399} \end{APACrefDOI}
\PrintBackRefs{\CurrentBib}

\bibitem [\protect \citeauthoryear {%
Behnke%
, Thomas%
, Krehbiel%
\BCBL {}\ \BBA {} Rison%
}{%
Behnke%
\ \protect \BOthers {.}}{%
{\protect \APACyear {2005}}%
}]{%
behnke2005}
\APACinsertmetastar {%
behnke2005}%
\begin{APACrefauthors}%
Behnke, S\BPBI A.%
, Thomas, R\BPBI J.%
, Krehbiel, P\BPBI R.%
\BCBL {}\ \BBA {} Rison, W.%
\end{APACrefauthors}%
\unskip\
\newblock
\APACrefYearMonthDay{2005}{}{}.
\newblock
{\BBOQ}\APACrefatitle {Initial leader velocities during intracloud lightning:
  Possible evidence for a runaway breakdown effect} {Initial leader velocities
  during intracloud lightning: Possible evidence for a runaway breakdown
  effect}.{\BBCQ}
\newblock
\APACjournalVolNumPages{Journal of Geophysical Research:
  Atmospheres}{110}{D10}{}.
\newblock
\begin{APACrefDOI} \doi{10.1029/2004JD005312} \end{APACrefDOI}
\PrintBackRefs{\CurrentBib}

\bibitem [\protect \citeauthoryear {%
Biagi%
\ \protect \BOthers {.}}{%
Biagi%
\ \protect \BOthers {.}}{%
{\protect \APACyear {2010}}%
}]{%
biagi2010}
\APACinsertmetastar {%
biagi2010}%
\begin{APACrefauthors}%
Biagi, C\BPBI J.%
, Uman, M\BPBI A.%
, Hill, J\BPBI D.%
, Jordan, D\BPBI M.%
, Rakov, V\BPBI A.%
\BCBL {}\ \BBA {} Dwyer, J.%
\end{APACrefauthors}%
\unskip\
\newblock
\APACrefYearMonthDay{2010}{}{}.
\newblock
{\BBOQ}\APACrefatitle {Observations of stepping mechanisms in a rocket-and-wire
  triggered lightning flash} {Observations of stepping mechanisms in a
  rocket-and-wire triggered lightning flash}.{\BBCQ}
\newblock
\APACjournalVolNumPages{Journal of Geophysical Research:
  Atmospheres}{115}{D23}{}.
\newblock
\begin{APACrefDOI} \doi{10.1029/2010JD014616} \end{APACrefDOI}
\PrintBackRefs{\CurrentBib}

\bibitem [\protect \citeauthoryear {%
Briggs%
\ \protect \BOthers {.}}{%
Briggs%
\ \protect \BOthers {.}}{%
{\protect \APACyear {2010}}%
}]{%
briggs2010}
\APACinsertmetastar {%
briggs2010}%
\begin{APACrefauthors}%
Briggs, M\BPBI S.%
\BCBT {}\ \BOthersPeriod {.}
\end{APACrefauthors}%
\unskip\
\newblock
\APACrefYearMonthDay{2010}{}{}.
\newblock
{\BBOQ}\APACrefatitle {First results on terrestrial gamma ray flashes from the
  {F}ermi Gamma{-}ray Burst Monitor} {First results on terrestrial gamma ray
  flashes from the {F}ermi gamma{-}ray burst monitor}.{\BBCQ}
\newblock
\APACjournalVolNumPages{Journal of Geophysical Research: Space
  Physics}{115}{A7}{}.
\newblock
\begin{APACrefDOI} \doi{10.1029/2009JA015242} \end{APACrefDOI}
\PrintBackRefs{\CurrentBib}

\bibitem [\protect \citeauthoryear {%
Briggs%
\ \protect \BOthers {.}}{%
Briggs%
\ \protect \BOthers {.}}{%
{\protect \APACyear {2013}}%
}]{%
briggs2013}
\APACinsertmetastar {%
briggs2013}%
\begin{APACrefauthors}%
Briggs, M\BPBI S.%
\BCBT {}\ \BOthersPeriod {.}
\end{APACrefauthors}%
\unskip\
\newblock
\APACrefYearMonthDay{2013}{}{}.
\newblock
{\BBOQ}\APACrefatitle {Terrestrial gamm{a‐}ray flashes in the {F}ermi era:
  Improved observations and analysis methods} {Terrestrial gamm{a‐}ray
  flashes in the {F}ermi era: Improved observations and analysis
  methods}.{\BBCQ}
\newblock
\APACjournalVolNumPages{Journal of Geophysical Research: Space
  Physics}{118}{6}{3805-3830}.
\newblock
\begin{APACrefDOI} \doi{10.1002/jgra.50205} \end{APACrefDOI}
\PrintBackRefs{\CurrentBib}

\bibitem [\protect \citeauthoryear {%
Carlson%
, Lehtinen%
\BCBL {}\ \BBA {} Inan%
}{%
Carlson%
\ \protect \BOthers {.}}{%
{\protect \APACyear {2007}}%
}]{%
carlson2007}
\APACinsertmetastar {%
carlson2007}%
\begin{APACrefauthors}%
Carlson, B\BPBI E.%
, Lehtinen, N\BPBI G.%
\BCBL {}\ \BBA {} Inan, U\BPBI S.%
\end{APACrefauthors}%
\unskip\
\newblock
\APACrefYearMonthDay{2007}{}{}.
\newblock
{\BBOQ}\APACrefatitle {Constraints on terrestrial gamma ray flash production
  from satellite observation} {Constraints on terrestrial gamma ray flash
  production from satellite observation}.{\BBCQ}
\newblock
\APACjournalVolNumPages{Geophysical Research Letters}{34}{8}{}.
\newblock
\begin{APACrefDOI} \doi{10.1029/2006GL029229} \end{APACrefDOI}
\PrintBackRefs{\CurrentBib}

\bibitem [\protect \citeauthoryear {%
Celestin%
\ \BBA {} Pasko%
}{%
Celestin%
\ \BBA {} Pasko%
}{%
{\protect \APACyear {2011}}%
}]{%
celestin2011}
\APACinsertmetastar {%
celestin2011}%
\begin{APACrefauthors}%
Celestin, S.%
\BCBT {}\ \BBA {} Pasko, V.%
\end{APACrefauthors}%
\unskip\
\newblock
\APACrefYearMonthDay{2011}{}{}.
\newblock
{\BBOQ}\APACrefatitle {Energy and fluxes of thermal runaway electrons produced
  by exponential growth of streamers during the stepping of lightning leaders
  and in transient luminous events} {Energy and fluxes of thermal runaway
  electrons produced by exponential growth of streamers during the stepping of
  lightning leaders and in transient luminous events}.{\BBCQ}
\newblock
\APACjournalVolNumPages{Journal of Geophysical Research: Space
  Physics}{116}{A3}{}.
\newblock
\begin{APACrefDOI} \doi{10.1029/2010JA016260} \end{APACrefDOI}
\PrintBackRefs{\CurrentBib}

\bibitem [\protect \citeauthoryear {%
Celestin%
\ \BBA {} Pasko%
}{%
Celestin%
\ \BBA {} Pasko%
}{%
{\protect \APACyear {2012}}%
}]{%
celestin2012}
\APACinsertmetastar {%
celestin2012}%
\begin{APACrefauthors}%
Celestin, S.%
\BCBT {}\ \BBA {} Pasko, V.%
\end{APACrefauthors}%
\unskip\
\newblock
\APACrefYearMonthDay{2012}{}{}.
\newblock
{\BBOQ}\APACrefatitle {Compton scattering effects on the duration of
  terrestrial gamma-ray flashes} {Compton scattering effects on the duration of
  terrestrial gamma-ray flashes}.{\BBCQ}
\newblock
\APACjournalVolNumPages{Geophysical Research Letters}{39}{2}{}.
\newblock
\begin{APACrefDOI} \doi{10.1029/2011GL050342} \end{APACrefDOI}
\PrintBackRefs{\CurrentBib}

\bibitem [\protect \citeauthoryear {%
Celestin%
, Xu%
\BCBL {}\ \BBA {} Pasko%
}{%
Celestin%
\ \protect \BOthers {.}}{%
{\protect \APACyear {2012}}%
}]{%
celestin2012b}
\APACinsertmetastar {%
celestin2012b}%
\begin{APACrefauthors}%
Celestin, S.%
, Xu, W.%
\BCBL {}\ \BBA {} Pasko, V.%
\end{APACrefauthors}%
\unskip\
\newblock
\APACrefYearMonthDay{2012}{}{}.
\newblock
{\BBOQ}\APACrefatitle {Terrestrial gamma ray flashes with energies up to 100
  {M}e{V} produced by nonequilibrium acceleration of electrons in lightning}
  {Terrestrial gamma ray flashes with energies up to 100 {M}e{V} produced by
  nonequilibrium acceleration of electrons in lightning}.{\BBCQ}
\newblock
\APACjournalVolNumPages{Journal of Geophysical Research: Space
  Physics}{117}{A5}{}.
\newblock
\begin{APACrefDOI} \doi{10.1029/2012JA017535} \end{APACrefDOI}
\PrintBackRefs{\CurrentBib}

\bibitem [\protect \citeauthoryear {%
Celestin%
, Xu%
\BCBL {}\ \BBA {} Pasko%
}{%
Celestin%
\ \protect \BOthers {.}}{%
{\protect \APACyear {2015}}%
}]{%
celestin2015}
\APACinsertmetastar {%
celestin2015}%
\begin{APACrefauthors}%
Celestin, S.%
, Xu, W.%
\BCBL {}\ \BBA {} Pasko, V.%
\end{APACrefauthors}%
\unskip\
\newblock
\APACrefYearMonthDay{2015}{}{}.
\newblock
{\BBOQ}\APACrefatitle {Variability in fluence and spectrum of high-energy
  photon bursts produced by lightning leaders} {Variability in fluence and
  spectrum of high-energy photon bursts produced by lightning leaders}.{\BBCQ}
\newblock
\APACjournalVolNumPages{Journal of Geophysical Research: Space
  Physics}{120}{}{10712-10723}.
\newblock
\begin{APACrefDOI} \doi{10.1002/2015JA021410} \end{APACrefDOI}
\PrintBackRefs{\CurrentBib}

\bibitem [\protect \citeauthoryear {%
Cummer%
}{%
Cummer%
}{%
{\protect \APACyear {2020}}%
}]{%
cummer2020}
\APACinsertmetastar {%
cummer2020}%
\begin{APACrefauthors}%
Cummer, S\BPBI A.%
\end{APACrefauthors}%
\unskip\
\newblock
\APACrefYearMonthDay{2020}{}{}.
\newblock
{\BBOQ}\APACrefatitle {Indirectly measured ambient electric fields for
  lightning initiation in fast breakdown regions} {Indirectly measured ambient
  electric fields for lightning initiation in fast breakdown regions}.{\BBCQ}
\newblock
\APACjournalVolNumPages{Geophysical Research Letters}{47}{4}{}.
\newblock
\begin{APACrefDOI} \doi{10.1029/2019GL086089} \end{APACrefDOI}
\PrintBackRefs{\CurrentBib}

\bibitem [\protect \citeauthoryear {%
Cummer%
\ \protect \BOthers {.}}{%
Cummer%
\ \protect \BOthers {.}}{%
{\protect \APACyear {2011}}%
}]{%
cummer2011}
\APACinsertmetastar {%
cummer2011}%
\begin{APACrefauthors}%
Cummer, S\BPBI A.%
\BCBT {}\ \BOthersPeriod {.}
\end{APACrefauthors}%
\unskip\
\newblock
\APACrefYearMonthDay{2011}{}{}.
\newblock
{\BBOQ}\APACrefatitle {The lightning{-TGF} relationship on microsecond
  timescales} {The lightning{-TGF} relationship on microsecond
  timescales}.{\BBCQ}
\newblock
\APACjournalVolNumPages{Geophysical Research Letters}{38}{}{L14810}.
\newblock
\begin{APACrefDOI} \doi{10.1029/2011GL048099} \end{APACrefDOI}
\PrintBackRefs{\CurrentBib}

\bibitem [\protect \citeauthoryear {%
Cummer%
\ \protect \BOthers {.}}{%
Cummer%
\ \protect \BOthers {.}}{%
{\protect \APACyear {2015}}%
}]{%
cummer2015}
\APACinsertmetastar {%
cummer2015}%
\begin{APACrefauthors}%
Cummer, S\BPBI A.%
\BCBT {}\ \BOthersPeriod {.}
\end{APACrefauthors}%
\unskip\
\newblock
\APACrefYearMonthDay{2015}{}{}.
\newblock
{\BBOQ}\APACrefatitle {Lightning leader altitude progression in terrestrial
  gamma-ray flashes} {Lightning leader altitude progression in terrestrial
  gamma-ray flashes}.{\BBCQ}
\newblock
\APACjournalVolNumPages{Geophysical Research Letters}{42}{}{7792–7798}.
\newblock
\begin{APACrefDOI} \doi{10.1002/2015GL065228} \end{APACrefDOI}
\PrintBackRefs{\CurrentBib}

\bibitem [\protect \citeauthoryear {%
Cummer%
\ \protect \BOthers {.}}{%
Cummer%
\ \protect \BOthers {.}}{%
{\protect \APACyear {2017}}%
}]{%
cummer2017}
\APACinsertmetastar {%
cummer2017}%
\begin{APACrefauthors}%
Cummer, S\BPBI A.%
\BCBT {}\ \BOthersPeriod {.}
\end{APACrefauthors}%
\unskip\
\newblock
\APACrefYearMonthDay{2017}{{\APACmonth{12}}}{}.
\newblock
{\BBOQ}\APACrefatitle {The Connection Between Terrestrial Gamma-Ray Flashes and
  Energetic In-Cloud Lightning Pulses} {The connection between terrestrial
  gamma-ray flashes and energetic in-cloud lightning pulses}.{\BBCQ}
\newblock
\BIn{} \APACrefbtitle {American {G}eophysical {U}nion {F}all {M}eeting 2017
  {A}bstracts.} {American {G}eophysical {U}nion {F}all {M}eeting 2017
  {A}bstracts.}
\newblock
\APACrefnote{AE33B-2547}
\PrintBackRefs{\CurrentBib}

\bibitem [\protect \citeauthoryear {%
da Silva%
\ \BBA {} Pasko%
}{%
da Silva%
\ \BBA {} Pasko%
}{%
{\protect \APACyear {2013}}%
}]{%
dasilva2013}
\APACinsertmetastar {%
dasilva2013}%
\begin{APACrefauthors}%
da Silva, C.%
\BCBT {}\ \BBA {} Pasko, V.%
\end{APACrefauthors}%
\unskip\
\newblock
\APACrefYearMonthDay{2013}{}{}.
\newblock
{\BBOQ}\APACrefatitle {Dynamics of streamer-to-leader transition at reduced air
  densities and its implications for propagation of lightning leaders and
  gigantic jets} {Dynamics of streamer-to-leader transition at reduced air
  densities and its implications for propagation of lightning leaders and
  gigantic jets}.{\BBCQ}
\newblock
\APACjournalVolNumPages{Journal of Geophysical Research:
  Atmospheres}{118}{24}{13,561-13,590}.
\newblock
\begin{APACrefDOI} \doi{10.1002/2013JD020618} \end{APACrefDOI}
\PrintBackRefs{\CurrentBib}

\bibitem [\protect \citeauthoryear {%
da Silva%
\ \BBA {} Pasko%
}{%
da Silva%
\ \BBA {} Pasko%
}{%
{\protect \APACyear {2015}}%
}]{%
dasilva2015}
\APACinsertmetastar {%
dasilva2015}%
\begin{APACrefauthors}%
da Silva, C.%
\BCBT {}\ \BBA {} Pasko, V.%
\end{APACrefauthors}%
\unskip\
\newblock
\APACrefYearMonthDay{2015}{}{}.
\newblock
{\BBOQ}\APACrefatitle {Physical mechanism of initial breakdown pulses and
  narrow bipolar events in lightning discharges} {Physical mechanism of initial
  breakdown pulses and narrow bipolar events in lightning discharges}.{\BBCQ}
\newblock
\APACjournalVolNumPages{Journal of Geophysical Research:
  Atmospheres}{120}{}{4989-5009}.
\newblock
\begin{APACrefDOI} \doi{10.1002/2015JD023209} \end{APACrefDOI}
\PrintBackRefs{\CurrentBib}

\bibitem [\protect \citeauthoryear {%
da Silva%
\ \protect \BOthers {.}}{%
da Silva%
\ \protect \BOthers {.}}{%
{\protect \APACyear {2019}}%
}]{%
dasilva2019}
\APACinsertmetastar {%
dasilva2019}%
\begin{APACrefauthors}%
da Silva, C.%
, Sonnenfeld, R\BPBI G.%
, Edens, H\BPBI E.%
, Krehbiel, P\BPBI R.%
, Quick, M\BPBI G.%
\BCBL {}\ \BBA {} Koshak, W\BPBI J.%
\end{APACrefauthors}%
\unskip\
\newblock
\APACrefYearMonthDay{2019}{}{}.
\newblock
{\BBOQ}\APACrefatitle {The Plasma Nature of Lightning Channels and the
  Resulting Nonlinear Resistance} {The plasma nature of lightning channels and
  the resulting nonlinear resistance}.{\BBCQ}
\newblock
\APACjournalVolNumPages{Journal of Geophysical Research:
  Atmospheres}{124}{16}{9442-9463}.
\newblock
\begin{APACrefDOI} \doi{10.1029/2019JD030693} \end{APACrefDOI}
\PrintBackRefs{\CurrentBib}

\bibitem [\protect \citeauthoryear {%
Dwyer%
}{%
Dwyer%
}{%
{\protect \APACyear {2003}}%
}]{%
dwyer2003}
\APACinsertmetastar {%
dwyer2003}%
\begin{APACrefauthors}%
Dwyer, J.%
\end{APACrefauthors}%
\unskip\
\newblock
\APACrefYearMonthDay{2003}{}{}.
\newblock
{\BBOQ}\APACrefatitle {A fundamental limit on electric fields in air} {A
  fundamental limit on electric fields in air}.{\BBCQ}
\newblock
\APACjournalVolNumPages{Geophysical Research Letters}{30}{20}{}.
\newblock
\begin{APACrefDOI} \doi{10.1029/2003GL017781} \end{APACrefDOI}
\PrintBackRefs{\CurrentBib}

\bibitem [\protect \citeauthoryear {%
Dwyer%
}{%
Dwyer%
}{%
{\protect \APACyear {2004}}%
}]{%
dwyer2004}
\APACinsertmetastar {%
dwyer2004}%
\begin{APACrefauthors}%
Dwyer, J.%
\end{APACrefauthors}%
\unskip\
\newblock
\APACrefYearMonthDay{2004}{}{}.
\newblock
{\BBOQ}\APACrefatitle {Implications of x-ray emission from lightning}
  {Implications of x-ray emission from lightning}.{\BBCQ}
\newblock
\APACjournalVolNumPages{Geophysical Research Letters}{31}{}{L12102}.
\newblock
\begin{APACrefDOI} \doi{10.1029/2004GL019795} \end{APACrefDOI}
\PrintBackRefs{\CurrentBib}

\bibitem [\protect \citeauthoryear {%
Dwyer%
}{%
Dwyer%
}{%
{\protect \APACyear {2012}}%
}]{%
dwyer2012a}
\APACinsertmetastar {%
dwyer2012a}%
\begin{APACrefauthors}%
Dwyer, J.%
\end{APACrefauthors}%
\unskip\
\newblock
\APACrefYearMonthDay{2012}{}{}.
\newblock
{\BBOQ}\APACrefatitle {The relativistic feedback discharge model of terrestrial
  gamma ray flashes} {The relativistic feedback discharge model of terrestrial
  gamma ray flashes}.{\BBCQ}
\newblock
\APACjournalVolNumPages{Journal of Geophysical Research: Space
  Physics}{117}{A2}{}.
\newblock
\begin{APACrefDOI} \doi{10.1029/2011JA017160} \end{APACrefDOI}
\PrintBackRefs{\CurrentBib}

\bibitem [\protect \citeauthoryear {%
Dwyer%
\ \protect \BOthers {.}}{%
Dwyer%
\ \protect \BOthers {.}}{%
{\protect \APACyear {2012}}%
}]{%
dwyer2012b}
\APACinsertmetastar {%
dwyer2012b}%
\begin{APACrefauthors}%
Dwyer, J.%
\BCBT {}\ \BOthersPeriod {.}
\end{APACrefauthors}%
\unskip\
\newblock
\APACrefYearMonthDay{2012}{}{}.
\newblock
{\BBOQ}\APACrefatitle {Observation of a gamma-ray flash at ground level in
  association with a cloud-to-ground lightning return stroke} {Observation of a
  gamma-ray flash at ground level in association with a cloud-to-ground
  lightning return stroke}.{\BBCQ}
\newblock
\APACjournalVolNumPages{Journal of Geophysical Research}{117}{A10303}{}.
\newblock
\begin{APACrefDOI} \doi{10.1029/2012JA017810} \end{APACrefDOI}
\PrintBackRefs{\CurrentBib}

\bibitem [\protect \citeauthoryear {%
Fishman%
\ \protect \BOthers {.}}{%
Fishman%
\ \protect \BOthers {.}}{%
{\protect \APACyear {1994}}%
}]{%
fishman1994}
\APACinsertmetastar {%
fishman1994}%
\begin{APACrefauthors}%
Fishman, G\BPBI J.%
\BCBT {}\ \BOthersPeriod {.}
\end{APACrefauthors}%
\unskip\
\newblock
\APACrefYearMonthDay{1994}{}{}.
\newblock
{\BBOQ}\APACrefatitle {Discovery of Intense Gamma-Ray Flashes of Atmospheric
  Origin} {Discovery of intense gamma-ray flashes of atmospheric
  origin}.{\BBCQ}
\newblock
\APACjournalVolNumPages{Science}{264}{5163}{1313--1316}.
\newblock
\begin{APACrefDOI} \doi{10.1126/science.264.5163.1313} \end{APACrefDOI}
\PrintBackRefs{\CurrentBib}

\bibitem [\protect \citeauthoryear {%
Foley%
\ \protect \BOthers {.}}{%
Foley%
\ \protect \BOthers {.}}{%
{\protect \APACyear {2014}}%
}]{%
foley2014}
\APACinsertmetastar {%
foley2014}%
\begin{APACrefauthors}%
Foley, S.%
\BCBT {}\ \BOthersPeriod {.}
\end{APACrefauthors}%
\unskip\
\newblock
\APACrefYearMonthDay{2014}{}{}.
\newblock
{\BBOQ}\APACrefatitle {{Pulse properties of terrestrial gamma-ray flashes
  detected by the Fermi Gamma-Ray Burst Monitor}} {{Pulse properties of
  terrestrial gamma-ray flashes detected by the Fermi Gamma-Ray Burst
  Monitor}}.{\BBCQ}
\newblock
\APACjournalVolNumPages{Journal of Geophysical Research: Space
  Physics}{119}{}{5931-5942}.
\newblock
\begin{APACrefDOI} \doi{10.1002/2014JA019805} \end{APACrefDOI}
\PrintBackRefs{\CurrentBib}

\bibitem [\protect \citeauthoryear {%
Gjesteland%
\ \protect \BOthers {.}}{%
Gjesteland%
\ \protect \BOthers {.}}{%
{\protect \APACyear {2012}}%
}]{%
gjesteland2012}
\APACinsertmetastar {%
gjesteland2012}%
\begin{APACrefauthors}%
Gjesteland, T.%
\BCBT {}\ \BOthersPeriod {.}
\end{APACrefauthors}%
\unskip\
\newblock
\APACrefYearMonthDay{2012}{}{}.
\newblock
{\BBOQ}\APACrefatitle {A new method reveals more {TGF}s in the {RHESSI} data}
  {A new method reveals more {TGF}s in the {RHESSI} data}.{\BBCQ}
\newblock
\APACjournalVolNumPages{Geophysical Research Letters}{39}{5}{}.
\newblock
\begin{APACrefDOI} \doi{10.1029/2012GL050899} \end{APACrefDOI}
\PrintBackRefs{\CurrentBib}

\bibitem [\protect \citeauthoryear {%
Grefenstette%
, Smith%
, Hazelton%
\BCBL {}\ \BBA {} Lopez%
}{%
Grefenstette%
\ \protect \BOthers {.}}{%
{\protect \APACyear {2009}}%
}]{%
grefenstette2009}
\APACinsertmetastar {%
grefenstette2009}%
\begin{APACrefauthors}%
Grefenstette, B\BPBI W.%
, Smith, D\BPBI M.%
, Hazelton, B\BPBI J.%
\BCBL {}\ \BBA {} Lopez, L\BPBI I.%
\end{APACrefauthors}%
\unskip\
\newblock
\APACrefYearMonthDay{2009}{}{}.
\newblock
{\BBOQ}\APACrefatitle {First {RHESSI} terrestrial gamma ray flash catalog}
  {First {RHESSI} terrestrial gamma ray flash catalog}.{\BBCQ}
\newblock
\APACjournalVolNumPages{Journal of Geophysical Research: Space
  Physics}{114}{A2}{}.
\newblock
\begin{APACrefDOI} \doi{10.1029/2008JA013721} \end{APACrefDOI}
\PrintBackRefs{\CurrentBib}

\bibitem [\protect \citeauthoryear {%
Hare%
\ \protect \BOthers {.}}{%
Hare%
\ \protect \BOthers {.}}{%
{\protect \APACyear {2016}}%
}]{%
hare2016}
\APACinsertmetastar {%
hare2016}%
\begin{APACrefauthors}%
Hare, B.%
\BCBT {}\ \BOthersPeriod {.}
\end{APACrefauthors}%
\unskip\
\newblock
\APACrefYearMonthDay{2016}{}{}.
\newblock
{\BBOQ}\APACrefatitle {Ground-level observation of a terrestrial gamma-ray
  flash initiated by triggered lightning} {Ground-level observation of a
  terrestrial gamma-ray flash initiated by triggered lightning}.{\BBCQ}
\newblock
\APACjournalVolNumPages{Journal of Geophysical Research:
  Atmospheres}{121}{}{6511}.
\newblock
\begin{APACrefDOI} \doi{10.1002/2015JD024426} \end{APACrefDOI}
\PrintBackRefs{\CurrentBib}

\bibitem [\protect \citeauthoryear {%
Karunarathne%
, Karunarathne%
, Marshall%
\BCBL {}\ \BBA {} Stolzenburg%
}{%
Karunarathne%
\ \protect \BOthers {.}}{%
{\protect \APACyear {2014}}%
}]{%
karunarathne2014}
\APACinsertmetastar {%
karunarathne2014}%
\begin{APACrefauthors}%
Karunarathne, N.%
, Karunarathne, S.%
, Marshall, T.%
\BCBL {}\ \BBA {} Stolzenburg, M.%
\end{APACrefauthors}%
\unskip\
\newblock
\APACrefYearMonthDay{2014}{}{}.
\newblock
{\BBOQ}\APACrefatitle {Modeling initial breakdown pulses of {CG} lightning
  flashes} {Modeling initial breakdown pulses of {CG} lightning
  flashes}.{\BBCQ}
\newblock
\APACjournalVolNumPages{Journal of Geophysical Research:
  Atmospheres}{119}{}{9003}.
\newblock
\begin{APACrefDOI} \doi{10.1002/2014JD021553} \end{APACrefDOI}
\PrintBackRefs{\CurrentBib}

\bibitem [\protect \citeauthoryear {%
Krehbiel%
}{%
Krehbiel%
}{%
{\protect \APACyear {2018}}%
}]{%
krehbiel2018}
\APACinsertmetastar {%
krehbiel2018}%
\begin{APACrefauthors}%
Krehbiel, P.%
\end{APACrefauthors}%
\unskip\
\newblock
\APACrefYearMonthDay{2018}{}{}.
\newblock
{\BBOQ}\APACrefatitle {The Initial Development of Intracloud Flashes} {The
  initial development of intracloud flashes}.{\BBCQ}
\newblock
\BIn{} \APACrefbtitle {American {G}eophysical {U}nion {F}all {M}eeting 2018.}
  {American {G}eophysical {U}nion {F}all {M}eeting 2018.}
\newblock
\APACrefnote{AE11A-01}
\PrintBackRefs{\CurrentBib}

\bibitem [\protect \citeauthoryear {%
Krehbiel%
\ \protect \BOthers {.}}{%
Krehbiel%
\ \protect \BOthers {.}}{%
{\protect \APACyear {2008}}%
}]{%
krehbiel2008}
\APACinsertmetastar {%
krehbiel2008}%
\begin{APACrefauthors}%
Krehbiel, P.%
\BCBT {}\ \BOthersPeriod {.}
\end{APACrefauthors}%
\unskip\
\newblock
\APACrefYearMonthDay{2008}{}{}.
\newblock
{\BBOQ}\APACrefatitle {Upward electrical discharges from thunderstorms} {Upward
  electrical discharges from thunderstorms}.{\BBCQ}
\newblock
\APACjournalVolNumPages{Nature Geoscience}{1}{}{}.
\newblock
\begin{APACrefDOI} \doi{10.1038/ngeo162} \end{APACrefDOI}
\PrintBackRefs{\CurrentBib}

\bibitem [\protect \citeauthoryear {%
LeVine%
}{%
LeVine%
}{%
{\protect \APACyear {1980}}%
}]{%
levine1980}
\APACinsertmetastar {%
levine1980}%
\begin{APACrefauthors}%
LeVine, D.%
\end{APACrefauthors}%
\unskip\
\newblock
\APACrefYearMonthDay{1980}{}{}.
\newblock
{\BBOQ}\APACrefatitle {Sources of the Strongest {RF} Radiation From Lightning}
  {Sources of the strongest {RF} radiation from lightning}.{\BBCQ}
\newblock
\APACjournalVolNumPages{Journal of Geophysical Research}{85}{C7}{4091-4095}.
\newblock
\begin{APACrefDOI} \doi{10.1029/JC085iC07p04091} \end{APACrefDOI}
\PrintBackRefs{\CurrentBib}

\bibitem [\protect \citeauthoryear {%
Lu%
\ \protect \BOthers {.}}{%
Lu%
\ \protect \BOthers {.}}{%
{\protect \APACyear {2011}}%
}]{%
lu2011}
\APACinsertmetastar {%
lu2011}%
\begin{APACrefauthors}%
Lu, G.%
, Cummer, S\BPBI A.%
, Li, J.%
, Han, F.%
, Smith, D\BPBI M.%
\BCBL {}\ \BBA {} Grefenstette, B\BPBI W.%
\end{APACrefauthors}%
\unskip\
\newblock
\APACrefYearMonthDay{2011}{}{}.
\newblock
{\BBOQ}\APACrefatitle {Characteristics of broadband lightning emissions
  associated with terrestrial gamma ray flashes} {Characteristics of broadband
  lightning emissions associated with terrestrial gamma ray flashes}.{\BBCQ}
\newblock
\APACjournalVolNumPages{Journal of Geophysical Research: Space
  Physics}{116}{A3}{}.
\PrintBackRefs{\CurrentBib}

\bibitem [\protect \citeauthoryear {%
Lu%
\ \protect \BOthers {.}}{%
Lu%
\ \protect \BOthers {.}}{%
{\protect \APACyear {2010}}%
}]{%
lu2010}
\APACinsertmetastar {%
lu2010}%
\begin{APACrefauthors}%
Lu, G.%
\BCBT {}\ \BOthersPeriod {.}
\end{APACrefauthors}%
\unskip\
\newblock
\APACrefYearMonthDay{2010}{}{}.
\newblock
{\BBOQ}\APACrefatitle {Lightning mapping observation of a terrestrial gamma ray
  flash} {Lightning mapping observation of a terrestrial gamma ray
  flash}.{\BBCQ}
\newblock
\APACjournalVolNumPages{Geophysical Research Letters}{37}{}{L11806}.
\newblock
\begin{APACrefDOI} \doi{10.1029/2010GL043494} \end{APACrefDOI}
\PrintBackRefs{\CurrentBib}

\bibitem [\protect \citeauthoryear {%
Lyu%
, Cummer%
\BCBL {}\ \BBA {} McTague%
}{%
Lyu%
\ \protect \BOthers {.}}{%
{\protect \APACyear {2015}}%
}]{%
lyu2015}
\APACinsertmetastar {%
lyu2015}%
\begin{APACrefauthors}%
Lyu, F.%
, Cummer, S\BPBI A.%
\BCBL {}\ \BBA {} McTague, L.%
\end{APACrefauthors}%
\unskip\
\newblock
\APACrefYearMonthDay{2015}{}{}.
\newblock
{\BBOQ}\APACrefatitle {Insights into high peak current in-cloud lightning
  events during thunderstorms} {Insights into high peak current in-cloud
  lightning events during thunderstorms}.{\BBCQ}
\newblock
\APACjournalVolNumPages{Geophysical Research Letters}{42}{16}{6836-6843}.
\newblock
\begin{APACrefDOI} \doi{10.1002/2015GL065047} \end{APACrefDOI}
\PrintBackRefs{\CurrentBib}

\bibitem [\protect \citeauthoryear {%
Lyu%
\ \protect \BOthers {.}}{%
Lyu%
\ \protect \BOthers {.}}{%
{\protect \APACyear {2016}}%
}]{%
lyu2016}
\APACinsertmetastar {%
lyu2016}%
\begin{APACrefauthors}%
Lyu, F.%
\BCBT {}\ \BOthersPeriod {.}
\end{APACrefauthors}%
\unskip\
\newblock
\APACrefYearMonthDay{2016}{}{}.
\newblock
{\BBOQ}\APACrefatitle {Ground detection of terrestrial gamma ray flashes from
  distant radio signals} {Ground detection of terrestrial gamma ray flashes
  from distant radio signals}.{\BBCQ}
\newblock
\APACjournalVolNumPages{Geophysical Research Letters}{43}{16}{8728-8734}.
\newblock
\begin{APACrefDOI} \doi{10.1002/2016GL070154} \end{APACrefDOI}
\PrintBackRefs{\CurrentBib}

\bibitem [\protect \citeauthoryear {%
Lyu%
\ \protect \BOthers {.}}{%
Lyu%
\ \protect \BOthers {.}}{%
{\protect \APACyear {2018}}%
}]{%
lyu2018a}
\APACinsertmetastar {%
lyu2018a}%
\begin{APACrefauthors}%
Lyu, F.%
\BCBT {}\ \BOthersPeriod {.}
\end{APACrefauthors}%
\unskip\
\newblock
\APACrefYearMonthDay{2018}{}{}.
\newblock
{\BBOQ}\APACrefatitle {Very High Frequency Radio Emissions Associated With the
  Production of Terrestrial Gamma-Ray Flashes} {Very high frequency radio
  emissions associated with the production of terrestrial gamma-ray
  flashes}.{\BBCQ}
\newblock
\APACjournalVolNumPages{Geophysical Research Letters}{45}{4}{2097-2105}.
\newblock
\begin{APACrefDOI} \doi{10.1002/2018GL077102} \end{APACrefDOI}
\PrintBackRefs{\CurrentBib}

\bibitem [\protect \citeauthoryear {%
MacGorman%
, MacGorman%
, Rust%
\BCBL {}\ \BBA {} Rust%
}{%
MacGorman%
\ \protect \BOthers {.}}{%
{\protect \APACyear {1998}}%
}]{%
macgorman1998}
\APACinsertmetastar {%
macgorman1998}%
\begin{APACrefauthors}%
MacGorman, D\BPBI R.%
, MacGorman, R.%
, Rust, W\BPBI D.%
\BCBL {}\ \BBA {} Rust, W.%
\end{APACrefauthors}%
\unskip\
\newblock
\APACrefYear{1998}.
\newblock
\APACrefbtitle {The Electrical Nature of Storms} {The electrical nature of
  storms}.
\newblock
\APACaddressPublisher{}{Oxford University Press}.
\newblock
\begin{APACrefURL} \url{https://books.google.com/books?id=\_NbHNj7KJecC}
  \end{APACrefURL}
\PrintBackRefs{\CurrentBib}

\bibitem [\protect \citeauthoryear {%
Mailyan%
\ \protect \BOthers {.}}{%
Mailyan%
\ \protect \BOthers {.}}{%
{\protect \APACyear {2016}}%
}]{%
mailyan2016}
\APACinsertmetastar {%
mailyan2016}%
\begin{APACrefauthors}%
Mailyan, B\BPBI G.%
\BCBT {}\ \BOthersPeriod {.}
\end{APACrefauthors}%
\unskip\
\newblock
\APACrefYearMonthDay{2016}{}{}.
\newblock
{\BBOQ}\APACrefatitle {The spectroscopy of individual terrestrial gamma-ray
  flashes: Constraining the source properties} {The spectroscopy of individual
  terrestrial gamma-ray flashes: Constraining the source properties}.{\BBCQ}
\newblock
\APACjournalVolNumPages{Journal of Geophysical Research: Space
  Physics}{121}{11}{11,346-11,363}.
\newblock
\begin{APACrefDOI} \doi{10.1002/2016JA022702} \end{APACrefDOI}
\PrintBackRefs{\CurrentBib}

\bibitem [\protect \citeauthoryear {%
Mailyan%
\ \protect \BOthers {.}}{%
Mailyan%
\ \protect \BOthers {.}}{%
{\protect \APACyear {2018}}%
}]{%
mailyan2018}
\APACinsertmetastar {%
mailyan2018}%
\begin{APACrefauthors}%
Mailyan, B\BPBI G.%
\BCBT {}\ \BOthersPeriod {.}
\end{APACrefauthors}%
\unskip\
\newblock
\APACrefYearMonthDay{2018}{}{}.
\newblock
{\BBOQ}\APACrefatitle {Characteristics of Radio Emissions Associated With
  Terrestrial Gamma-Ray Flashes} {Characteristics of radio emissions associated
  with terrestrial gamma-ray flashes}.{\BBCQ}
\newblock
\APACjournalVolNumPages{Journal of Geophysical Research: Space
  Physics}{123}{7}{5933-5948}.
\newblock
\begin{APACrefDOI} \doi{10.1029/2018JA025450} \end{APACrefDOI}
\PrintBackRefs{\CurrentBib}

\bibitem [\protect \citeauthoryear {%
Mailyan%
\ \protect \BOthers {.}}{%
Mailyan%
\ \protect \BOthers {.}}{%
{\protect \APACyear {2019}}%
}]{%
mailyan2019}
\APACinsertmetastar {%
mailyan2019}%
\begin{APACrefauthors}%
Mailyan, B\BPBI G.%
\BCBT {}\ \BOthersPeriod {.}
\end{APACrefauthors}%
\unskip\
\newblock
\APACrefYearMonthDay{2019}{}{}.
\newblock
{\BBOQ}\APACrefatitle {Analysis of Individual Terrestrial Gamma-Ray Flashes
  With Lightning Leader Models and Fermi Gamma-Ray Burst Monitor Data}
  {Analysis of individual terrestrial gamma-ray flashes with lightning leader
  models and fermi gamma-ray burst monitor data}.{\BBCQ}
\newblock
\APACjournalVolNumPages{Journal of Geophysical Research:
  Atmospheres}{124}{8}{7170-7183}.
\newblock
\begin{APACrefDOI} \doi{10.1029/2019JA026912} \end{APACrefDOI}
\PrintBackRefs{\CurrentBib}

\bibitem [\protect \citeauthoryear {%
Malagon-Romero%
\ \BBA {} Luque%
}{%
Malagon-Romero%
\ \BBA {} Luque%
}{%
{\protect \APACyear {2019}}%
}]{%
malagon2019}
\APACinsertmetastar {%
malagon2019}%
\begin{APACrefauthors}%
Malagon-Romero, A.%
\BCBT {}\ \BBA {} Luque, A.%
\end{APACrefauthors}%
\unskip\
\newblock
\APACrefYearMonthDay{2019}{}{}.
\newblock
{\BBOQ}\APACrefatitle {Spontaneous emergence of space stems ahead of negative
  leaders in lightning and long sparks} {Spontaneous emergence of space stems
  ahead of negative leaders in lightning and long sparks}.{\BBCQ}
\newblock
\APACjournalVolNumPages{Geophysical Research Letters}{46}{7}{4029-4038}.
\newblock
\begin{APACrefDOI} \doi{10.1029/2019GL082063} \end{APACrefDOI}
\PrintBackRefs{\CurrentBib}

\bibitem [\protect \citeauthoryear {%
Marisaldi%
\ \protect \BOthers {.}}{%
Marisaldi%
\ \protect \BOthers {.}}{%
{\protect \APACyear {2014}}%
}]{%
marisaldi2014}
\APACinsertmetastar {%
marisaldi2014}%
\begin{APACrefauthors}%
Marisaldi, M.%
\BCBT {}\ \BOthersPeriod {.}
\end{APACrefauthors}%
\unskip\
\newblock
\APACrefYearMonthDay{2014}{}{}.
\newblock
{\BBOQ}\APACrefatitle {Properties of terrestrial gamma ray flashes detected by
  {AGILE} {MCAL} below 30{ M}e{V}} {Properties of terrestrial gamma ray flashes
  detected by {AGILE} {MCAL} below 30{ M}e{V}}.{\BBCQ}
\newblock
\APACjournalVolNumPages{Journal of Geophysical Research: Space
  Physics}{119}{2}{1337-1355}.
\newblock
\begin{APACrefDOI} \doi{10.1002/2013JA019301} \end{APACrefDOI}
\PrintBackRefs{\CurrentBib}

\bibitem [\protect \citeauthoryear {%
Marshall%
\ \protect \BOthers {.}}{%
Marshall%
\ \protect \BOthers {.}}{%
{\protect \APACyear {2013}}%
}]{%
marshall2013}
\APACinsertmetastar {%
marshall2013}%
\begin{APACrefauthors}%
Marshall, T.%
\BCBT {}\ \BOthersPeriod {.}
\end{APACrefauthors}%
\unskip\
\newblock
\APACrefYearMonthDay{2013}{}{}.
\newblock
{\BBOQ}\APACrefatitle {Initial breakdown pulses in intracloud lightning flashes
  and their relation to terrestrial gamma ray flashes} {Initial breakdown
  pulses in intracloud lightning flashes and their relation to terrestrial
  gamma ray flashes}.{\BBCQ}
\newblock
\APACjournalVolNumPages{Journal of Geophysical Research:
  Atmospheres}{118}{19}{10,907-10,925}.
\newblock
\begin{APACrefDOI} \doi{10.1002/jgrd.50866} \end{APACrefDOI}
\PrintBackRefs{\CurrentBib}

\bibitem [\protect \citeauthoryear {%
Moss%
, Pasko%
, Liu%
\BCBL {}\ \BBA {} Veronis%
}{%
Moss%
\ \protect \BOthers {.}}{%
{\protect \APACyear {2006}}%
}]{%
moss2006}
\APACinsertmetastar {%
moss2006}%
\begin{APACrefauthors}%
Moss, G.%
, Pasko, V.%
, Liu, N.%
\BCBL {}\ \BBA {} Veronis, G.%
\end{APACrefauthors}%
\unskip\
\newblock
\APACrefYearMonthDay{2006}{}{}.
\newblock
{\BBOQ}\APACrefatitle {Monte Carlo model for analysis of thermal runaway
  electrons in streamer tips in transient luminous events and streamer zones of
  lightning leaders} {Monte carlo model for analysis of thermal runaway
  electrons in streamer tips in transient luminous events and streamer zones of
  lightning leaders}.{\BBCQ}
\newblock
\APACjournalVolNumPages{Journal of Geophysical Research}{111}{A02307}{}.
\newblock
\begin{APACrefDOI} \doi{10.1029/2005JA011350} \end{APACrefDOI}
\PrintBackRefs{\CurrentBib}

\bibitem [\protect \citeauthoryear {%
Nag%
, DeCarlo%
\BCBL {}\ \BBA {} Rakov%
}{%
Nag%
\ \protect \BOthers {.}}{%
{\protect \APACyear {2009}}%
}]{%
nag2009}
\APACinsertmetastar {%
nag2009}%
\begin{APACrefauthors}%
Nag, A.%
, DeCarlo, B.%
\BCBL {}\ \BBA {} Rakov, V.%
\end{APACrefauthors}%
\unskip\
\newblock
\APACrefYearMonthDay{2009}{}{}.
\newblock
{\BBOQ}\APACrefatitle {Analysis of microsecond- and submicrosecond-scale
  electric field pulses produced by cloud and ground lightning discharges}
  {Analysis of microsecond- and submicrosecond-scale electric field pulses
  produced by cloud and ground lightning discharges}.{\BBCQ}
\newblock
\APACjournalVolNumPages{Atmospheric Research}{91}{}{316}.
\newblock
\begin{APACrefDOI} \doi{10.1016/j.atmosres.2008.01.014} \end{APACrefDOI}
\PrintBackRefs{\CurrentBib}

\bibitem [\protect \citeauthoryear {%
Nijdam%
, Teunissen%
\BCBL {}\ \BBA {} Ebert%
}{%
Nijdam%
\ \protect \BOthers {.}}{%
{\protect \APACyear {2020}}%
}]{%
nijdam2020}
\APACinsertmetastar {%
nijdam2020}%
\begin{APACrefauthors}%
Nijdam, S.%
, Teunissen, J.%
\BCBL {}\ \BBA {} Ebert, U.%
\end{APACrefauthors}%
\unskip\
\newblock
\APACrefYearMonthDay{2020}{}{}.
\newblock
{\BBOQ}\APACrefatitle {The physics of streamer discharge phenomena} {The
  physics of streamer discharge phenomena}.{\BBCQ}
\newblock
\APACjournalVolNumPages{Plasma Sources Science and Technology}{}{}{}.
\newblock
\begin{APACrefDOI} \doi{10.1088/1361-6595/abaa05} \end{APACrefDOI}
\PrintBackRefs{\CurrentBib}

\bibitem [\protect \citeauthoryear {%
{\O}stgaard%
\ \protect \BOthers {.}}{%
{\O}stgaard%
\ \protect \BOthers {.}}{%
{\protect \APACyear {2019}}%
}]{%
ostgaard2019}
\APACinsertmetastar {%
ostgaard2019}%
\begin{APACrefauthors}%
{\O}stgaard, N.%
\BCBT {}\ \BOthersPeriod {.}
\end{APACrefauthors}%
\unskip\
\newblock
\APACrefYearMonthDay{2019}{}{}.
\newblock
{\BBOQ}\APACrefatitle {First 10 Months of {TGF} Observations by {ASIM}} {First
  10 months of {TGF} observations by {ASIM}}.{\BBCQ}
\newblock
\APACjournalVolNumPages{Journal of Geophysical Research:
  Atmospheres}{124}{24}{14024-14036}.
\newblock
\begin{APACrefDOI} \doi{10.1029/2019JD031214} \end{APACrefDOI}
\PrintBackRefs{\CurrentBib}

\bibitem [\protect \citeauthoryear {%
Petersen%
, Bailey%
, Beasley%
\BCBL {}\ \BBA {} Hallett%
}{%
Petersen%
\ \protect \BOthers {.}}{%
{\protect \APACyear {2008}}%
}]{%
petersen2008}
\APACinsertmetastar {%
petersen2008}%
\begin{APACrefauthors}%
Petersen, D.%
, Bailey, M.%
, Beasley, W.%
\BCBL {}\ \BBA {} Hallett, J.%
\end{APACrefauthors}%
\unskip\
\newblock
\APACrefYearMonthDay{2008}{}{}.
\newblock
{\BBOQ}\APACrefatitle {A brief review of the problem of lightning initiation
  and a hypothesis of initial lightning leader formation} {A brief review of
  the problem of lightning initiation and a hypothesis of initial lightning
  leader formation}.{\BBCQ}
\newblock
\APACjournalVolNumPages{Journal of Geophysical Research:
  Atmospheres}{113}{D17}{}.
\newblock
\begin{APACrefDOI} \doi{10.1029/2007JD009036} \end{APACrefDOI}
\PrintBackRefs{\CurrentBib}

\bibitem [\protect \citeauthoryear {%
Pu%
\ \protect \BOthers {.}}{%
Pu%
\ \protect \BOthers {.}}{%
{\protect \APACyear {2019}}%
}]{%
pu2019}
\APACinsertmetastar {%
pu2019}%
\begin{APACrefauthors}%
Pu, Y.%
\BCBT {}\ \BOthersPeriod {.}
\end{APACrefauthors}%
\unskip\
\newblock
\APACrefYearMonthDay{2019}{}{}.
\newblock
{\BBOQ}\APACrefatitle {Low frequency radio pulses produced by terrestrial gamma
  ray flashes} {Low frequency radio pulses produced by terrestrial gamma ray
  flashes}.{\BBCQ}
\newblock
\APACjournalVolNumPages{Geophysical Research Letters}{46}{}{6990-6997}.
\newblock
\begin{APACrefDOI} \doi{10.1029/2019GL082743} \end{APACrefDOI}
\PrintBackRefs{\CurrentBib}

\bibitem [\protect \citeauthoryear {%
Ringuette%
\ \protect \BOthers {.}}{%
Ringuette%
\ \protect \BOthers {.}}{%
{\protect \APACyear {2013}}%
}]{%
ringuette2013}
\APACinsertmetastar {%
ringuette2013}%
\begin{APACrefauthors}%
Ringuette, R.%
\BCBT {}\ \BOthersPeriod {.}
\end{APACrefauthors}%
\unskip\
\newblock
\APACrefYearMonthDay{2013}{}{}.
\newblock
{\BBOQ}\APACrefatitle {{TETRA} observation of gamma-rays at ground level
  associated with nearby thunderstorms} {{TETRA} observation of gamma-rays at
  ground level associated with nearby thunderstorms}.{\BBCQ}
\newblock
\APACjournalVolNumPages{Journal of Geophysical Research: Space
  Physics}{118}{}{7841}.
\newblock
\begin{APACrefDOI} \doi{10.1002/jgra.50712} \end{APACrefDOI}
\PrintBackRefs{\CurrentBib}

\bibitem [\protect \citeauthoryear {%
Rison%
\ \protect \BOthers {.}}{%
Rison%
\ \protect \BOthers {.}}{%
{\protect \APACyear {2016}}%
}]{%
rison2016}
\APACinsertmetastar {%
rison2016}%
\begin{APACrefauthors}%
Rison, W.%
\BCBT {}\ \BOthersPeriod {.}
\end{APACrefauthors}%
\unskip\
\newblock
\APACrefYearMonthDay{2016}{}{}.
\newblock
{\BBOQ}\APACrefatitle {Observations of narrow bipolar events reveal how
  lightning is initiated in thunderstorms} {Observations of narrow bipolar
  events reveal how lightning is initiated in thunderstorms}.{\BBCQ}
\newblock
\APACjournalVolNumPages{Nature Communications}{7}{}{}.
\newblock
\begin{APACrefDOI} \doi{10.1038/ncomms10721} \end{APACrefDOI}
\PrintBackRefs{\CurrentBib}

\bibitem [\protect \citeauthoryear {%
Rison%
, Thomas%
, Krehbiel%
, Hamlin%
\BCBL {}\ \BBA {} Harlin%
}{%
Rison%
\ \protect \BOthers {.}}{%
{\protect \APACyear {1999}}%
}]{%
rison1999}
\APACinsertmetastar {%
rison1999}%
\begin{APACrefauthors}%
Rison, W.%
, Thomas, R.%
, Krehbiel, P.%
, Hamlin, T.%
\BCBL {}\ \BBA {} Harlin, J.%
\end{APACrefauthors}%
\unskip\
\newblock
\APACrefYearMonthDay{1999}{}{}.
\newblock
{\BBOQ}\APACrefatitle {A {GPS}-based three-dimensional lightning mapping
  system: Initial observations in central {N}ew {M}exico} {A {GPS}-based
  three-dimensional lightning mapping system: Initial observations in central
  {N}ew {M}exico}.{\BBCQ}
\newblock
\APACjournalVolNumPages{Geophysical Research Letters}{26}{23}{3573-3576}.
\newblock
\begin{APACrefDOI} \doi{10.1029/1999GL010856} \end{APACrefDOI}
\PrintBackRefs{\CurrentBib}

\bibitem [\protect \citeauthoryear {%
Roberts%
\ \protect \BOthers {.}}{%
Roberts%
\ \protect \BOthers {.}}{%
{\protect \APACyear {2017}}%
}]{%
roberts2017}
\APACinsertmetastar {%
roberts2017}%
\begin{APACrefauthors}%
Roberts, O.%
\BCBT {}\ \BOthersPeriod {.}
\end{APACrefauthors}%
\unskip\
\newblock
\APACrefYearMonthDay{2017}{}{}.
\newblock
{\BBOQ}\APACrefatitle {Terrestrial gamma ray flashes due to particle
  acceleration in tropical storm systems} {Terrestrial gamma ray flashes due to
  particle acceleration in tropical storm systems}.{\BBCQ}
\newblock
\APACjournalVolNumPages{Journal of Geophysical Research:
  Atmospheres}{122}{}{3374-3395}.
\newblock
\begin{APACrefDOI} \doi{10.1002/2016JD025799} \end{APACrefDOI}
\PrintBackRefs{\CurrentBib}

\bibitem [\protect \citeauthoryear {%
Shao%
, Hamlin%
\BCBL {}\ \BBA {} Smith%
}{%
Shao%
\ \protect \BOthers {.}}{%
{\protect \APACyear {2010}}%
}]{%
shao2010}
\APACinsertmetastar {%
shao2010}%
\begin{APACrefauthors}%
Shao, X.%
, Hamlin, T.%
\BCBL {}\ \BBA {} Smith, D.%
\end{APACrefauthors}%
\unskip\
\newblock
\APACrefYearMonthDay{2010}{}{}.
\newblock
{\BBOQ}\APACrefatitle {A closer examination of terrestrial gamma ray flash
  related lightning processes} {A closer examination of terrestrial gamma ray
  flash related lightning processes}.{\BBCQ}
\newblock
\APACjournalVolNumPages{Journal of Geophysical Research: Space
  Physics}{115}{}{A00E30}.
\newblock
\begin{APACrefDOI} \doi{10.1029/2009JA014835} \end{APACrefDOI}
\PrintBackRefs{\CurrentBib}

\bibitem [\protect \citeauthoryear {%
D.~Shi%
, Wang%
, Wu%
\BCBL {}\ \BBA {} Takagi%
}{%
D.~Shi%
\ \protect \BOthers {.}}{%
{\protect \APACyear {2019}}%
}]{%
shi2019}
\APACinsertmetastar {%
shi2019}%
\begin{APACrefauthors}%
Shi, D.%
, Wang, D.%
, Wu, T.%
\BCBL {}\ \BBA {} Takagi, N.%
\end{APACrefauthors}%
\unskip\
\newblock
\APACrefYearMonthDay{2019}{}{}.
\newblock
{\BBOQ}\APACrefatitle {Temporal and Spatial Characteristics of Preliminary
  Breakdown Pulses in Intracloud Lightning Flashes} {Temporal and spatial
  characteristics of preliminary breakdown pulses in intracloud lightning
  flashes}.{\BBCQ}
\newblock
\APACjournalVolNumPages{Journal of Geophysical Research:
  Atmospheres}{124}{23}{12901-12914}.
\newblock
\begin{APACrefDOI} \doi{10.1029/2019JD031130} \end{APACrefDOI}
\PrintBackRefs{\CurrentBib}

\bibitem [\protect \citeauthoryear {%
F.~Shi%
, Liu%
\BCBL {}\ \BBA {} Rassoul%
}{%
F.~Shi%
\ \protect \BOthers {.}}{%
{\protect \APACyear {2016}}%
}]{%
shi2016}
\APACinsertmetastar {%
shi2016}%
\begin{APACrefauthors}%
Shi, F.%
, Liu, N.%
\BCBL {}\ \BBA {} Rassoul, H\BPBI K.%
\end{APACrefauthors}%
\unskip\
\newblock
\APACrefYearMonthDay{2016}{}{}.
\newblock
{\BBOQ}\APACrefatitle {Properties of relatively long streamers initiated from
  an isolated hydrometeor} {Properties of relatively long streamers initiated
  from an isolated hydrometeor}.{\BBCQ}
\newblock
\APACjournalVolNumPages{Journal of Geophysical Research:
  Atmospheres}{121}{}{7284-7295}.
\newblock
\begin{APACrefDOI} \doi{10.1002/2015JD024580} \end{APACrefDOI}
\PrintBackRefs{\CurrentBib}

\bibitem [\protect \citeauthoryear {%
Smith%
\ \protect \BOthers {.}}{%
Smith%
\ \protect \BOthers {.}}{%
{\protect \APACyear {2018}}%
}]{%
smith2018}
\APACinsertmetastar {%
smith2018}%
\begin{APACrefauthors}%
Smith, D\BPBI M.%
\BCBT {}\ \BOthersPeriod {.}
\end{APACrefauthors}%
\unskip\
\newblock
\APACrefYearMonthDay{2018}{}{}.
\newblock
{\BBOQ}\APACrefatitle {Characterizing Upward Lightning With and Without a
  Terrestrial Gamma Ray Flash} {Characterizing upward lightning with and
  without a terrestrial gamma ray flash}.{\BBCQ}
\newblock
\APACjournalVolNumPages{Journal of Geophysical Research:
  Atmospheres}{123}{20}{}.
\newblock
\begin{APACrefDOI} \doi{10.1029/2018JD029105} \end{APACrefDOI}
\PrintBackRefs{\CurrentBib}

\bibitem [\protect \citeauthoryear {%
Stanley%
\ \protect \BOthers {.}}{%
Stanley%
\ \protect \BOthers {.}}{%
{\protect \APACyear {2006}}%
}]{%
Stanley2006}
\APACinsertmetastar {%
Stanley2006}%
\begin{APACrefauthors}%
Stanley, M\BPBI A.%
\BCBT {}\ \BOthersPeriod {.}
\end{APACrefauthors}%
\unskip\
\newblock
\APACrefYearMonthDay{2006}{}{}.
\newblock
{\BBOQ}\APACrefatitle {A link between terrestrial gamma-ray flashes and
  intracloud lightning discharges} {A link between terrestrial gamma-ray
  flashes and intracloud lightning discharges}.{\BBCQ}
\newblock
\APACjournalVolNumPages{Geophysical Research Letters}{33}{6}{}.
\newblock
\APACrefnote{L06803}
\newblock
\begin{APACrefDOI} \doi{10.1029/2005GL025537} \end{APACrefDOI}
\PrintBackRefs{\CurrentBib}

\bibitem [\protect \citeauthoryear {%
Stock%
\ \protect \BOthers {.}}{%
Stock%
\ \protect \BOthers {.}}{%
{\protect \APACyear {2014}}%
}]{%
stock2014a}
\APACinsertmetastar {%
stock2014a}%
\begin{APACrefauthors}%
Stock, M\BPBI G.%
\BCBT {}\ \BOthersPeriod {.}
\end{APACrefauthors}%
\unskip\
\newblock
\APACrefYearMonthDay{2014}{}{}.
\newblock
{\BBOQ}\APACrefatitle {Continuous broadband digital interferometry of lightning
  using a generalized cross-correlation algorithm} {Continuous broadband
  digital interferometry of lightning using a generalized cross-correlation
  algorithm}.{\BBCQ}
\newblock
\APACjournalVolNumPages{Journal of Geophysical Research:
  Atmospheres}{119}{6}{3134-3165}.
\newblock
\begin{APACrefDOI} \doi{10.1002/2013JD020217} \end{APACrefDOI}
\PrintBackRefs{\CurrentBib}

\bibitem [\protect \citeauthoryear {%
Stolzenburg%
, Marshall%
, Karunarathne%
\BCBL {}\ \BBA {} Orville%
}{%
Stolzenburg%
\ \protect \BOthers {.}}{%
{\protect \APACyear {2016}}%
}]{%
stolzenburg2016}
\APACinsertmetastar {%
stolzenburg2016}%
\begin{APACrefauthors}%
Stolzenburg, M.%
, Marshall, T\BPBI C.%
, Karunarathne, S.%
\BCBL {}\ \BBA {} Orville, R\BPBI E.%
\end{APACrefauthors}%
\unskip\
\newblock
\APACrefYearMonthDay{2016}{}{}.
\newblock
{\BBOQ}\APACrefatitle {Luminosity with intracloud-type lightning initial
  breakdown pulses and terrestrial gamma-ray flash candidates} {Luminosity with
  intracloud-type lightning initial breakdown pulses and terrestrial gamma-ray
  flash candidates}.{\BBCQ}
\newblock
\APACjournalVolNumPages{Journal of Geophysical Research:
  Atmospheres}{121}{}{10919-10936}.
\newblock
\begin{APACrefDOI} \doi{10.1002/2016JD025202} \end{APACrefDOI}
\PrintBackRefs{\CurrentBib}

\bibitem [\protect \citeauthoryear {%
Stolzenburg%
\ \protect \BOthers {.}}{%
Stolzenburg%
\ \protect \BOthers {.}}{%
{\protect \APACyear {2013}}%
}]{%
stolzenburg2013}
\APACinsertmetastar {%
stolzenburg2013}%
\begin{APACrefauthors}%
Stolzenburg, M.%
\BCBT {}\ \BOthersPeriod {.}
\end{APACrefauthors}%
\unskip\
\newblock
\APACrefYearMonthDay{2013}{}{}.
\newblock
{\BBOQ}\APACrefatitle {Luminosity of initial breakdown in lightning}
  {Luminosity of initial breakdown in lightning}.{\BBCQ}
\newblock
\APACjournalVolNumPages{Journal of Geophysical Research}{118}{}{2918-2937}.
\newblock
\begin{APACrefDOI} \doi{10.1002/jgrd.50276} \end{APACrefDOI}
\PrintBackRefs{\CurrentBib}

\bibitem [\protect \citeauthoryear {%
Stolzenburg%
\ \protect \BOthers {.}}{%
Stolzenburg%
\ \protect \BOthers {.}}{%
{\protect \APACyear {2014}}%
}]{%
stolzenburg2014}
\APACinsertmetastar {%
stolzenburg2014}%
\begin{APACrefauthors}%
Stolzenburg, M.%
\BCBT {}\ \BOthersPeriod {.}
\end{APACrefauthors}%
\unskip\
\newblock
\APACrefYearMonthDay{2014}{}{}.
\newblock
{\BBOQ}\APACrefatitle {Leader observations during the initial breakdown stage
  of a lightning flash} {Leader observations during the initial breakdown stage
  of a lightning flash}.{\BBCQ}
\newblock
\APACjournalVolNumPages{Journal of Geophysical Research:
  Atmospheres}{119}{21}{12,198-12,221}.
\newblock
\begin{APACrefDOI} \doi{10.1002/2014JD021994} \end{APACrefDOI}
\PrintBackRefs{\CurrentBib}

\bibitem [\protect \citeauthoryear {%
Thomas%
\ \protect \BOthers {.}}{%
Thomas%
\ \protect \BOthers {.}}{%
{\protect \APACyear {2004}}%
}]{%
thomas2004}
\APACinsertmetastar {%
thomas2004}%
\begin{APACrefauthors}%
Thomas, R\BPBI J.%
\BCBT {}\ \BOthersPeriod {.}
\end{APACrefauthors}%
\unskip\
\newblock
\APACrefYearMonthDay{2004}{}{}.
\newblock
{\BBOQ}\APACrefatitle {Accuracy of the {L}ightning {M}apping {A}rray} {Accuracy
  of the {L}ightning {M}apping {A}rray}.{\BBCQ}
\newblock
\APACjournalVolNumPages{Journal of Geophysical Research:
  Atmospheres}{109}{D14}{}.
\newblock
\begin{APACrefDOI} \doi{10.1029/2004JD004549} \end{APACrefDOI}
\PrintBackRefs{\CurrentBib}

\bibitem [\protect \citeauthoryear {%
Tilles%
}{%
Tilles%
}{%
{\protect \APACyear {2020}}%
}]{%
tilles2020}
\APACinsertmetastar {%
tilles2020}%
\begin{APACrefauthors}%
Tilles, J.%
\end{APACrefauthors}%
\unskip\
\newblock
\APACrefYearMonthDay{2020}{}{}.
\newblock
\APACrefbtitle {Broadband Radio Mapping and Imaging of Lightning Processes.}
  {Broadband radio mapping and imaging of lightning processes.}
\newblock
\APACaddressPublisher{}{PhD thesis, University of New Hampshire}.
\PrintBackRefs{\CurrentBib}

\bibitem [\protect \citeauthoryear {%
Tilles%
\ \protect \BOthers {.}}{%
Tilles%
\ \protect \BOthers {.}}{%
{\protect \APACyear {2019}}%
}]{%
tilles2019a}
\APACinsertmetastar {%
tilles2019a}%
\begin{APACrefauthors}%
Tilles, J.%
\BCBT {}\ \BOthersPeriod {.}
\end{APACrefauthors}%
\unskip\
\newblock
\APACrefYearMonthDay{2019}{}{}.
\newblock
{\BBOQ}\APACrefatitle {Fast negative breakdown in thunderstorms} {Fast negative
  breakdown in thunderstorms}.{\BBCQ}
\newblock
\APACjournalVolNumPages{Nature Communications}{10}{}{1648}.
\newblock
\begin{APACrefDOI} \doi{10.1038/s41467-019-09621-z} \end{APACrefDOI}
\PrintBackRefs{\CurrentBib}

\bibitem [\protect \citeauthoryear {%
Tilles%
\ \protect \BOthers {.}}{%
Tilles%
\ \protect \BOthers {.}}{%
{\protect \APACyear {2020}}%
}]{%
tilles2020b}
\APACinsertmetastar {%
tilles2020b}%
\begin{APACrefauthors}%
Tilles, J.%
\BCBT {}\ \BOthersPeriod {.}
\end{APACrefauthors}%
\unskip\
\newblock
\APACrefYearMonthDay{2020}{}{}.
\newblock
{\BBOQ}\APACrefatitle {Radio interferometer observations of an energetic
  in-cloud pulse reveal large currents generated by relativistic discharges}
  {Radio interferometer observations of an energetic in-cloud pulse reveal
  large currents generated by relativistic discharges}.{\BBCQ}
\newblock
\APACjournalVolNumPages{Journal of Geophysical Research: Atmospheres}{}{}{}.
\newblock
\begin{APACrefDOI} \doi{10.1029/2020JD032603} \end{APACrefDOI}
\PrintBackRefs{\CurrentBib}

\bibitem [\protect \citeauthoryear {%
Tran%
\ \protect \BOthers {.}}{%
Tran%
\ \protect \BOthers {.}}{%
{\protect \APACyear {2015}}%
}]{%
tran2015}
\APACinsertmetastar {%
tran2015}%
\begin{APACrefauthors}%
Tran, M.%
\BCBT {}\ \BOthersPeriod {.}
\end{APACrefauthors}%
\unskip\
\newblock
\APACrefYearMonthDay{2015}{}{}.
\newblock
{\BBOQ}\APACrefatitle {A terrestrial gamma-ray flash recorded at the lightning
  observatory in {G}ainsville, {F}lorida} {A terrestrial gamma-ray flash
  recorded at the lightning observatory in {G}ainsville, {F}lorida}.{\BBCQ}
\newblock
\APACjournalVolNumPages{Journal of Atmospheric and Solar-Terrestrial
  Physics}{136}{}{86}.
\newblock
\begin{APACrefDOI} \doi{10.1016/j.jastp.2015.10.010} \end{APACrefDOI}
\PrintBackRefs{\CurrentBib}

\bibitem [\protect \citeauthoryear {%
Wada%
\ \protect \BOthers {.}}{%
Wada%
\ \protect \BOthers {.}}{%
{\protect \APACyear {2019}}%
}]{%
wada2019}
\APACinsertmetastar {%
wada2019}%
\begin{APACrefauthors}%
Wada, Y.%
\BCBT {}\ \BOthersPeriod {.}
\end{APACrefauthors}%
\unskip\
\newblock
\APACrefYearMonthDay{2019}{}{}.
\newblock
{\BBOQ}\APACrefatitle {Gamma-ray glow preceding downward terrestrial gamma-ray
  flash} {Gamma-ray glow preceding downward terrestrial gamma-ray
  flash}.{\BBCQ}
\newblock
\APACjournalVolNumPages{Physics Communications}{2}{67}{}.
\newblock
\begin{APACrefDOI} \doi{10.1038/s42005-019-0168-y} \end{APACrefDOI}
\PrintBackRefs{\CurrentBib}

\bibitem [\protect \citeauthoryear {%
Weidman%
\ \BBA {} Krider%
}{%
Weidman%
\ \BBA {} Krider%
}{%
{\protect \APACyear {1979}}%
}]{%
weidman1979}
\APACinsertmetastar {%
weidman1979}%
\begin{APACrefauthors}%
Weidman, C\BPBI D.%
\BCBT {}\ \BBA {} Krider, E\BPBI P.%
\end{APACrefauthors}%
\unskip\
\newblock
\APACrefYearMonthDay{1979}{}{}.
\newblock
{\BBOQ}\APACrefatitle {The radiation field waveforms produced by intracloud
  lightning discharge processes} {The radiation field waveforms produced by
  intracloud lightning discharge processes}.{\BBCQ}
\newblock
\APACjournalVolNumPages{Journal of Geophysical Research}{84}{}{3159}.
\newblock
\begin{APACrefDOI} \doi{10.1029/JC084iC06p03159} \end{APACrefDOI}
\PrintBackRefs{\CurrentBib}

\bibitem [\protect \citeauthoryear {%
Winn%
\ \protect \BOthers {.}}{%
Winn%
\ \protect \BOthers {.}}{%
{\protect \APACyear {2011}}%
}]{%
winn2011}
\APACinsertmetastar {%
winn2011}%
\begin{APACrefauthors}%
Winn, W\BPBI P.%
\BCBT {}\ \BOthersPeriod {.}
\end{APACrefauthors}%
\unskip\
\newblock
\APACrefYearMonthDay{2011}{}{}.
\newblock
{\BBOQ}\APACrefatitle {Lightning leader stepping, {K} changes, and other
  observations near an intracloud flash} {Lightning leader stepping, {K}
  changes, and other observations near an intracloud flash}.{\BBCQ}
\newblock
\APACjournalVolNumPages{Journal of Geophysical Research:
  Atmospheres}{116}{D23}{}.
\newblock
\begin{APACrefDOI} \doi{10.1029/2011JD015998} \end{APACrefDOI}
\PrintBackRefs{\CurrentBib}

\bibitem [\protect \citeauthoryear {%
Zyla%
\ \protect \BOthers {.}}{%
Zyla%
\ \protect \BOthers {.}}{%
{\protect \APACyear {2020}}%
}]{%
pdg2020}
\APACinsertmetastar {%
pdg2020}%
\begin{APACrefauthors}%
Zyla, P.%
\BCBT {}\ \BOthersPeriod {.}
\end{APACrefauthors}%
\unskip\
\newblock
\APACrefYearMonthDay{2020}{}{}.
\newblock
{\BBOQ}\APACrefatitle {Particle Data Group} {Particle data group}.{\BBCQ}
\newblock
\APACjournalVolNumPages{Progress of Theoretical and Experimental
  Physics}{}{083C01 (2020)}{}.
\newblock
\begin{APACrefURL}
  \url{http://pdg.lbl.gov/2020/reviews/rpp2020-rev-passage-particles-matter.pdf}
  \end{APACrefURL}
\PrintBackRefs{\CurrentBib}

\end{thebibliography}

%
%
%
%
%

\end{document}


%
%


\title{Supporting Information for ``Observations of the Origin of Downward Terrestrial Gamma-Ray Flashes''}
%
%

%
%




J.W.~Belz$^{1}$, P.R.~Krehbiel$^{2}$, J.~Remington$^{1}$, M.A.~Stanley$^{2}$, R.U.~Abbasi$^{3}$, R.~LeVon$^{1}$, W.~Rison$^{2}$, D.~Rodeheffer$^{2}$
\\ {\em and the Telescope Array Scientific Collaboration} \\
T.~Abu-Zayyad$^{1}$, M.~Allen$^{1}$
% ...


\affiliation{1}{Department of Physics and Astronomy, University of Utah, Salt Lake City, Utah, USA}
\affiliation{2}{Langmuir Laboratory for Atmospheric Research, New Mexico Institute of Mining and Technology, Socorro, NM, USA}
\affiliation{3}{Department of Physics, Loyola University Chicago, Chicago, Illinois, USA}
% ...


%
%

%

\begin{article}

%
%

\newpage
\noindent\textbf{Contents of this file}
\begin{enumerate}
\item Figures S1 to S31
\item Tables S1 to S3
\end{enumerate}

\noindent\textbf{Introduction}

The tables and figures in this supporting document provide further detail on the Terrestrial Gamma-Ray Flash (TGF) observations by the Telescope Array Surface Detector (TASD), Lightning Mapping Array (LMA), fast electric field change antenna (FA), and VHF interferometer (INTF).

\end{article}
\clearpage


%
%
%
%
%
%
%
\setstretch{1.2}

\begin{figure}
\centering
\begin{tabular}{@{}cc@{}}
\includegraphics[width=0.45\textwidth]{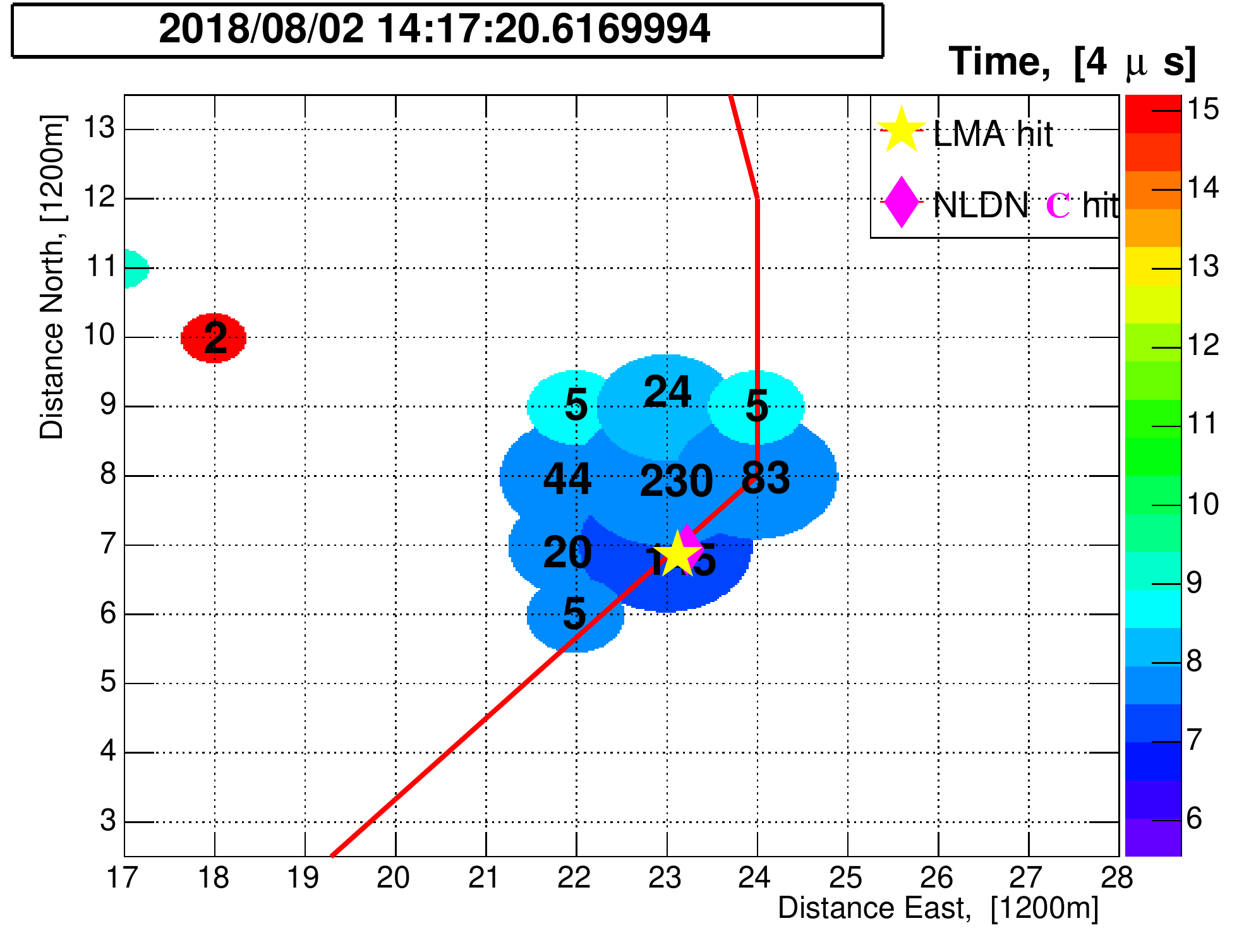} & 
\includegraphics[width=0.45\textwidth]{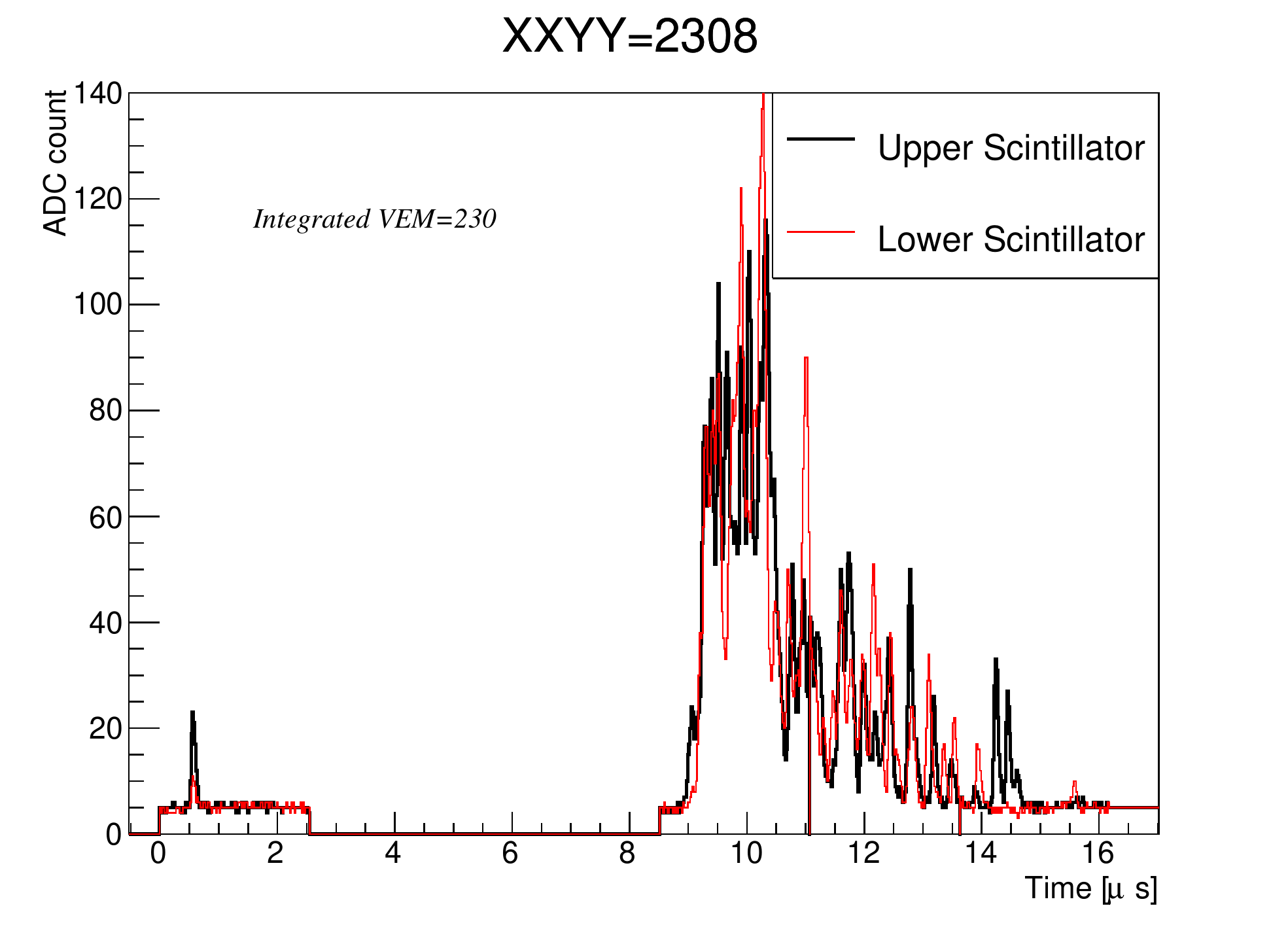} \\
\includegraphics[width=0.45\textwidth]{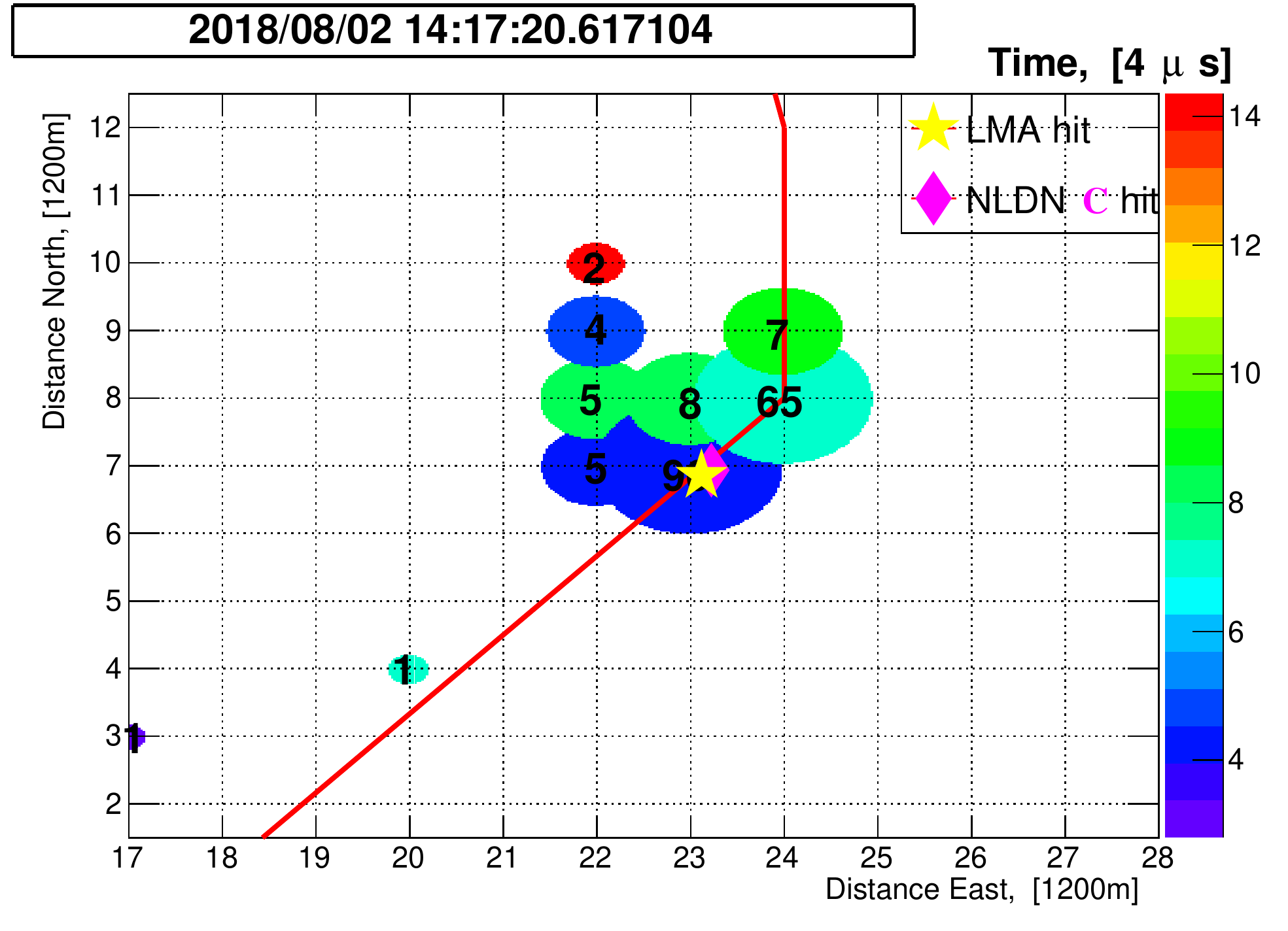} & 
\includegraphics[width=0.45\textwidth]{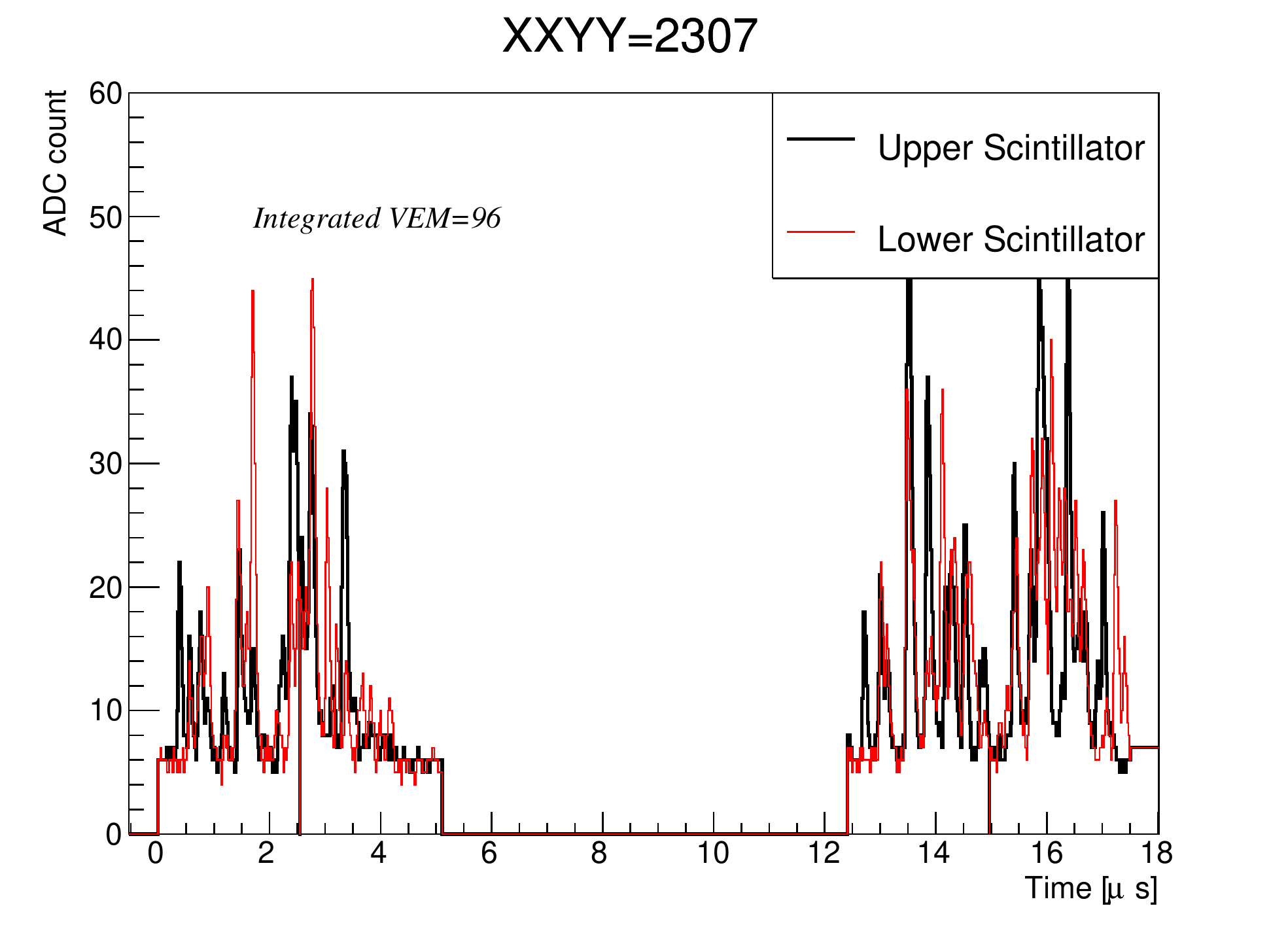} \\
\end{tabular}
\caption{ \textbf{TASD observations for TGF~A.} Footprints for the two
triggers of TGF~A and the waveforms at the TASD recording the strongest
energy deposit. Numbers in the footprint circles indicate the VEM energy
deposit, and color indicates the relative onset times (4~$\mu$s intervals).
The yellow star indicates the median plan location of the LMA sources within 
$\pm 1$~ms of the gamma burst. The magenta diamond indicates the NLDN event
corresponding the TGF sferic or the one closest in time prior to the TGF (see
Table~\ref{tab:events1}). Both events occurred over the edge of
southeastern TASD boundary (red line). }
\label{fig:flasha} 
\end{figure}

\begin{figure}
\centering
\begin{tabular}{@{}cc@{}}
\includegraphics[width=0.40\textwidth]{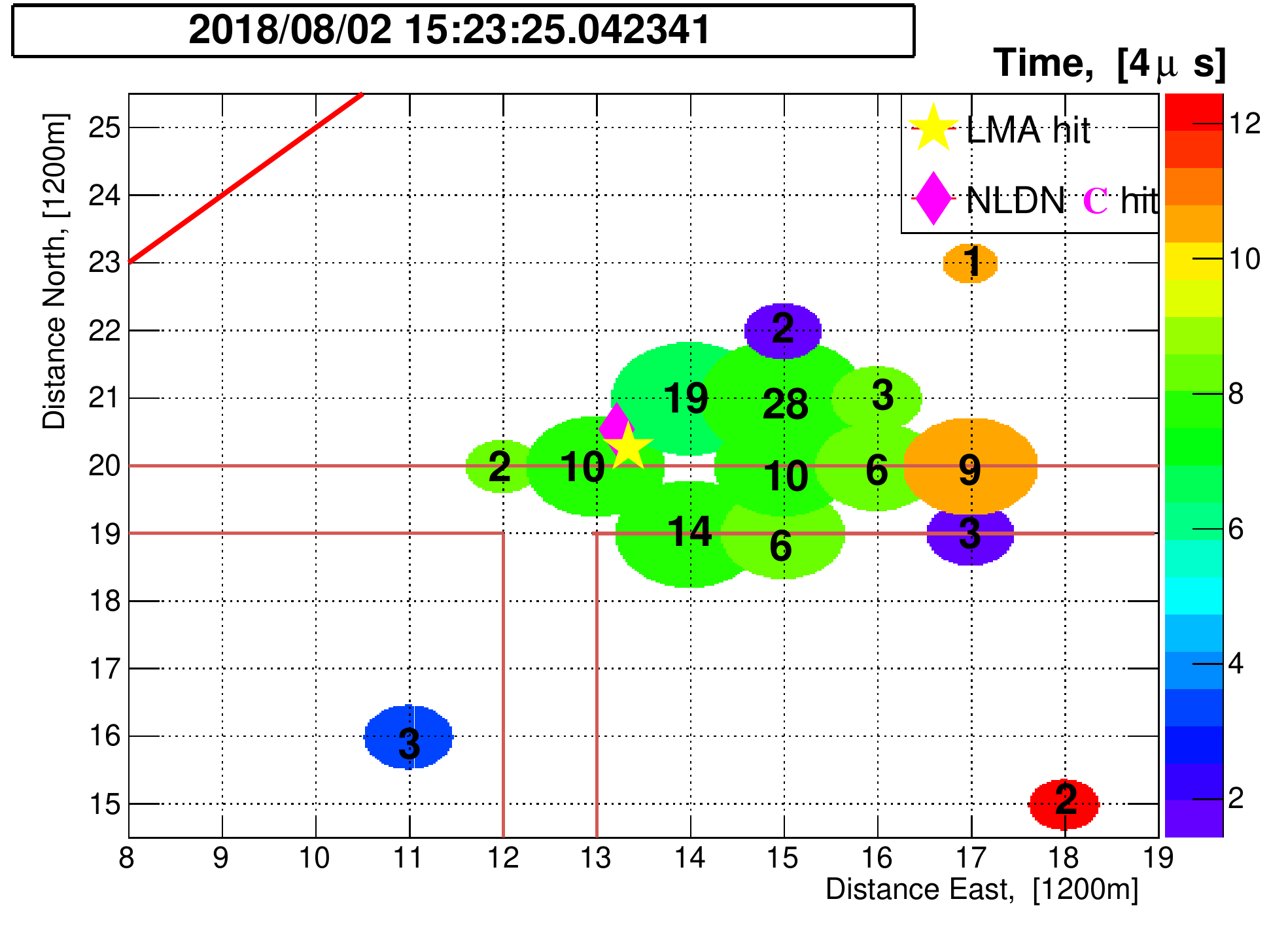} &
\includegraphics[width=0.43\textwidth]{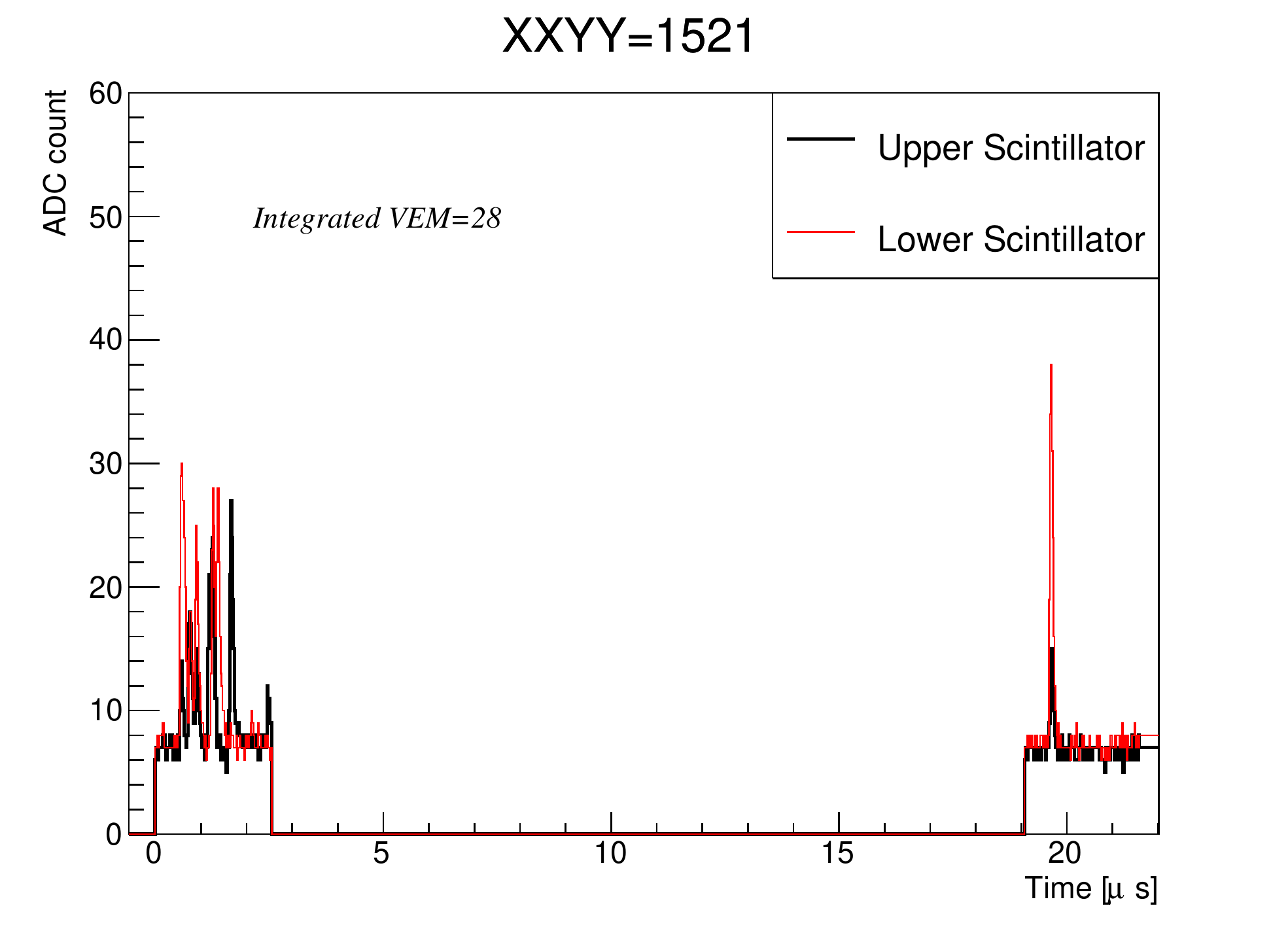} \\
\end{tabular}
\caption{\textbf{TASD observations for TGF~B.} Same as
Figure~\ref{fig:flasha}, except for the single trigger of TGF~B. The TGF was 
detected at four different TASDs or sets of TASDs at different onset times, beginning
with TASD 1421 immediately NE of the LMA- and NLDN-indicated source location,
and continuing in a rapid succession around a central annular hole to
the eastern, southern, and finally western station, finishing up almost
directly around the TGF's source. TASD 1420 associated with the annular hole did not record a
trigger, indicating the gamma bursts were relatively well-beamed (the
station was fully operational throughout the storm and was active for a cosmic ray event later in the day).}

\label{fig:flashb} 
\end{figure}

\begin{figure}
\centering
\begin{tabular}{@{}cc@{}}
\includegraphics[width=0.40\textwidth]{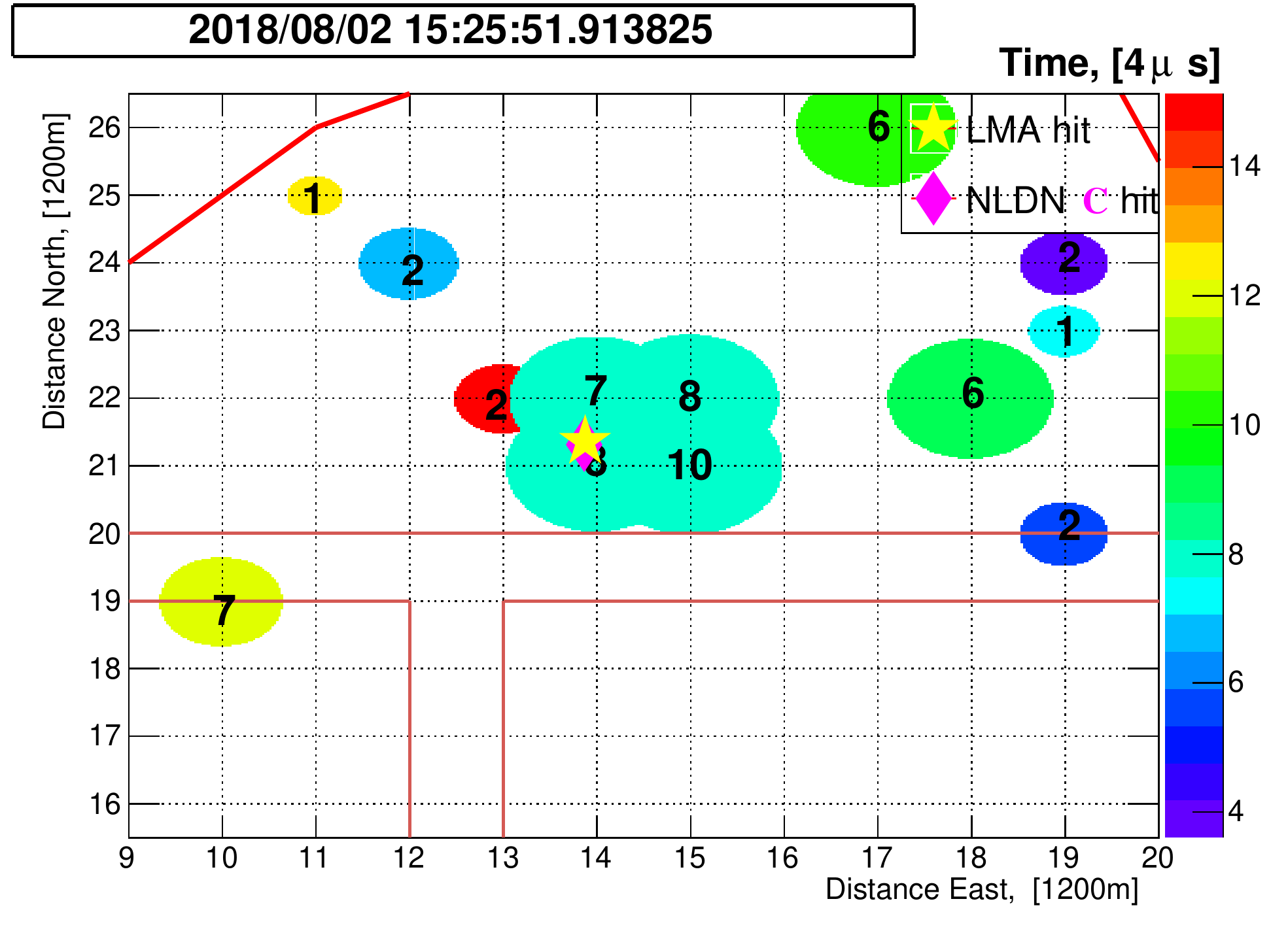} & 
\includegraphics[width=0.43\textwidth]{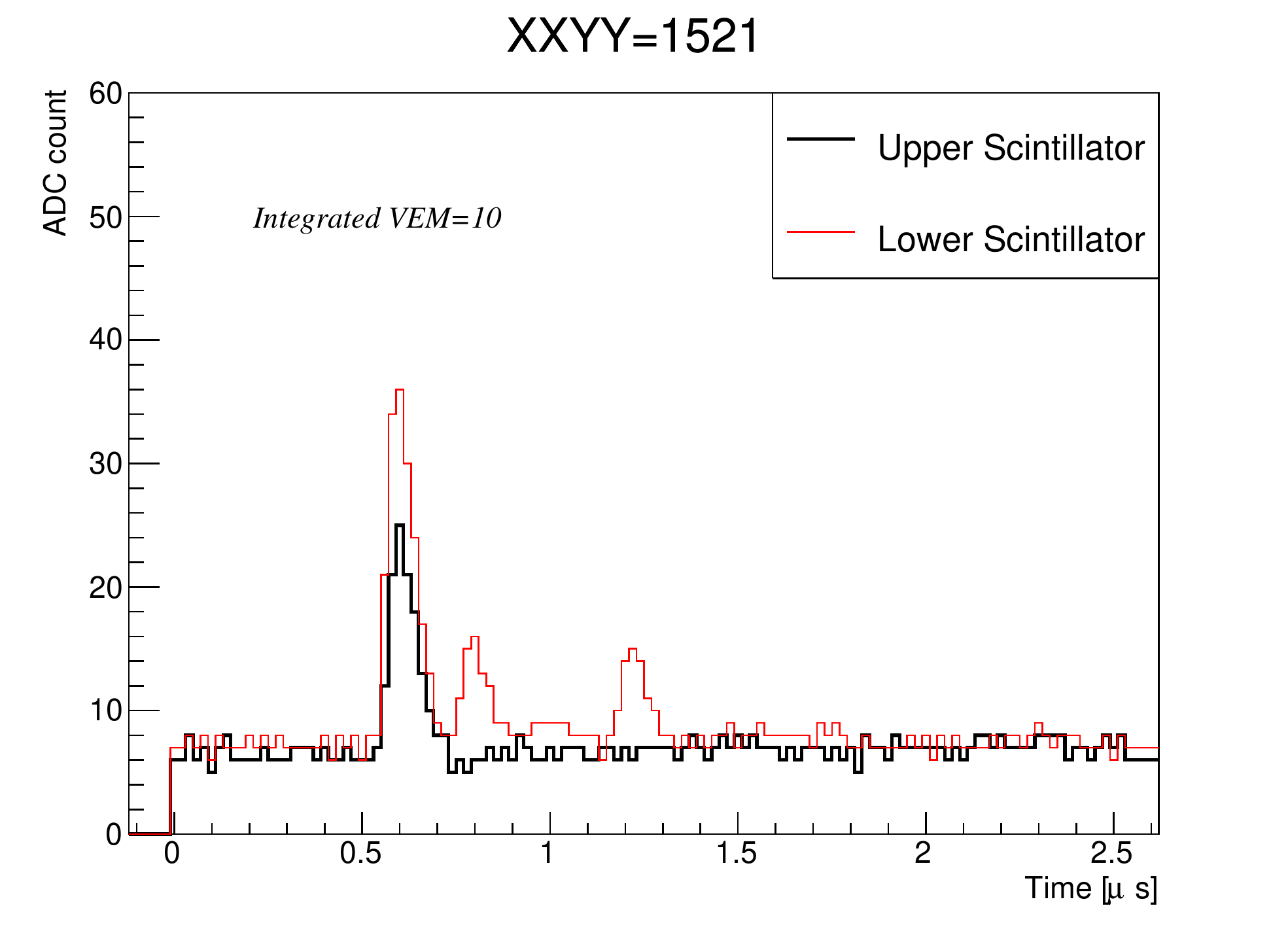} \\
\includegraphics[width=0.40\textwidth]{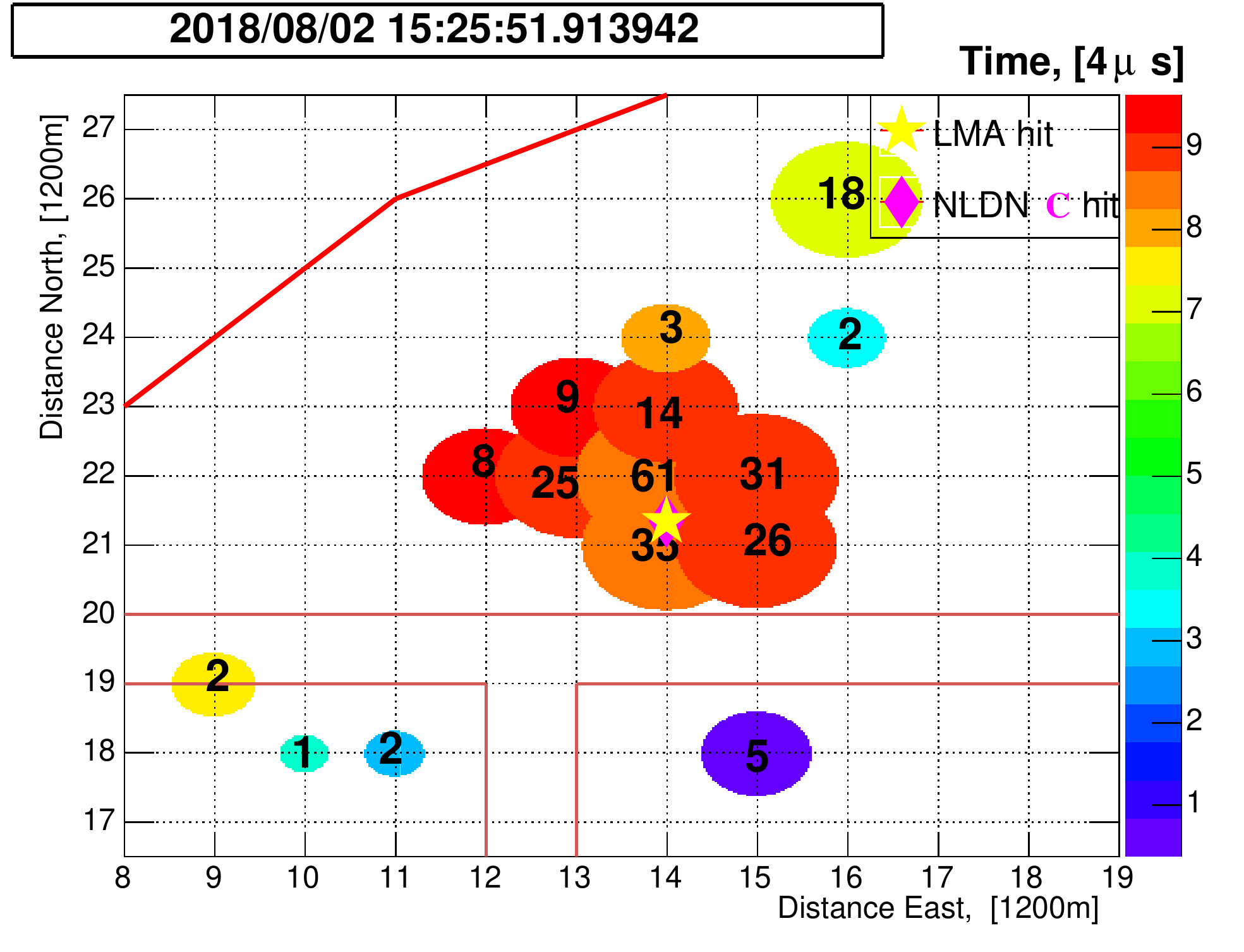} & 
\includegraphics[width=0.43\textwidth]{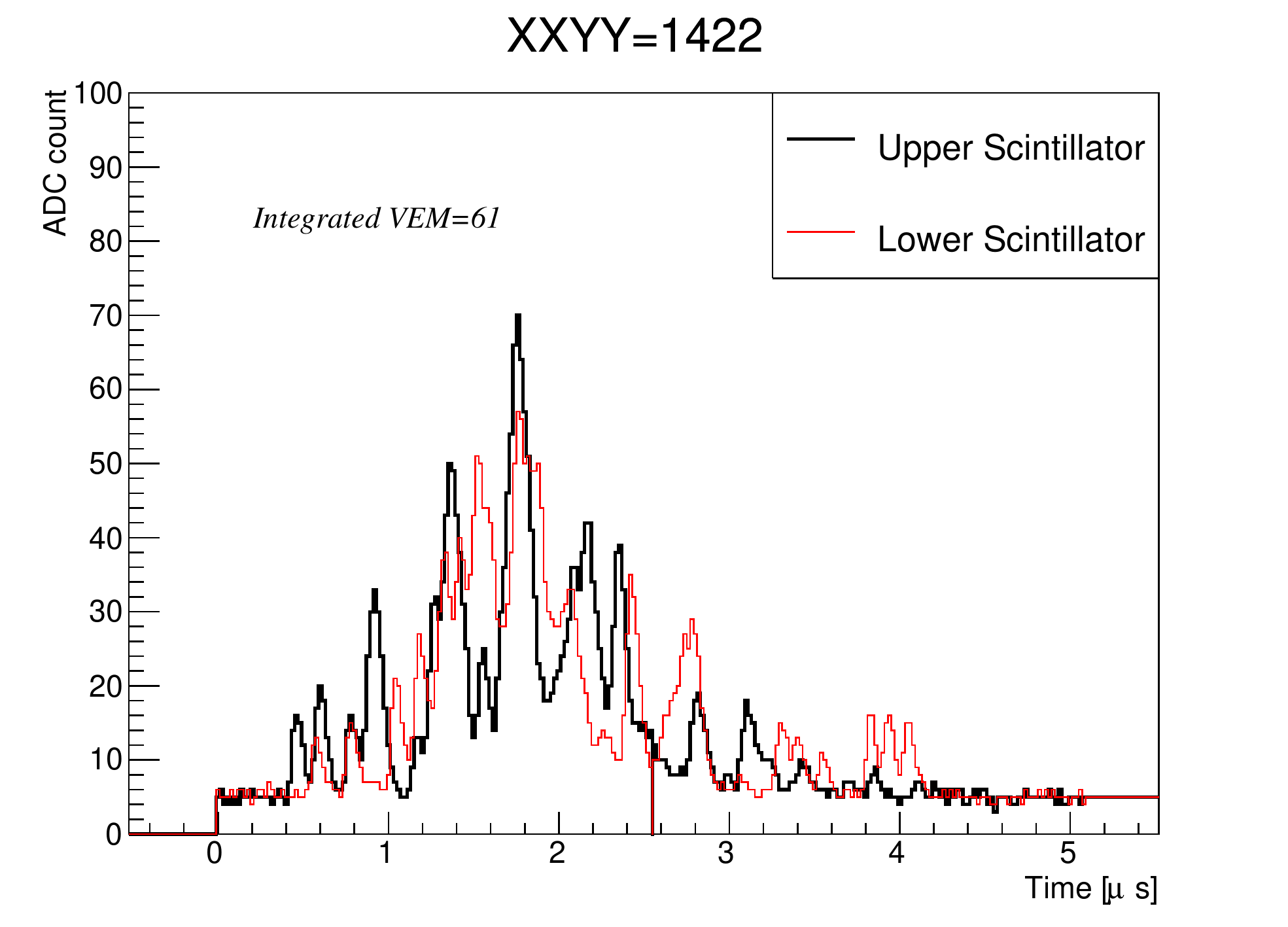} \\
\end{tabular}
\caption{\textbf{TASD observations for TGF~C.} Same as
Figure~\ref{fig:flasha}, except for two gamma bursts/triggers of TGF~C. Both
events were canonical examples of the basic processes involved in TGF
production (see text at end of Section~3.3). The zoomed-in views of the TASD signals in the right-hand panels illustrate
the fact that the SDs are sensing individual Compton electrons, with first event of the top right panel corresponding to an electron that penetrated both the upper and lower layers, and therefore was produced
by a gamma photon having a minimum energy of 6.4 MeV for rebounding
collisions and a most likely energy up to three times that for grazing
collisions (text at end of Section 3.1). The bottom right panel further
illustrates the individual nature of detections and their quantization both
in time and amplitude.The non-diagonal red lines of these footprints and the
footprint of TGF~B denote internal boundaries of different sub-sectors of the
TASD.}

\label{fig:flashc}
\end{figure} 

\begin{figure}
\centering
\begin{tabular}{@{}cc@{}}
\includegraphics[width=0.40\textwidth]{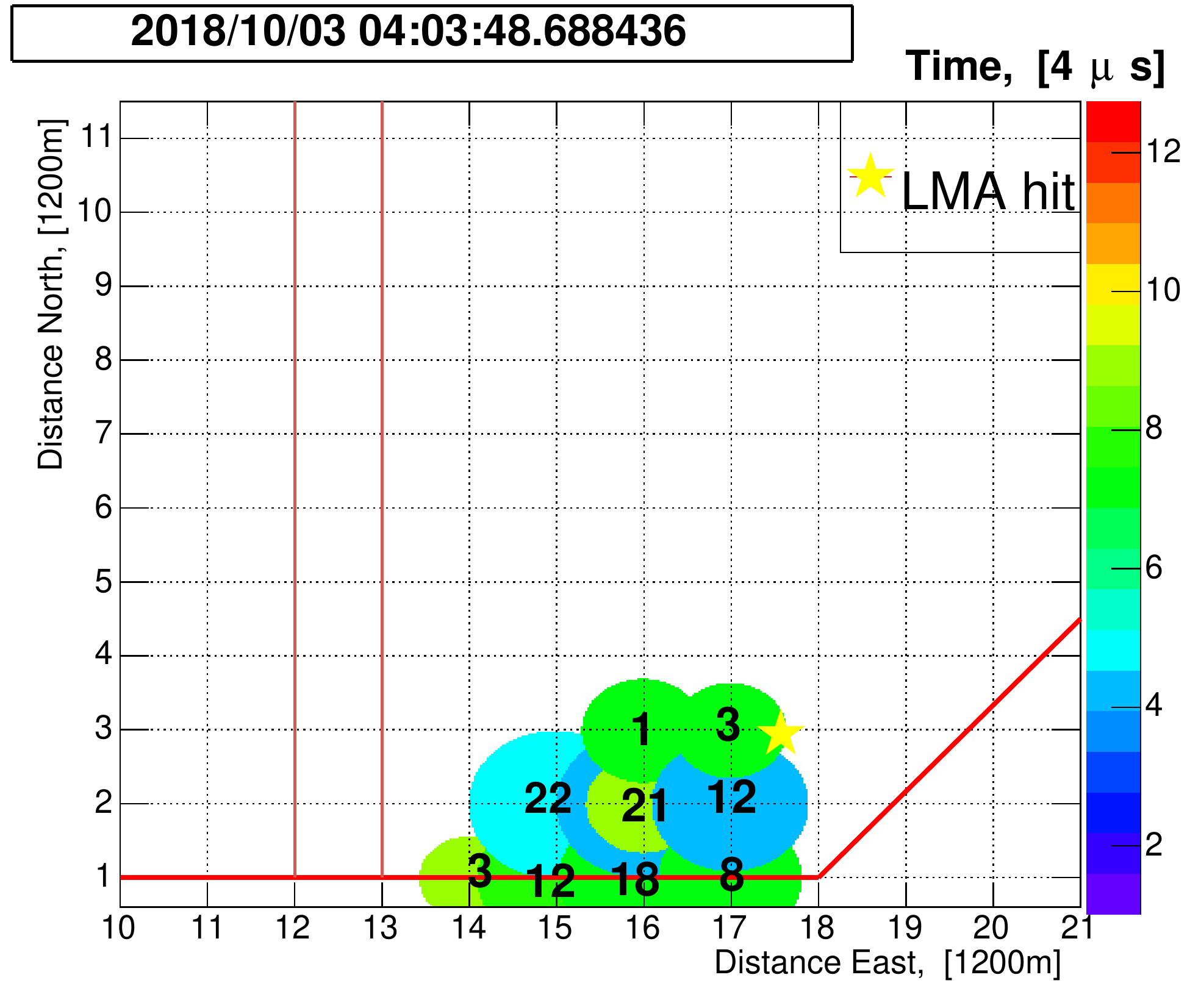} &
\includegraphics[width=0.43\textwidth]{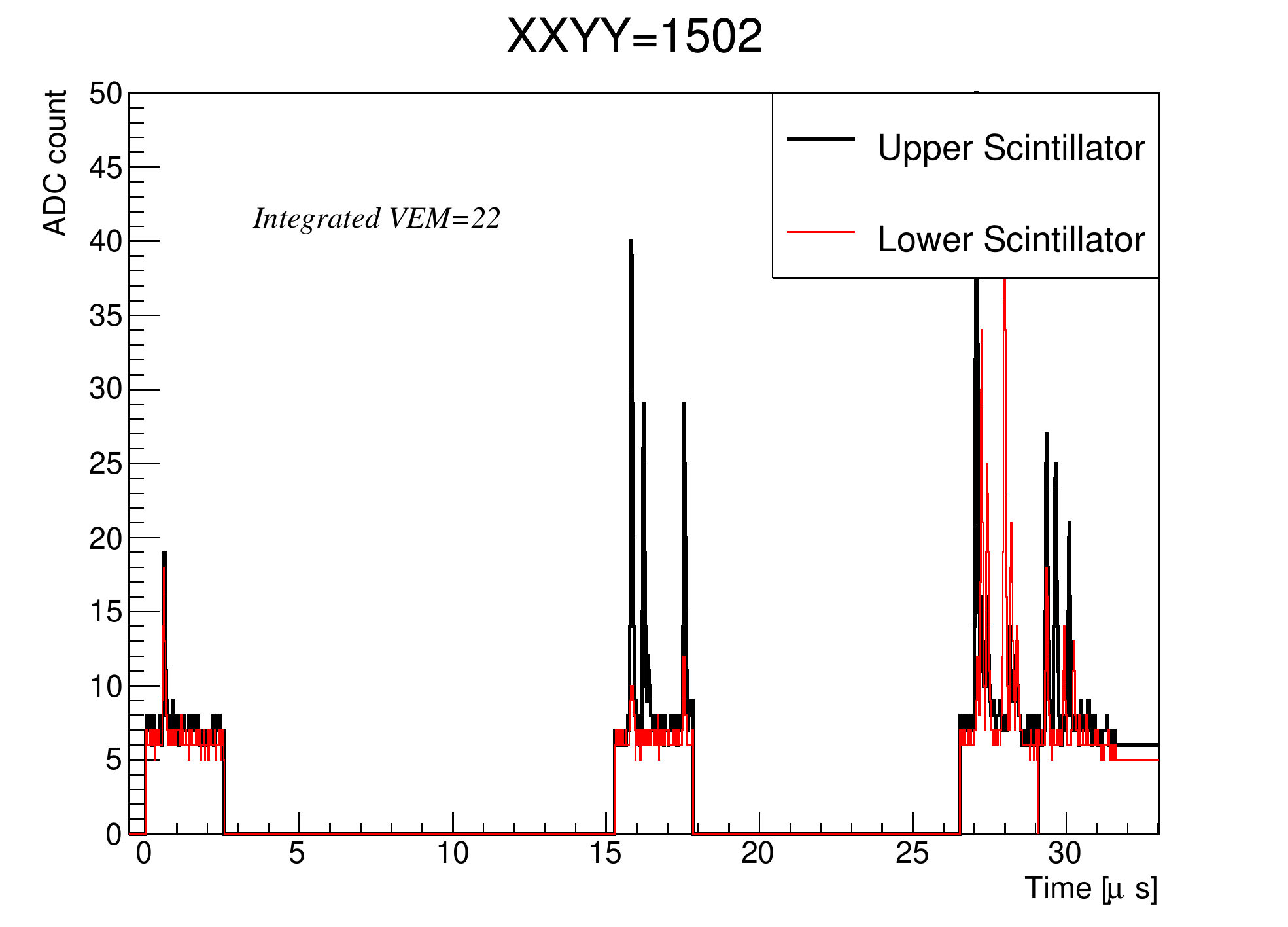} \\
\includegraphics[width=0.40\textwidth]{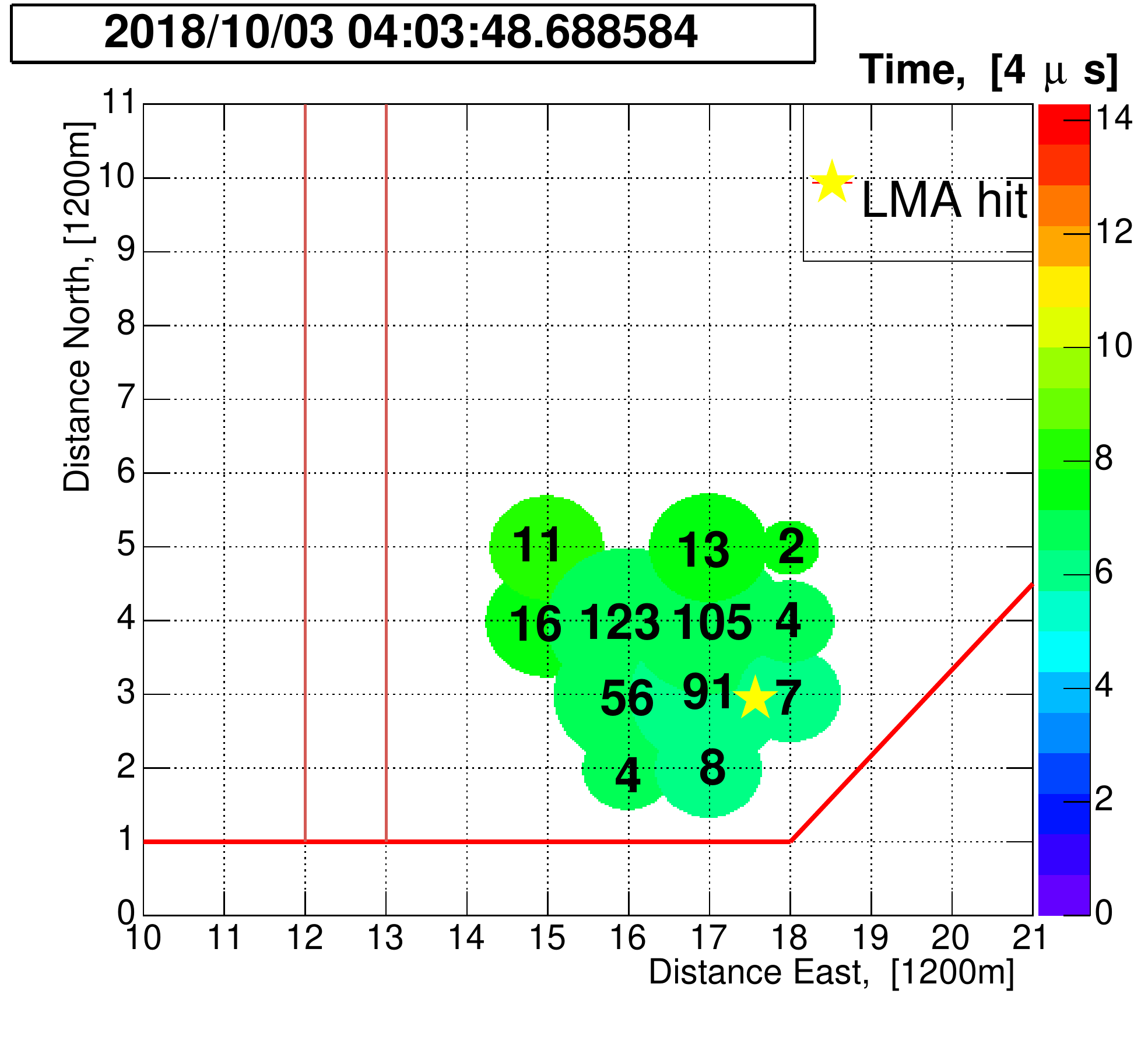} &
\includegraphics[width=0.43\textwidth]{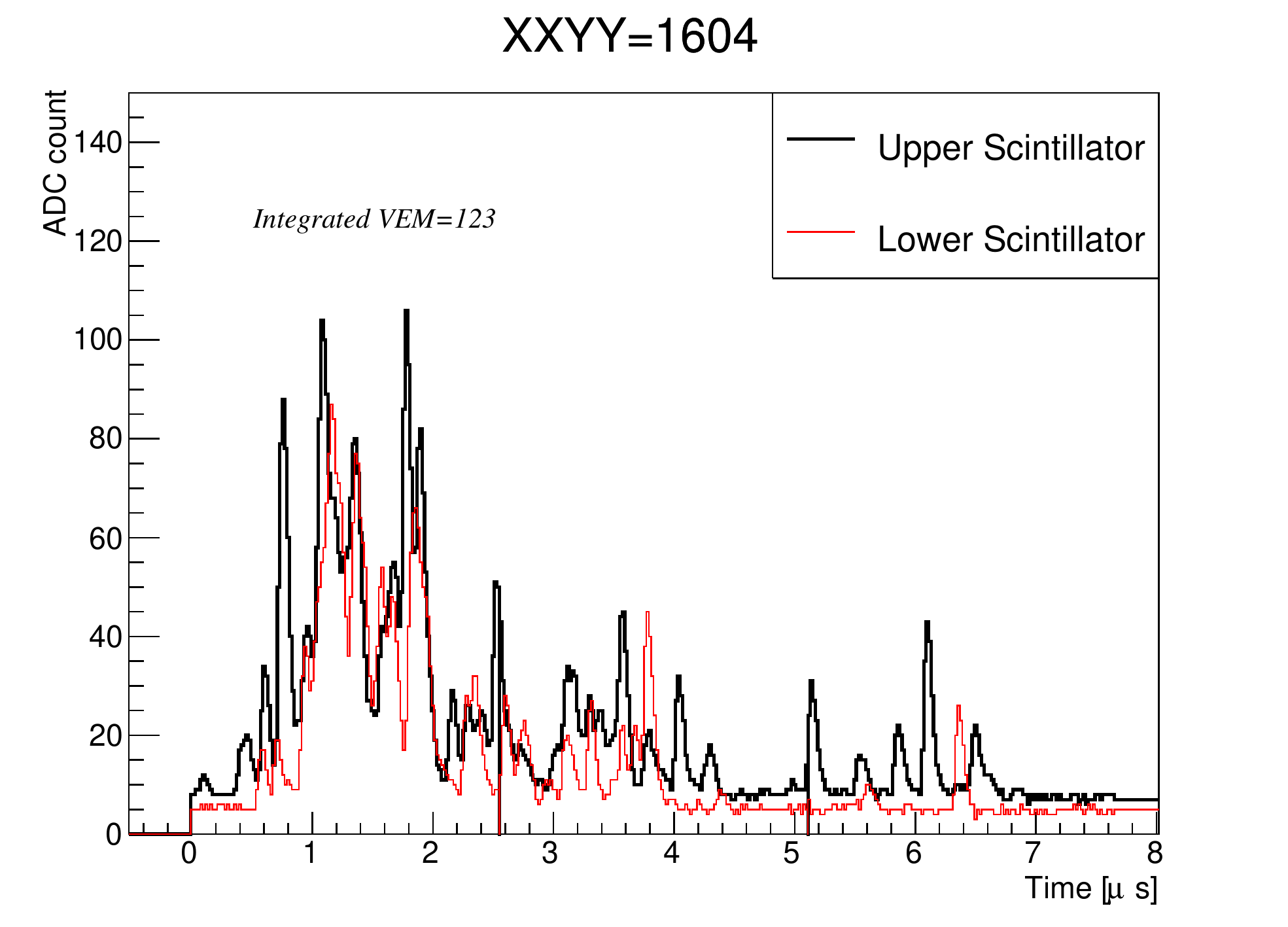} \\
\end{tabular}
\caption{\textbf{TASD observations for TGF~D.} Same as
Figure~\ref{fig:flasha}, except for two trigger events of TGF~D. This
event occurred two months later in the season (Oct. 3) in a nighttime storm
over the southern-most eastern corner of the TASD, and was less-well
located by the LMA. It was a low-altitude IC flash whose downward development
was strongly tilted from vertical (Figure~S22 and Figure~5d of the main text). Rather than
going to ground, the discharge terminated in a strong lower positive charge
region of the storm, which was displaced to the northwest of the
flash initiation point.  The tilted development was also reflected in the
TGF footprints, with the initial burst being southwest of the estimated source
of the gamma bursts (and partially outside the TASD boundary). The second
burst was tilted to the northwest, concomitant with the north-westward tilt of
the flash's development in the INTF and LMA observations.  The parent IBP of the main, second burst was the most extensive of the
four TGFs, lasting $\simeq$15~$\mu$s and propagating over a distance of 240~m (Table~S3).}

\label{fig:flashd} 
\end{figure}

\begin{figure}
\centering
\begin{tabular}{@{}cc@{}}
\includegraphics[width=0.33\textwidth]{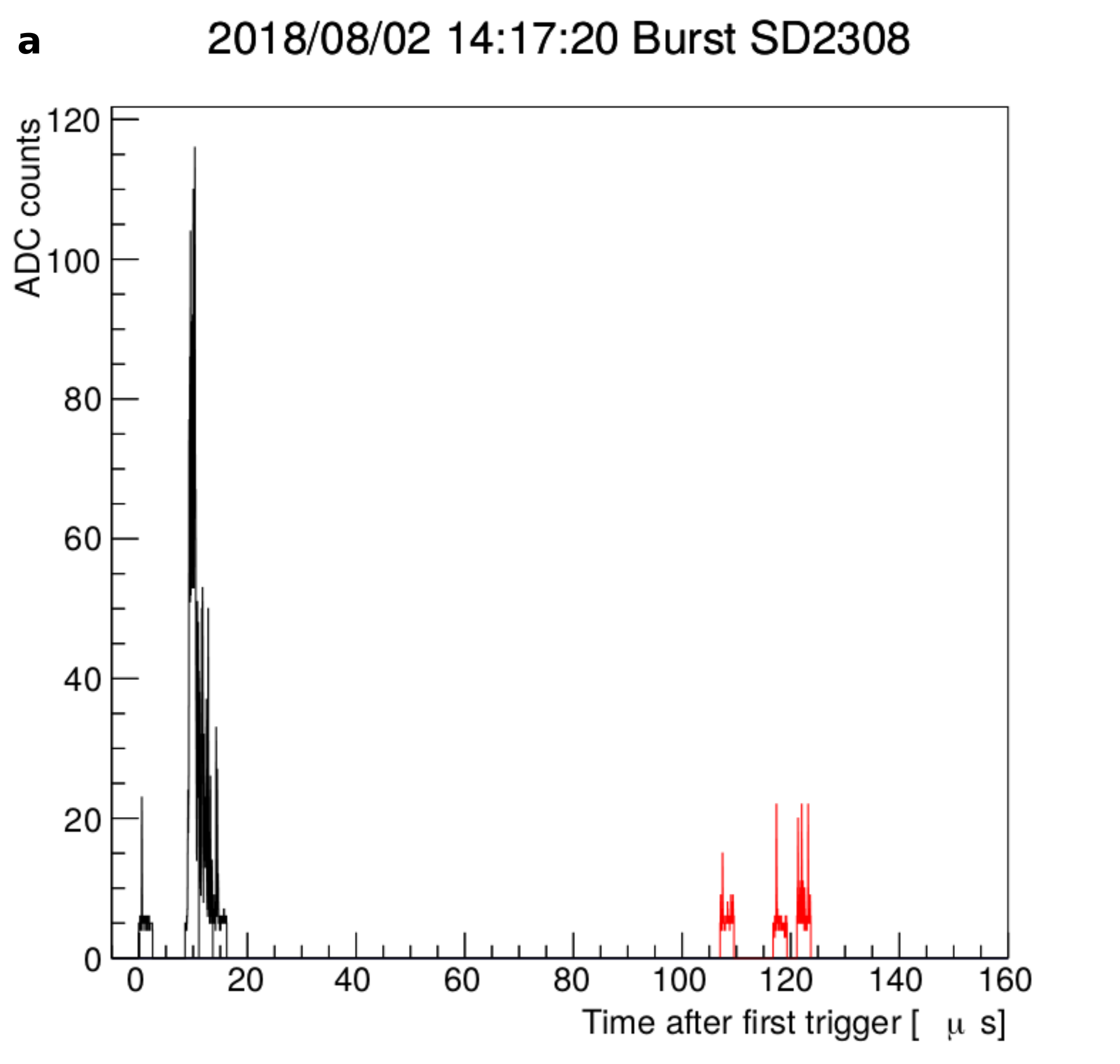} \\
\includegraphics[width=0.33\textwidth]{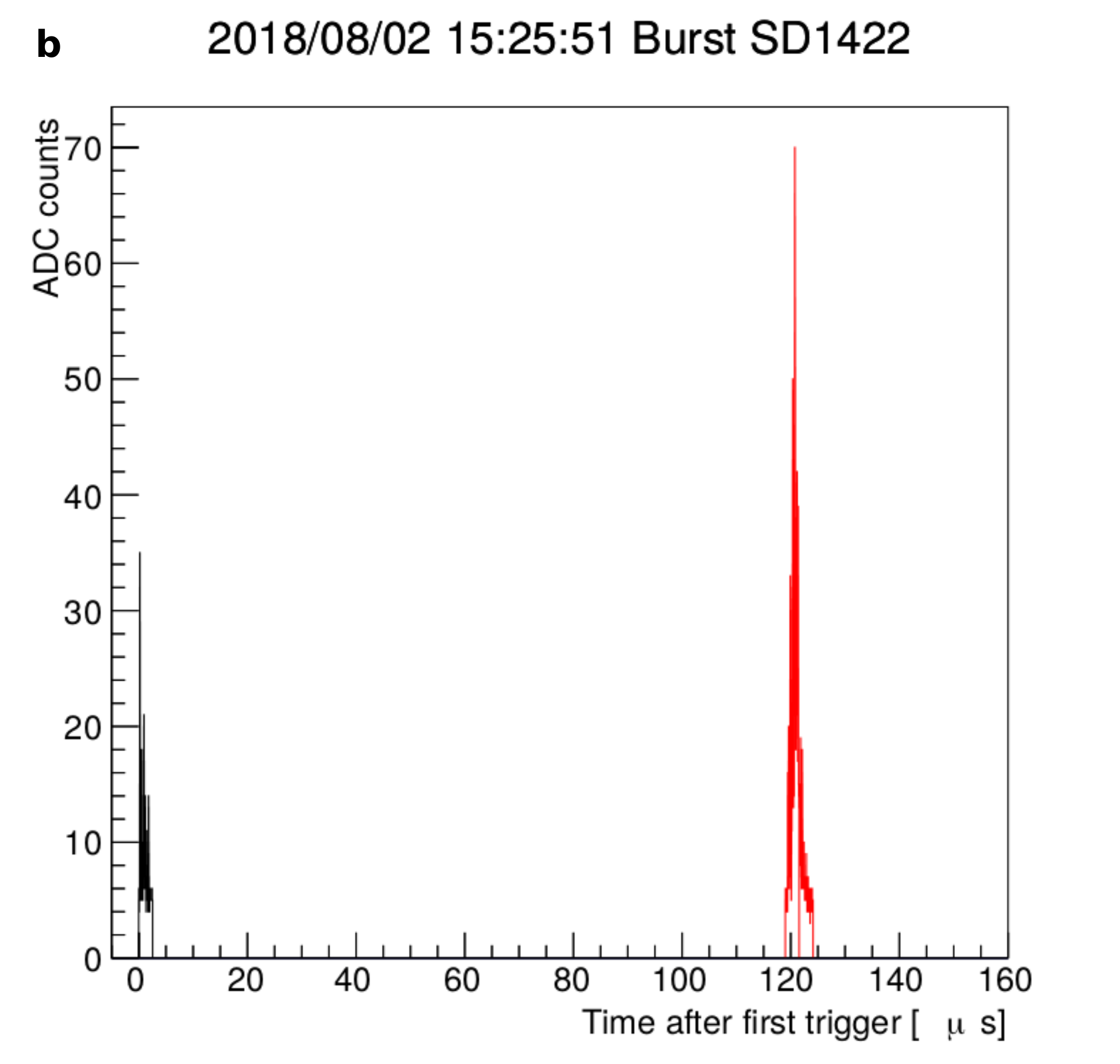} \\
\includegraphics[width=0.33\textwidth]{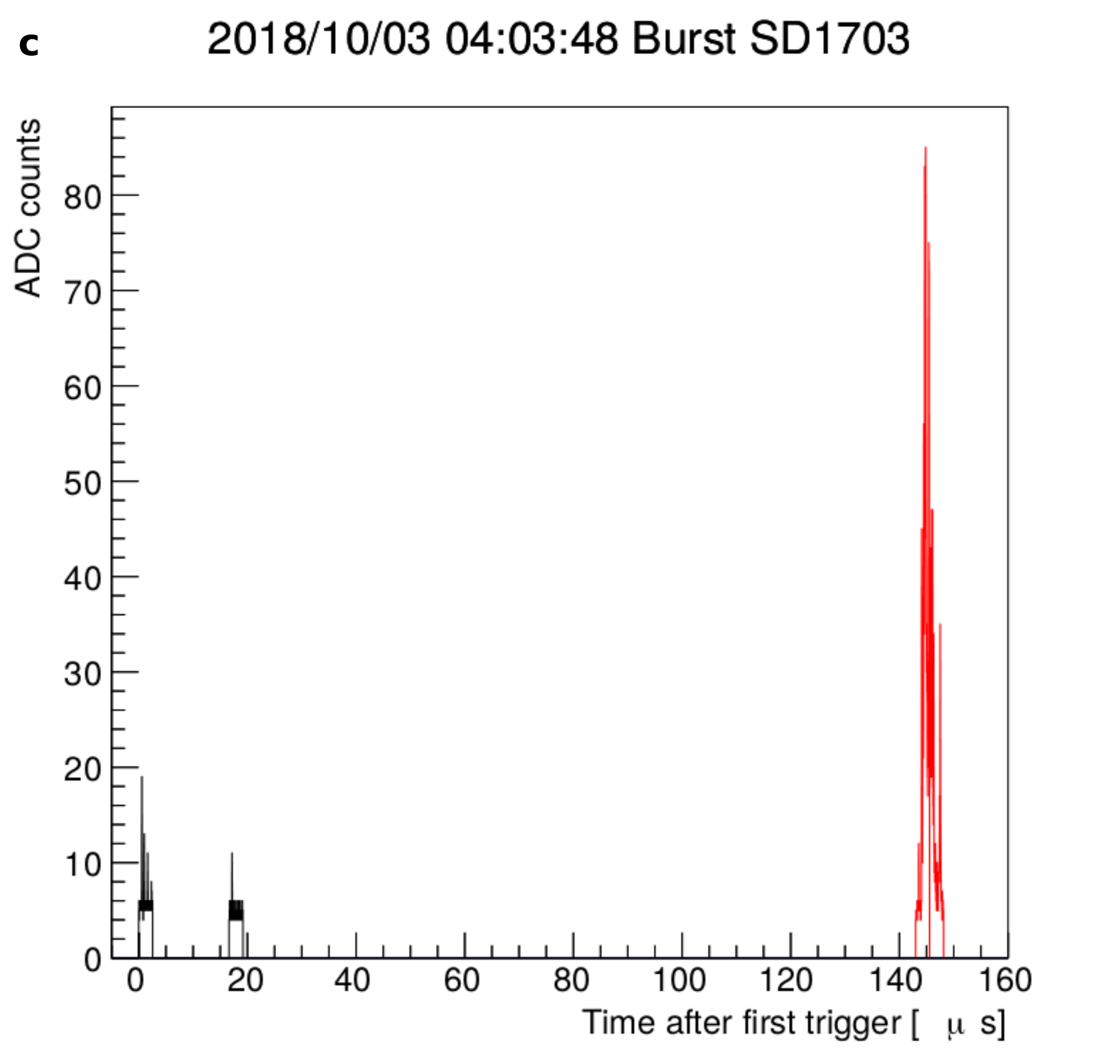}  
\end{tabular}
\caption{\textbf{Composite TASD waveforms.} Stitched-together waveforms for
each of the TGFs that produced multiple triggers (TGFs A,C,D), showing their
relative amplitudes and temporal separations.  Triggers 1 and 2 are colored in
black and red, respectively. The waveforms show the activity to be temporally
resolved into discrete few-microsecond long bursts over a time period of 
$\simeq$100--150~$\mu$s. Three bursts occurred for TGF~A, two for TGF~C, 
three for TGF~D, and one for TGF B (not shown).  The multiple sporadic nature
of the bursting is similar to that seen in the previous study by Abbasi et
al., 2018.  }

\label{fig:stitch} 
\end{figure}

\begin{figure}
\centering
\begin{tabular}{@{}cc@{}}
\includegraphics[width=0.9\textwidth]{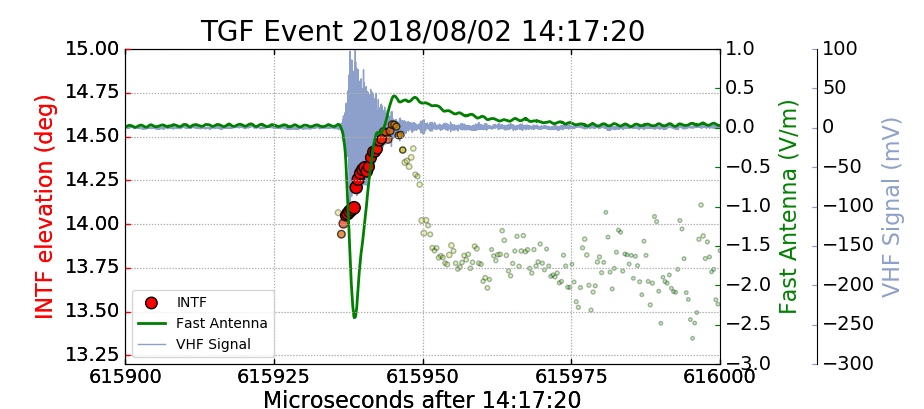} \\
\end{tabular}
\caption{\textbf{Initiating NBE of TGF~A flash.} Classic example of a narrow
bipolar event produced by fast positive breakdown that initiated the flash of
TGF~A.  The FPB propagated upward over a distance of 150~m in 11 $\mu$s,
corresponding to a speed of $1.3\times 10^7$~m/s. While the peak source powers of
the VHF radiation of the NBE and IBP were indistinguishable ($+$27.6 dBW vs.\
$+$27.7 dBW, respectively), the sferic amplitude was 24 times stronger for the
IBP than for the NBE (58 V/m vs.\ 2.4 V/m), being barely noticeable in
Figure~3a of the main text. As a result of this difference, the NLDN detected
the IBP sferic as having a peak current of --36.7~kA, but did not detect the
initiating NBE sferic, whose peak current would have been only $\simeq$2~kA.
}

\label{fig:nbe} 
\end{figure}

\begin{figure}
\centering
\begin{tabular}{@{}cc@{}}
\includegraphics[width=0.66\textwidth]{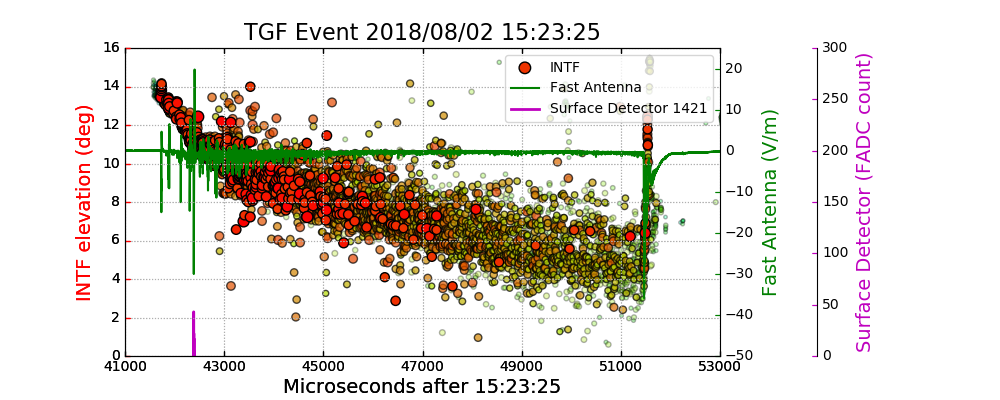} \\
\includegraphics[width=0.66\textwidth]{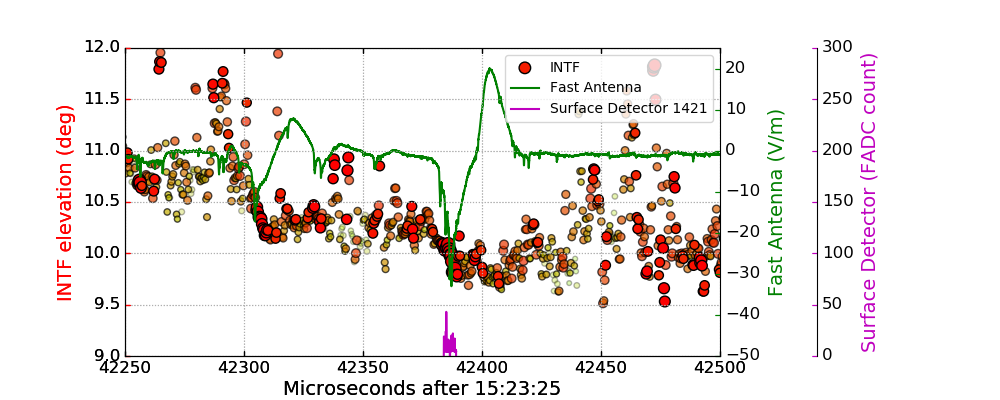} \\
\includegraphics[width=0.66\textwidth]{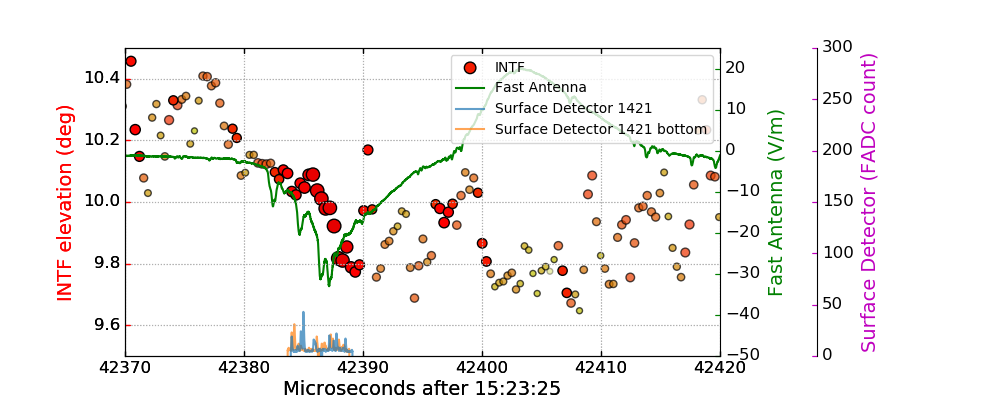} \\
\end{tabular}
\caption{\textbf{Observational data for TGF B.} Observations of TGF B on
2018/08/02 at 15:23:25 UT.  Panels show interferometer elevation versus time
(circular dots), fast electric field sferic waveform (green curve) and TASD
gamma detections (purple waveform). Top: Observations from initial breakdown
through time of --26.5~kA initial cloud-to-ground stroke. Gamma ray detections
occur in coincidence with strong (--30.1 kA) sferic pulse), 341~$\mu$s after
flash start (Table~S1).  Middle: 250~$\mu$s of observations around the time of
the gamma burst, showing the TGF's correlation with the largest amplitude
initial breakdown pulse (IBP) and episode of downward fast negative breakdown
(FNB).  Bottom: Expanded 50~$\mu$s view of the scintillator waveforms at TASD
1421, which detected the initial onset of the TGF, showing how the strong
gamma peak was associated with the second leading-edge, strong sub-pulse of the IBP.}

\label{fig:tgfbobs} 
\end{figure}

\begin{figure}
\centering
\begin{tabular}{@{}cc@{}}
\includegraphics[width=0.66\textwidth]{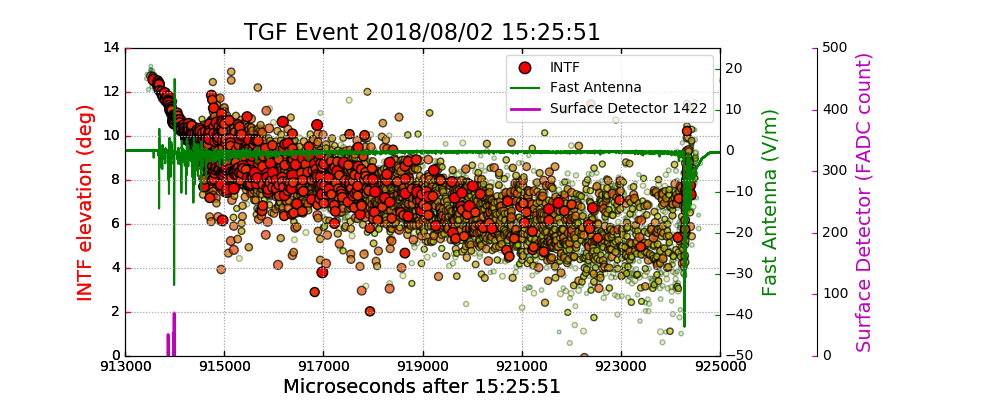} \\
\includegraphics[width=0.66\textwidth]{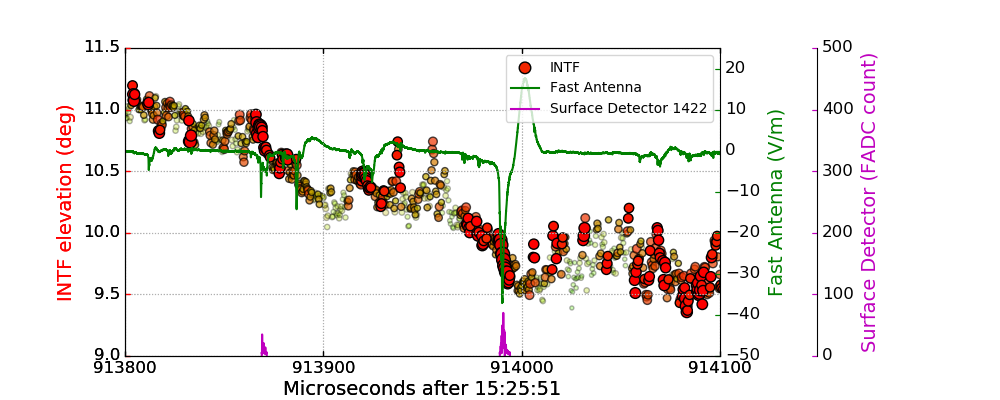} \\
\includegraphics[width=0.66\textwidth]{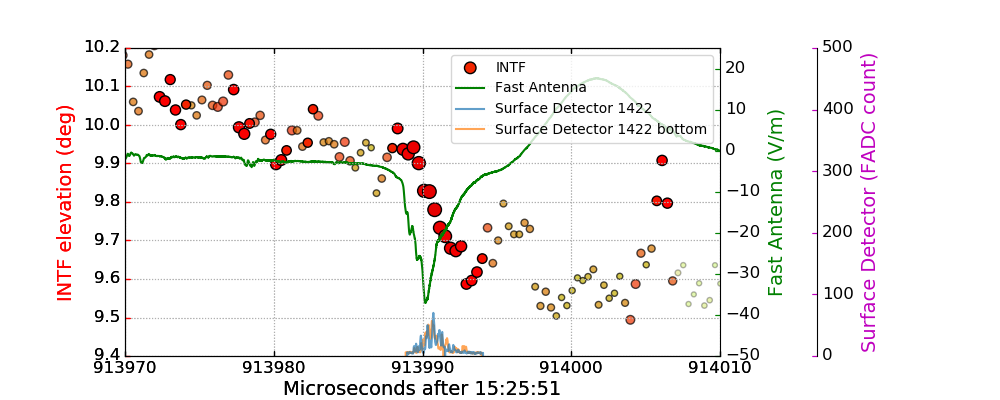} \\
\end{tabular}
\caption{\textbf{Observational data for TGF C.} Same as
Figure~\ref{fig:tgfbobs}, except for TGF~C on  2018/08/02 at 15:25:51 UT. 
In this case, two gamma bursts were produced 825 and 942~$\mu$s after flash
start. While the main, second burst was associated with a --21.7~kA sferic, again
comparable to the initial return stroke (--26.8~kA), the first burst was
associated with a weaker sferic, but with a similarly embedded episode of
enhanced-speed negative breakdown (Figure~S12f).  The correlation with FNB is seen in more
detail in the bottom panel, which illustrates the FNB being initiated by
upward positive VHF development at the beginning of the IBP---a characteristic feature of FNB
---and the gamma detection steadily increasing during the FNB.  }

\vspace{-3\baselineskip} 
\label{fig:tgfcobs}
\end{figure}

\begin{figure}
\centering
\begin{tabular}{@{}cc@{}}
\includegraphics[width=0.66\textwidth]{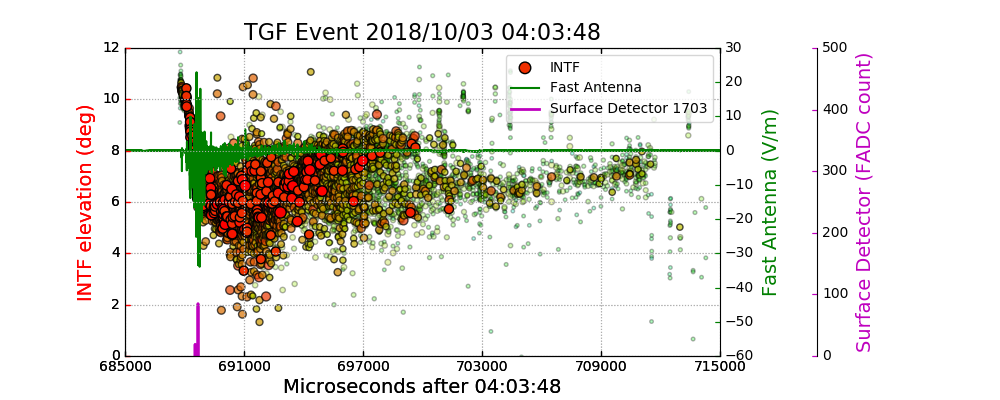} \\
\includegraphics[width=0.66\textwidth]{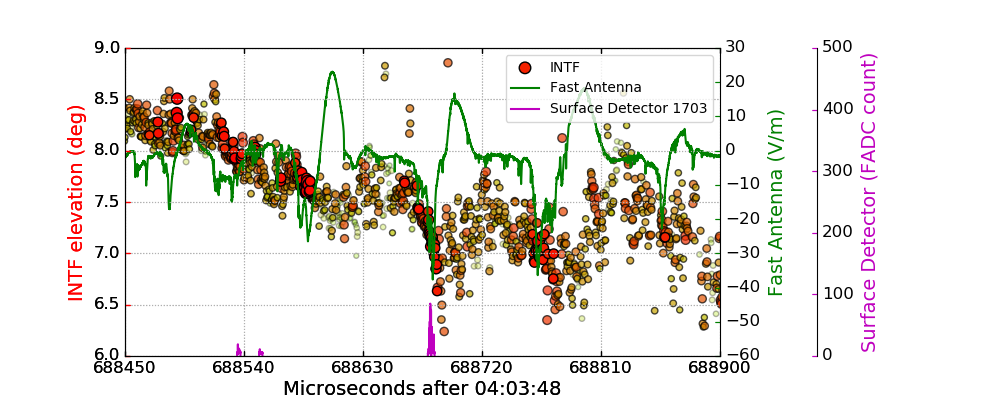} \\
\includegraphics[width=0.66\textwidth]{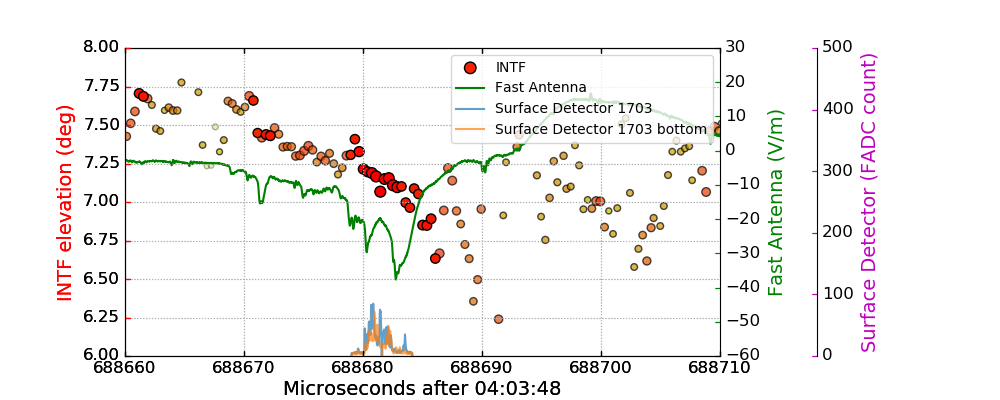} \\
\end{tabular}
\caption{\textbf{Observational data for TGF D.} Same as
Figure~\ref{fig:tgfbobs}, except for TGF D on 2018/10/03 at 04:03:48 UT.
As in TGF~C, two weak gamma bursts occurred in connection with relatively weak IBP sferics
(in this case, prior to the main burst, around 688,540~$\mu$s in middle panel), but not in stronger sferics both before and
  after the strongest sferic. The distinguishing
characteristic appears to
  have been that the other IBPs did not have embedded episodes of
enhanced-speed FNB. Again, the main gamma burst was associated with a
strong sub-pulse on the leading edge of its IBP sferic (bottom panel).}

\label{fig:tgfdobs} 
\end{figure}

\begin{figure}
\centering
\begin{tabular}{@{}cc@{}}
\includegraphics[width=1\textwidth]{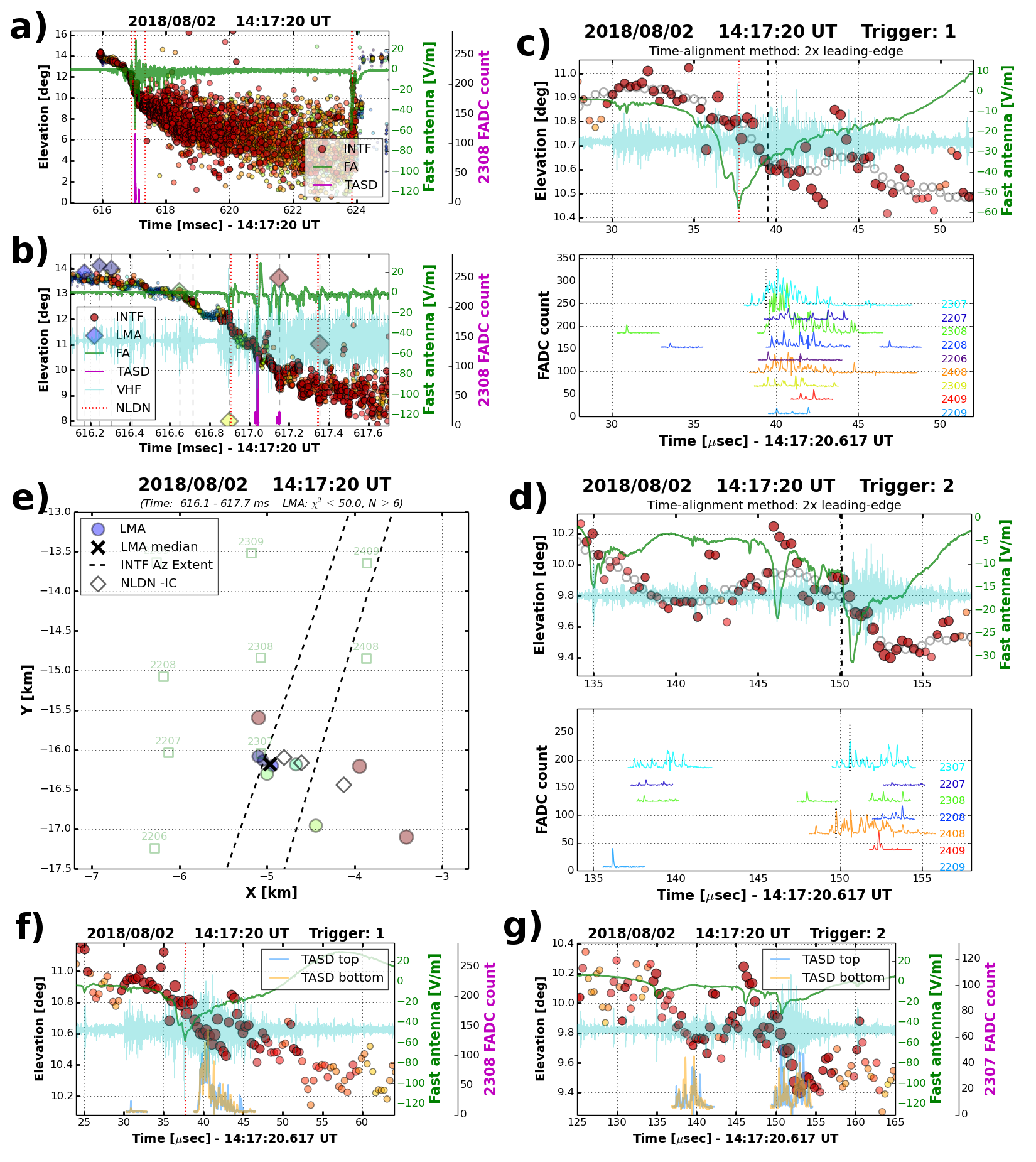} \\
\end{tabular}
\caption{(Caption next page.)}
\label{fig:tgfa_stanley} 
\end{figure}

\addtocounter{figure}{-1}
\begin{figure} [t!]
\caption{(Previous page.) \textbf{Additional data for TGF A.} More complete summary results for
TGF~A, including observations for the two triggers and three gamma bursts of
the TGF. {\bf (a,b)} Overview plots similar to the top and middle panels of
Figures~\ref{fig:tgfbobs}--\ref{fig:tgfdobs}, except showing the times of
NLDN detections (red dotted vertical lines) and (for panel b) times of
LMA-located VHF sources (colored diamonds, when within the displayed range of elevation values), and the
VHF time series waveform (cyan trace). {\bf (c,d)} Correlation results from
the alternative analysis process presented in the Methods section, illustrating time alignments obtained from the first point which
exceeds half-maximum at the two TASD stations with the strongest signals
(short vertical
dotted lines in the TASD waveform panels), with the average of the two results shown in the top panel
(dashed vertical line in the upper panel and in Figure A1b of the main
paper).  (e) Plan view of LMA sources within $\pm$0.8 ms of the TGF
(circular dots) and the median plan location of the sources used as the
estimate for the TGF location (bold black 'x').  Plot illustrates the
clustering of LMA sources and independently-determined NLDN locations
(diamonds) around the median.  Dashed lines indicate the 5--95\% boundaries of INTF azimuth angles within $\pm 0.8$~ms of the TGF. Light green squares 
are participating TASD locations. {\bf (f,g)} Detailed observations and
correlations for the two TASD triggers, showing waveforms from top and bottom
scintillators at the most active TASD station.  (Note perfect correlation of
time-shifted --36.7~kA NLDN event with sferic peak of initial trigger, and
consistent correlation with downward FNB episodes even for weak IBPs of second
trigger.) }
\end{figure}

\begin{figure}
\centering
\begin{tabular}{@{}cc@{}}
\includegraphics[width=1\textwidth]{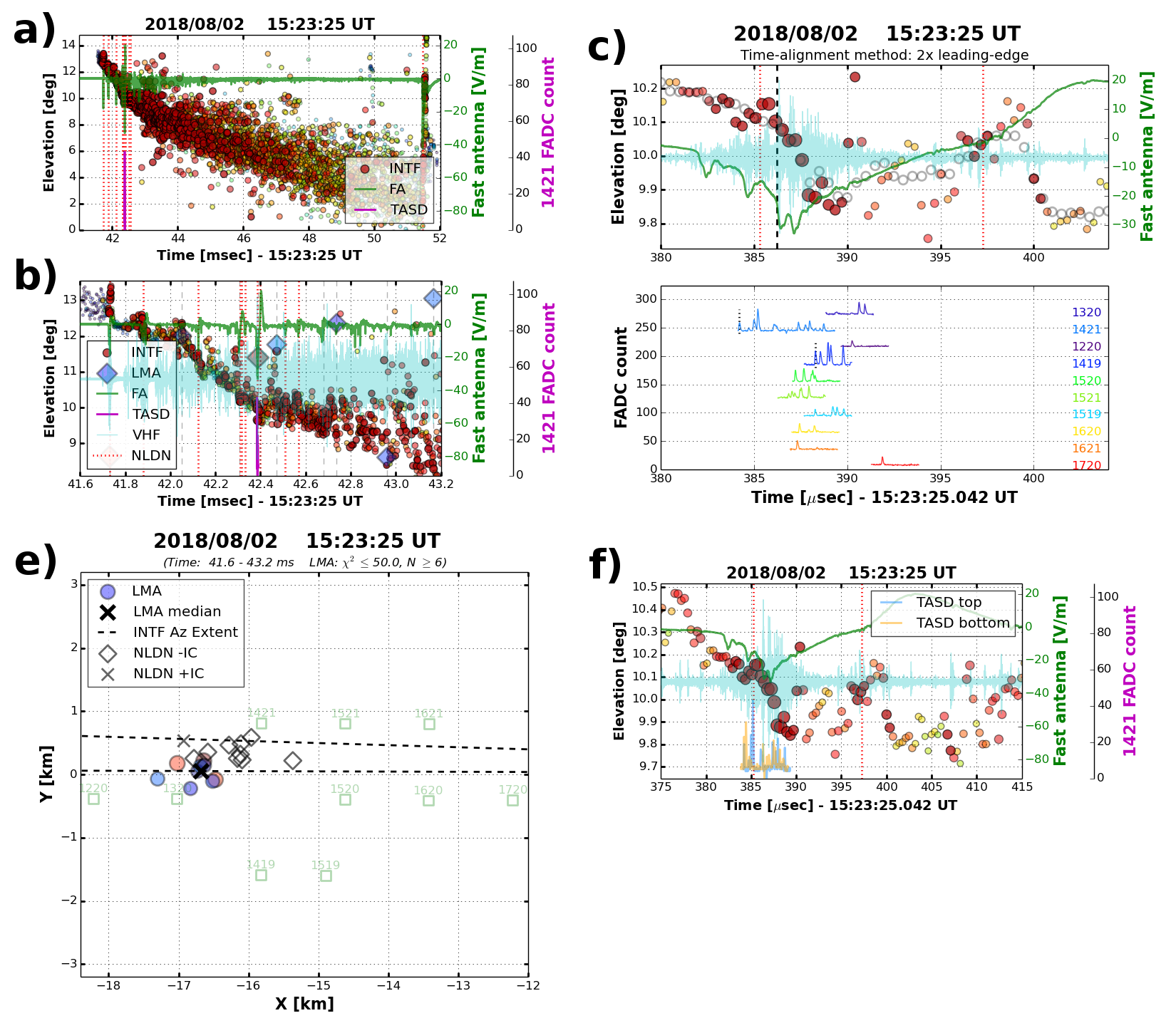} \\
\end{tabular}
\caption{(Caption next page.)}
\label{fig:tgfb_stanley} 
\end{figure}

\addtocounter{figure}{-1}
\begin{figure} [t!]
\caption{(Previous page.) \textbf{Additional data for TGF B.} Same as
Figure~\ref{fig:tgfa_stanley}, except for the single trigger and burst of
TGF B. 
Unfilled gray circles in panel c) indicate the 0.5~$\mu$s higher-time resolution
observations of the INTF elevation angle observations used in both the
iterative and alternative analysis procedures (see Methods Section~A2). 
The TASD signals are sorted top to bottom according to
increasing range from the source. Note the correlation in panel f) of the close 1421 TASD waveforms with
the sferic sub-pulses and sequence of upward and downward FPB and FNB for the
singular
1421-detected burst.  The non-aligned TASD waveforms in panel c)
reinforce multi-onset grouping seen in
Figure~4d of the main text and the full-page version in Figure~S16. Also note the large number of NLDN-located events
in the first ms of the flash in panel a), and their excellent correlation with
IBP sferics in panel b). The relatively large east-west variability of the
NLDN events in panel e) is not reflected in the LMA observations and is
presumably due to uncertainty in the NLDN location (caused by the particular NLDN stations used to locate the event).  Otherwise, the 
LMA and NLDN sources are closely clustered around the LMA-indicated median source location. }
\end{figure}

\begin{figure}
\centering
\begin{tabular}{@{}cc@{}}
\includegraphics[width=1\textwidth]{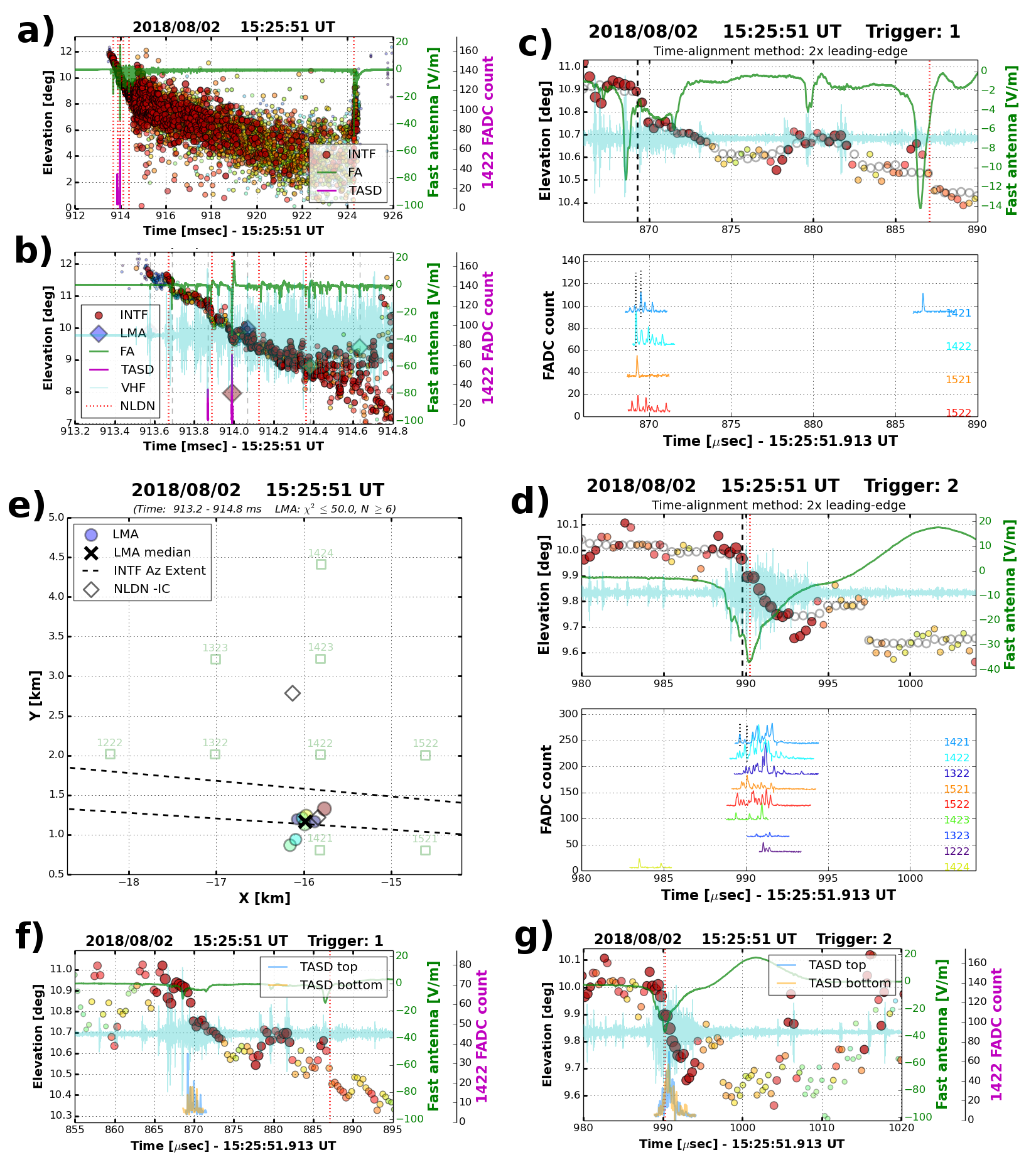} \\
\end{tabular}
\caption{(Caption next page.)}
\label{fig:tgfc_stanley} 
\end{figure}

\addtocounter{figure}{-1}
\begin{figure}
\caption{(Previous page.) \textbf{Additional data for TGF C.} Same as
Figure~\ref{fig:tgfa_stanley}, except for the two triggers and bursts of
TGF~C.  Again, the NLDN located a number of IBP events in the initial ms of
the flash (panels a and b). One NLDN event was noticeably mislocated in an
otherwise closely grouped set of LMA and NLDN events in the median plan
location plot of panel (e).  The main burst occurred during the second trigger
and is notable for its simplicity of interpretation (see
text).  The alignment time is slightly delayed relative to the onset of
the TASD 1422 signal by the use of a half-max threshold in the stepping
analysis (multi-waveform part of panel d).  The results are consistent
with the iterative approach utilizing median result of all TASD stations(Figure~4c and associated main text)
Accounting for the slight delay also better aligns the
initial burst of the first trigger with its relatively weak sferic and
associated VHF radiation (panel f).}
\end{figure}

\begin{figure}
\centering
\begin{tabular}{@{}cc@{}}
\includegraphics[width=1\textwidth]{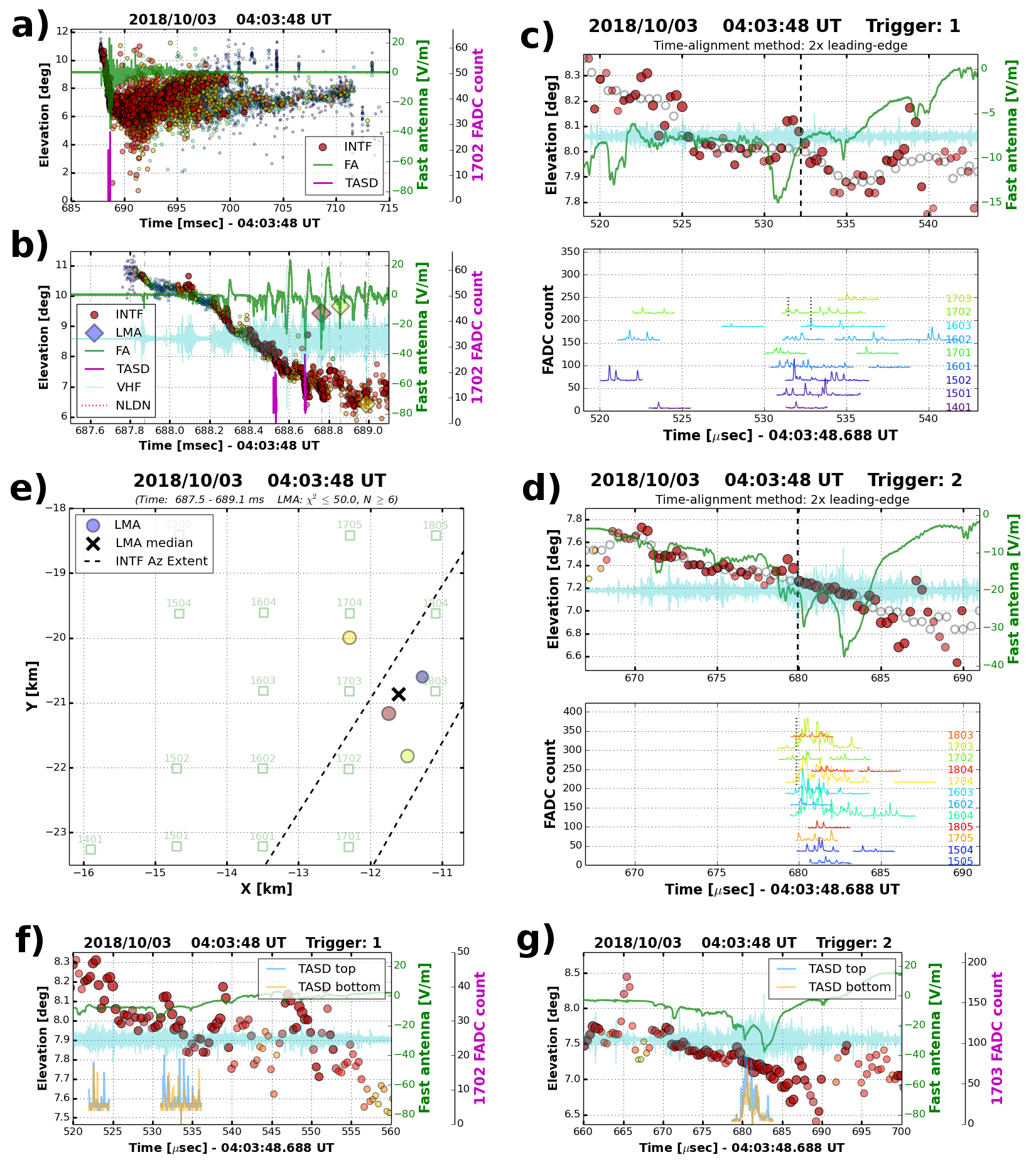} \\
\end{tabular}
\caption{Caption next page. }
\label{fig:tgfd_stanley} 
\end{figure}

\addtocounter{figure}{-1}
\begin{figure} [t!]
\caption{(Previous Page.) \textbf{Additional data for TGF D.} Same as
Figure~\ref{fig:tgfa_stanley}, except for the two triggers and three bursts of
TGF~D. As discussed in connection with Figure~\ref{fig:flashd}, the flash did
not go to ground and was not detected by the NLDN.  Otherwise its initial
breakdown was no different than that of a --CG discharge.  
Despite the increased uncertainty of the median plan location (panel e) and
slightly greater plan distance from the INTF (24~km), both analysis methods
gave essentially the same onset time for the main burst, which occurred
during trigger 2 (panel d).  In particular, the burst occurred in association
with FNB associated with one or both sub-pulses of the IBP leading up to the
main peak (panels d and g). The second of the two weaker bursts associated with
the initial trigger was also associated with a slightly noisy but clear
episode of downward FNB episode (and a weaker sferic) (panel f). The first
burst was associated with even weaker and noisier activity, but was still
capable of producing gamma radiation.  In contrast, intermediate to strong
IBPs between and after the two triggers did not produce detectable gamma
bursts.}
\end{figure}

\begin{figure}
\centering
\includegraphics[width=0.75\textwidth]{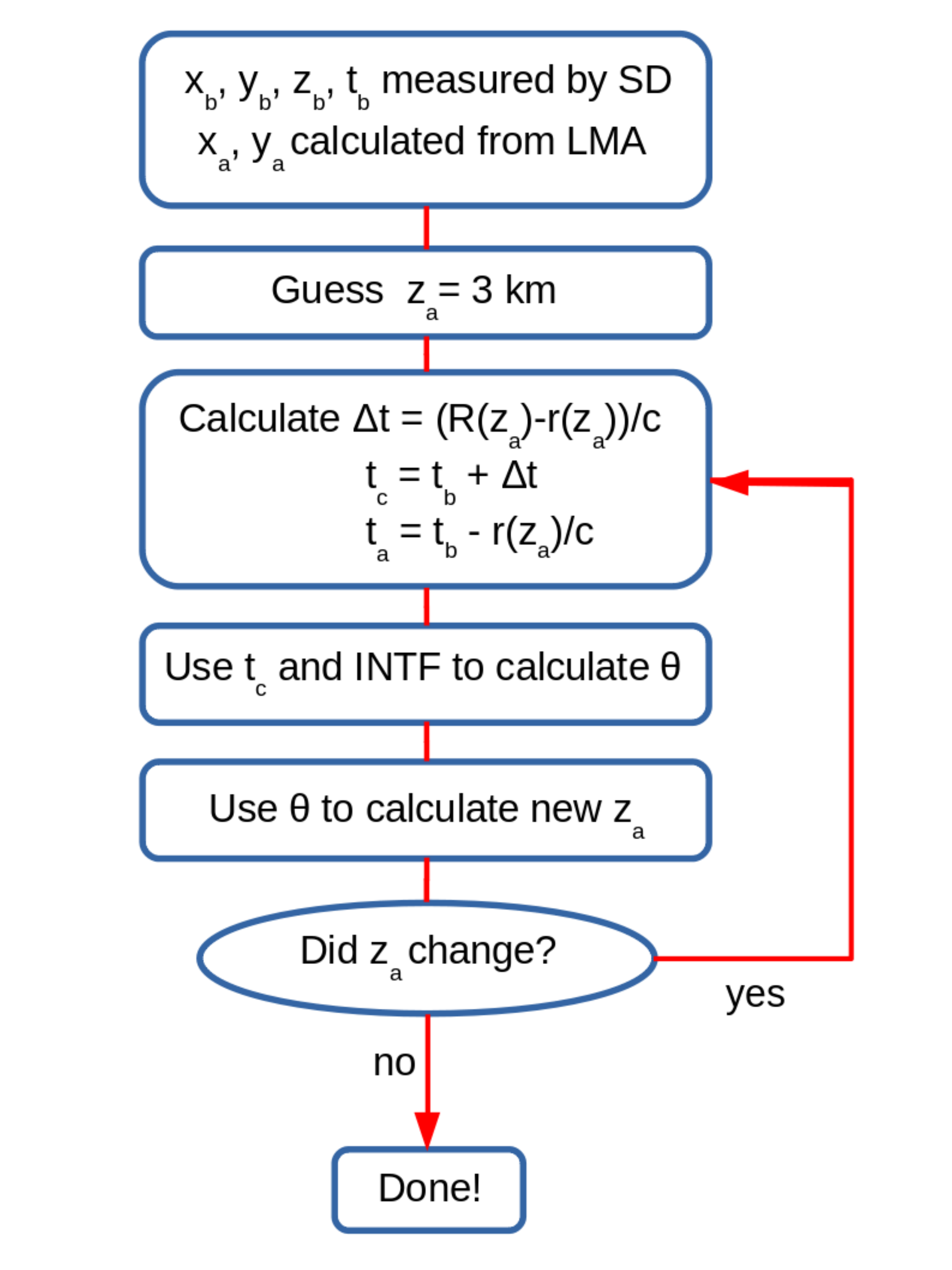} \\
\caption{
\textbf{Source Reconstruction Flowchart}.
Iterative procedure for determining $z_a$ and $t_a$,
the altitude and time of the TGF source. Note that this is performed individually for each participating TASD station. See full description in Section~2.2 and in Methods Section~A2.
}
\label{fig:flowchart} 
\end{figure}

\clearpage

\begin{figure}
\centering
\includegraphics[width=0.9\textwidth]{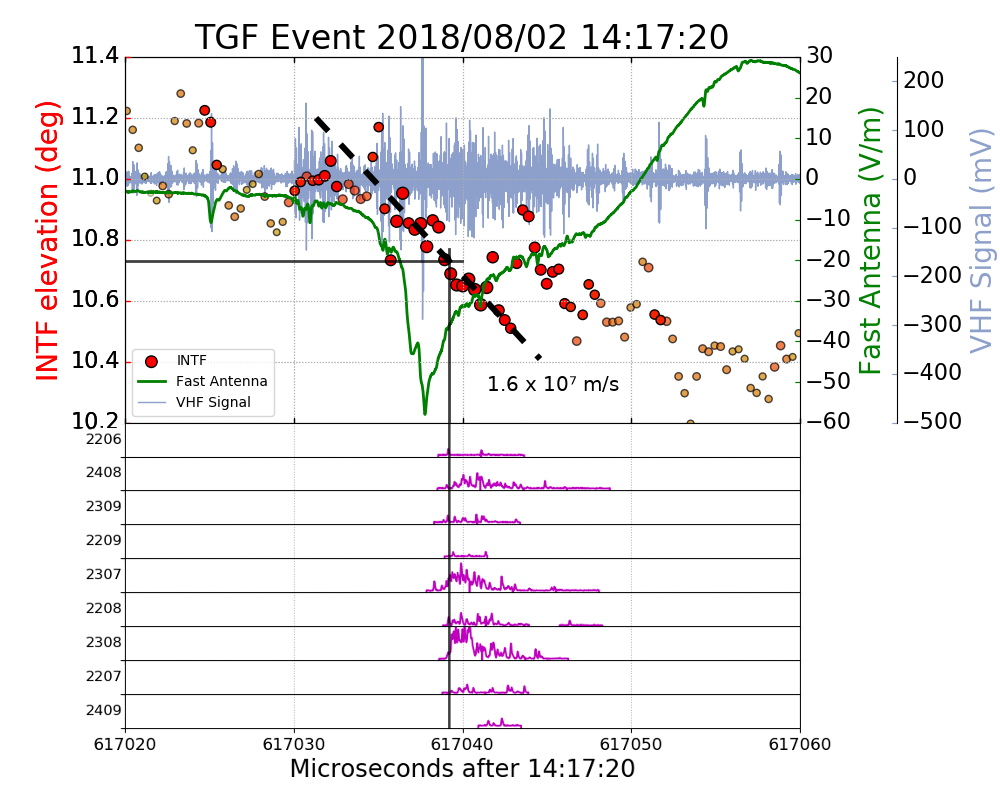} \\
\caption{
\textbf{TASD correlations for TGF~A}.
Enlarged view of panel~a) of Figure~4 showing data from all participating detectors, time-shifted relative to the INTF. The black vertical line shows the median onset time of the TGF relative to the INTF and fast antenna data, and the black horizontal line shows the corresponding elevation angle. The light blue trace shows the VHF time series waveform observed by the INTF, and dots represent VHF radiation sources with color and size representing relative power. Purple traces in the lower panels are particle detector responses, with station numbers XXYY identifying their easterly (XX) and northerly (YY) locations within the array in 1.2~km grid spacing units. The detection times are in good agreement with one another as well as the median, indicating the onset time of the TGF during the sferic and the VHF radiation development.}
\label{fig:4Panel-a} 
\end{figure}

\begin{figure}
\centering
\includegraphics[width=0.9\textwidth]{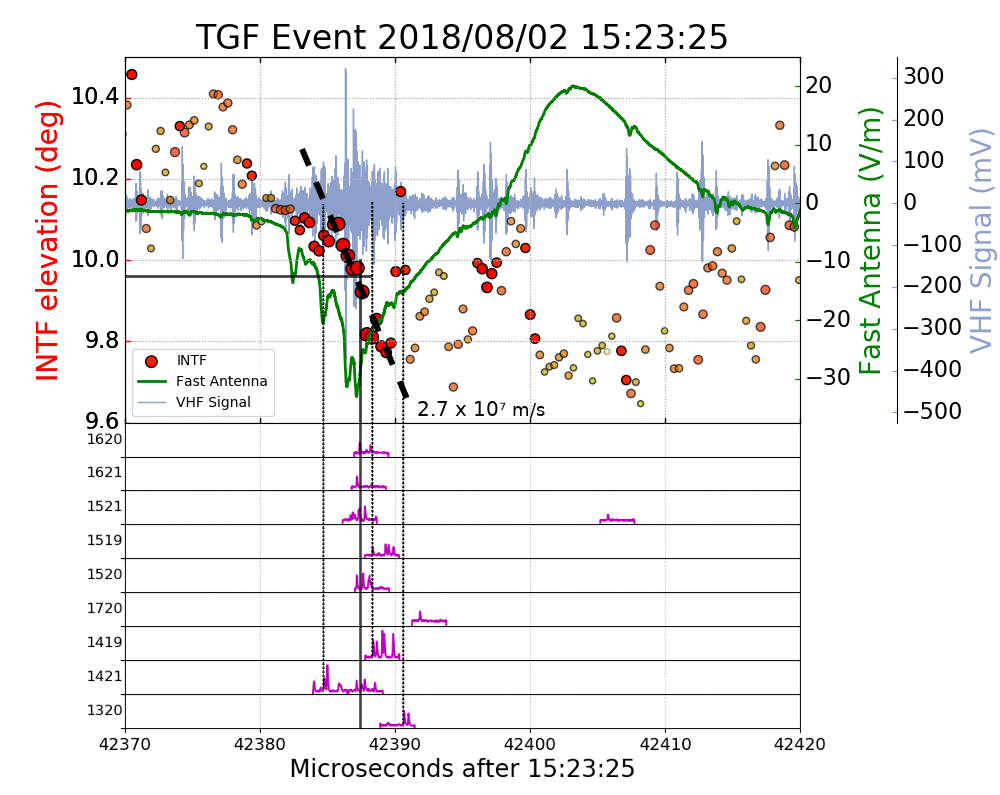} \\
\caption{
\textbf{TASD correlations for TGF~B}.
Same as Figure~\ref{fig:4Panel-a}, except for panel~b) of Figure~4. 
In contrast to the other TGFs, the TASD onset times are not all consistent with the median; TASD 1421 had a noticeably early onset time associated with the second strong sub-pulse, while the median onset was associated the peak of the IBP and with a step-discontinuity in the VHF radiation development. Slightly delayed TASD signals suggest additional onsets as discussed in Section~2.3.
}
\label{fig:4Panel-b} 
\end{figure}

\begin{figure}
\centering
\includegraphics[width=0.9\textwidth]{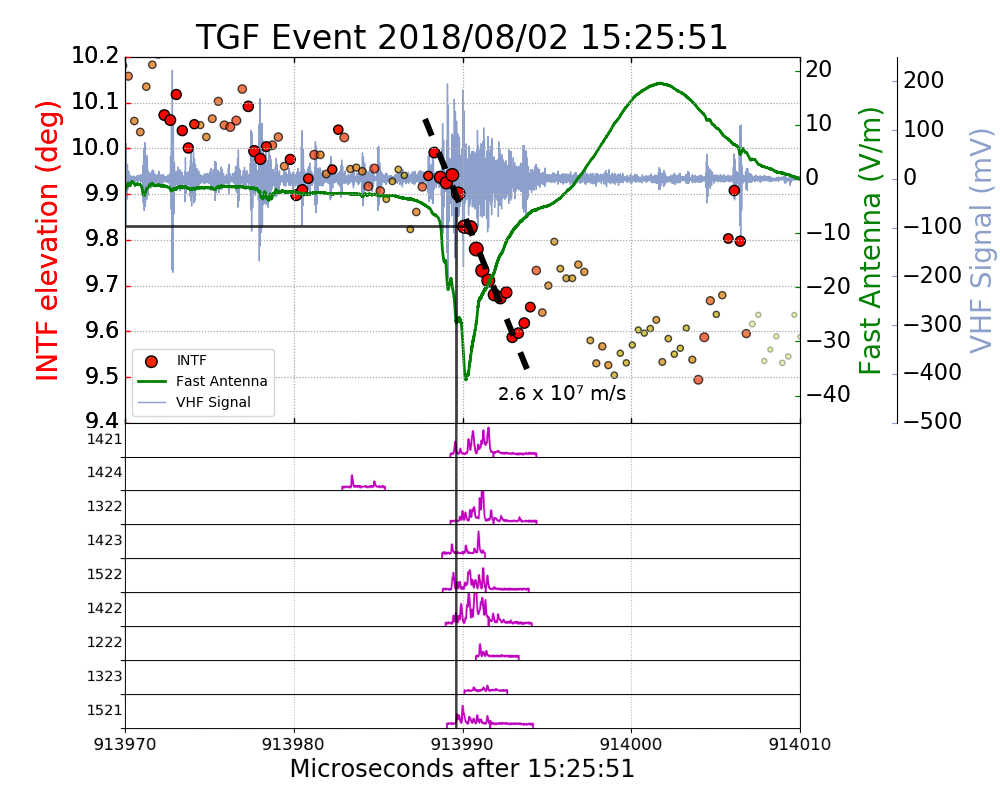} \\
\caption{
\textbf{TASD correlations for TGF~C}.
Same as Figure~\ref{fig:4Panel-a}, except for panel~c) of Figure~4.
This event is an example of a simple, canonical IBP. Fast positive breakdown briefly propagates upward before turning into downward fast negative breakdown during the IBP. The TASD onsets are mostly in good agreement and are associated with a sub-pulse and a step-discontinuity in the VHF radiation development.
}
\label{fig:4Panel-c} 
\end{figure}

\begin{figure}
\centering
\includegraphics[width=0.9\textwidth]{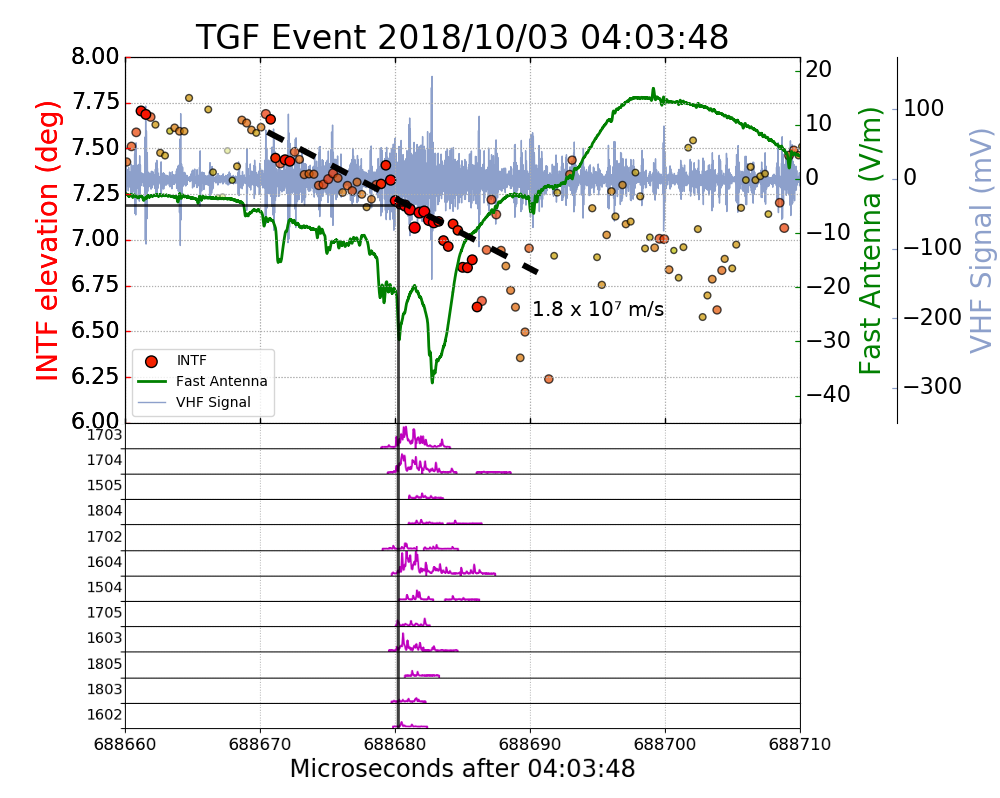} \\
\caption{
\textbf{TASD correlations for TGF~D}.
Same as Figure~\ref{fig:4Panel-a}, except for panel~d) of Figure~4.
The sferic of this TGF is similar to that of TGF~B, with a slower build-up and multiple embedded sub-pulses. The TASD onsets are closely correlated with one another and with a strong, impulsive sub-pulse before the main peak. The fast negative breakdown had a relatively long duration and extent (given in Table~S3) with a step-discontinuity occurring immediately before the median TASD onset.
}
\label{fig:4Panel-d} 
\end{figure}

\begin{figure}
\centering
\includegraphics[width=0.9\textwidth]{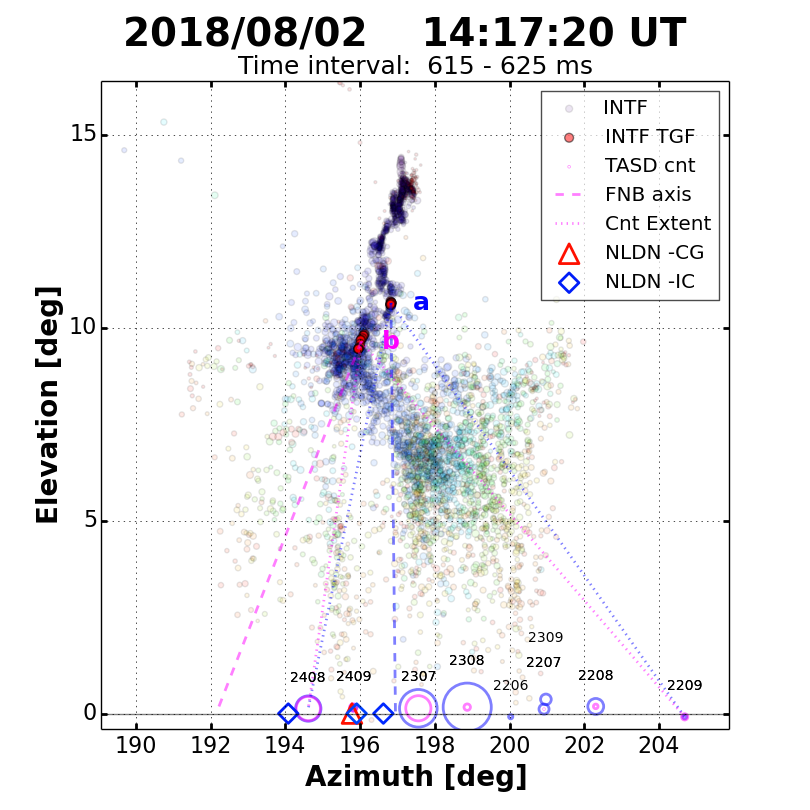} \\
\caption{
\textbf{Azimuth-elevation plots for TGF A}.
Enlarged view of panel~a) of
Figure~5 showing the initial downward development leading up to and following
the TGF occurrences. Red highlighted sources and $a$, $b$ labels indicate the
TGFs' calculated source locations.  Dashed red and blue lines indicate the
axes of the FNB associated with each TGF, while the finely dotted lines show
the angular extent of the TASD surface detections. (The plot has a 1:1 aspect
ratio so that the angular directions are faithfully replicated.) Baseline
circles indicate size-scaled relative energy deposit in each Surface Detector;
other baseline symbols indicate NLDN locations of CG and IC events, all as
viewed from the INTF site.
}
\label{fig:az-elA} 
\end{figure}

\begin{figure}
\centering
\includegraphics[width=0.9\textwidth]{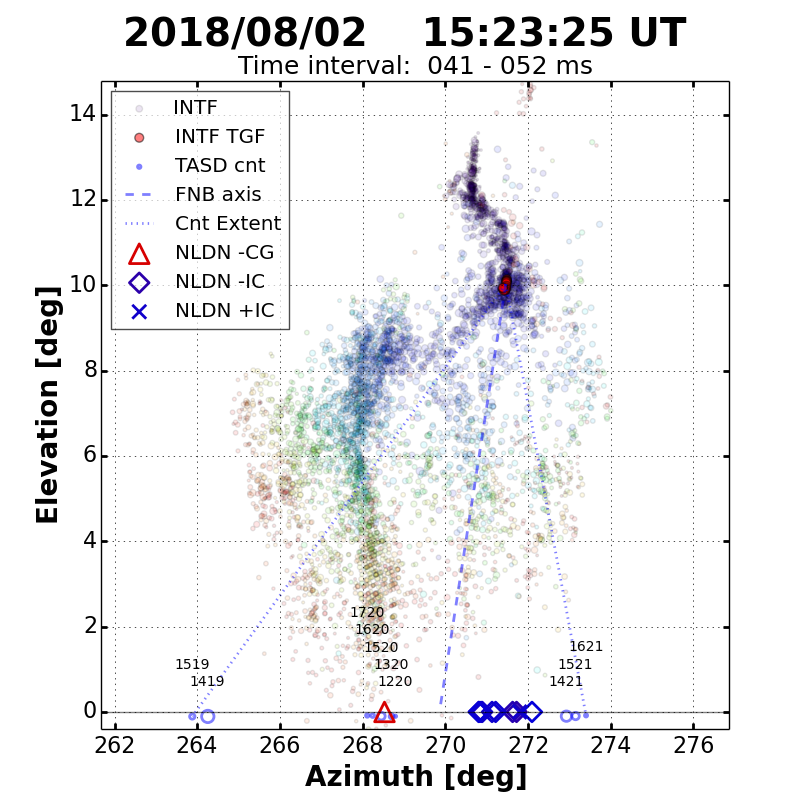} \\
\caption{
\textbf{Azimuth-elevation plots for TGF B}.
Same as Fig.\ S19, except for
TGF~B of Figure~5b.  This TGF had at least two onset times (and possibly one
or two more) at different SDs and sets of SDs, and therefore narrower beaming
than indicated by the overall angular extent (see Figure S16). This suggests
successively different orientations of the sub-pulses and FNB activity, which
would be consistent with INTF observations starting to broaden angular-wise at
the IPBs location.
}
\label{fig:az-elB} 
\end{figure}

\begin{figure}
\centering
\includegraphics[width=0.9\textwidth]{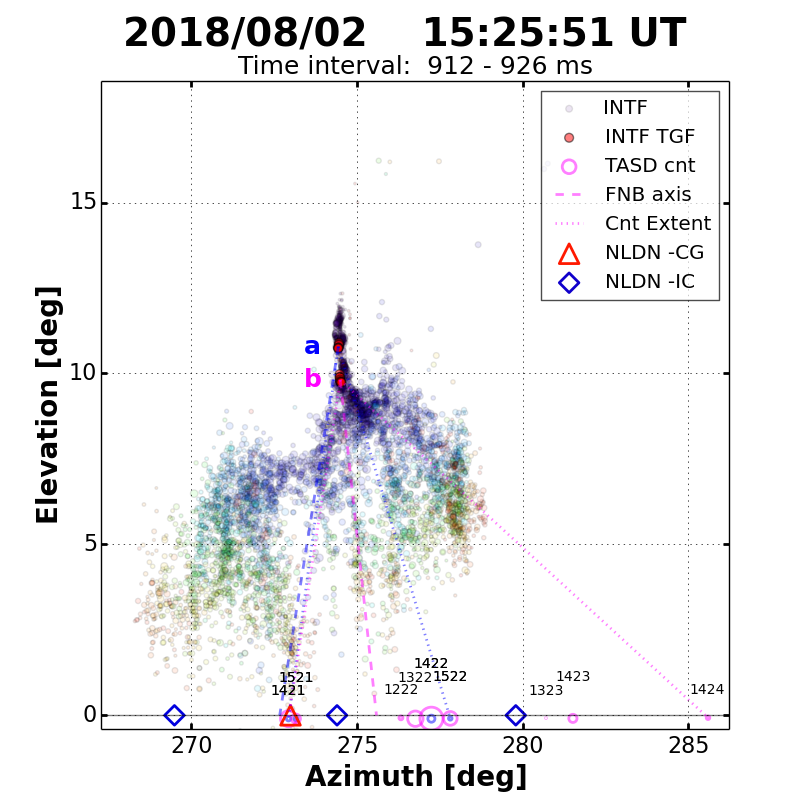} \\
\caption{
\textbf{Azimuth-elevation plots for TGF C}.
Same as Fig.\ S19, except for the
relatively simple and conical TGF~C of Figure~5c.  The simplicity is seen in
the vertically-downward development of the INTF sources. Of interest in
this and the other TGFs, the main TGF of each flash (event $a$ of TGF~A, the
only event of TGF~B, and event $b$ of the present TGF) are all produced by
the strongest IBP of the flashes, which occur as the INTF sources start to
broaden out, indicative of the onset of branching.  After that, the IBPs
begin to weaken, suggesting the IBPs are strongest up until branching
starts. The weakened IBPs can also produce gamma bursts, however, as seen in
TGF~A, where the second, weaker gamma event ($b$) occurred $\simeq$100~$\mu$s
after the main ($a$) trigger and further into the branching (Fig.\ S19).
}
\label{fig:az-elC} 
\end{figure}

\begin{figure}
\centering
\includegraphics[width=0.9\textwidth]{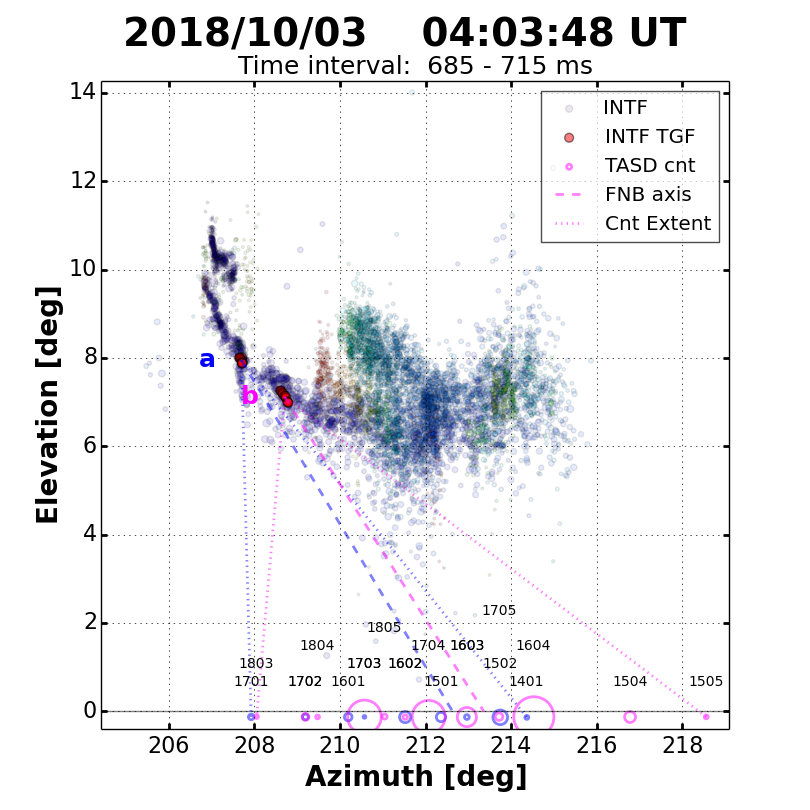} \\
\caption{
\textbf{Azimuth-elevation plots for TGF D}.
Same as Fig.\ S19, except for TGF~D
during the strongly-tilted, low-altitude IC flash of Figure~5d. Again, the
strongest IBP and TGF occurred at the lowest exent of the downward negative
breakdown before it started branching. The flash occurred in a late-season
nocturnal storm that had a more complex electrical structure, as indicated by
the disjointed nature of the downward negative breakdown before entering the
storm's offset lower positive charge region. The IBP that generated the TGF
had the longest duration ($\simeq$15~$\mu$s) and extent (240~m vertical
component---but longer due to being oriented $\simeq30^\circ$ from
vertical).
}
\label{fig:az-elD} 
\end{figure}

\begin{figure}
\centering
\includegraphics[width=0.99\textwidth]{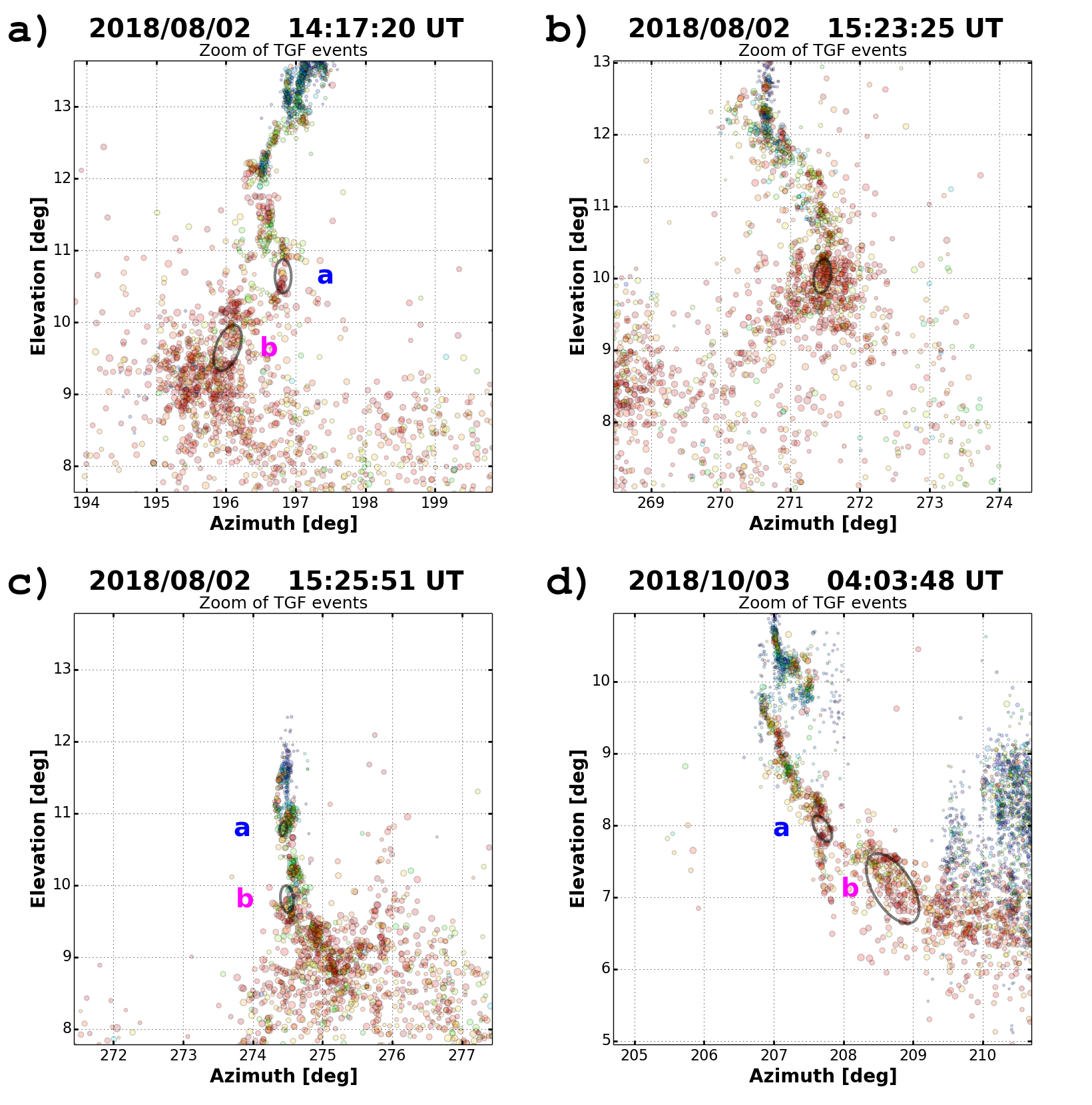} \\
\caption{
\textbf{Zoomed azimuth-elevation plots}.
6x6 degree view of TGFs~A--D. The location of the VHF sources for each TGF trigger are indicated by the ellipses.  All VHF sources are rainbow color-coded from blue to red according to increasing power, rather than according to time as in Figures~S19--S22.
}
\label{fig:az-el-zoom} 
\end{figure}

\begin{figure}
\centering
\includegraphics[width=0.7\textwidth]{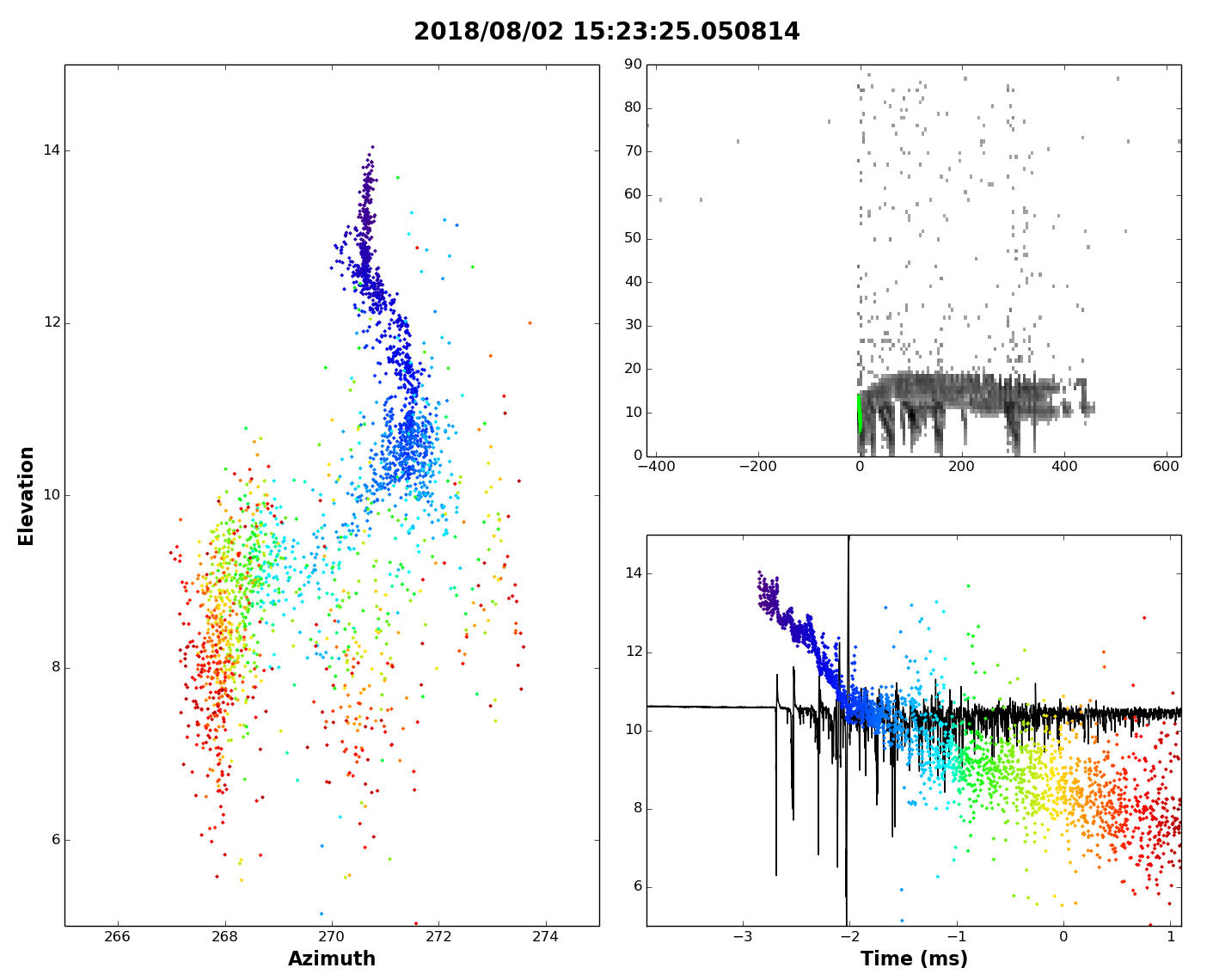} \\
\includegraphics[width=0.7\textwidth]{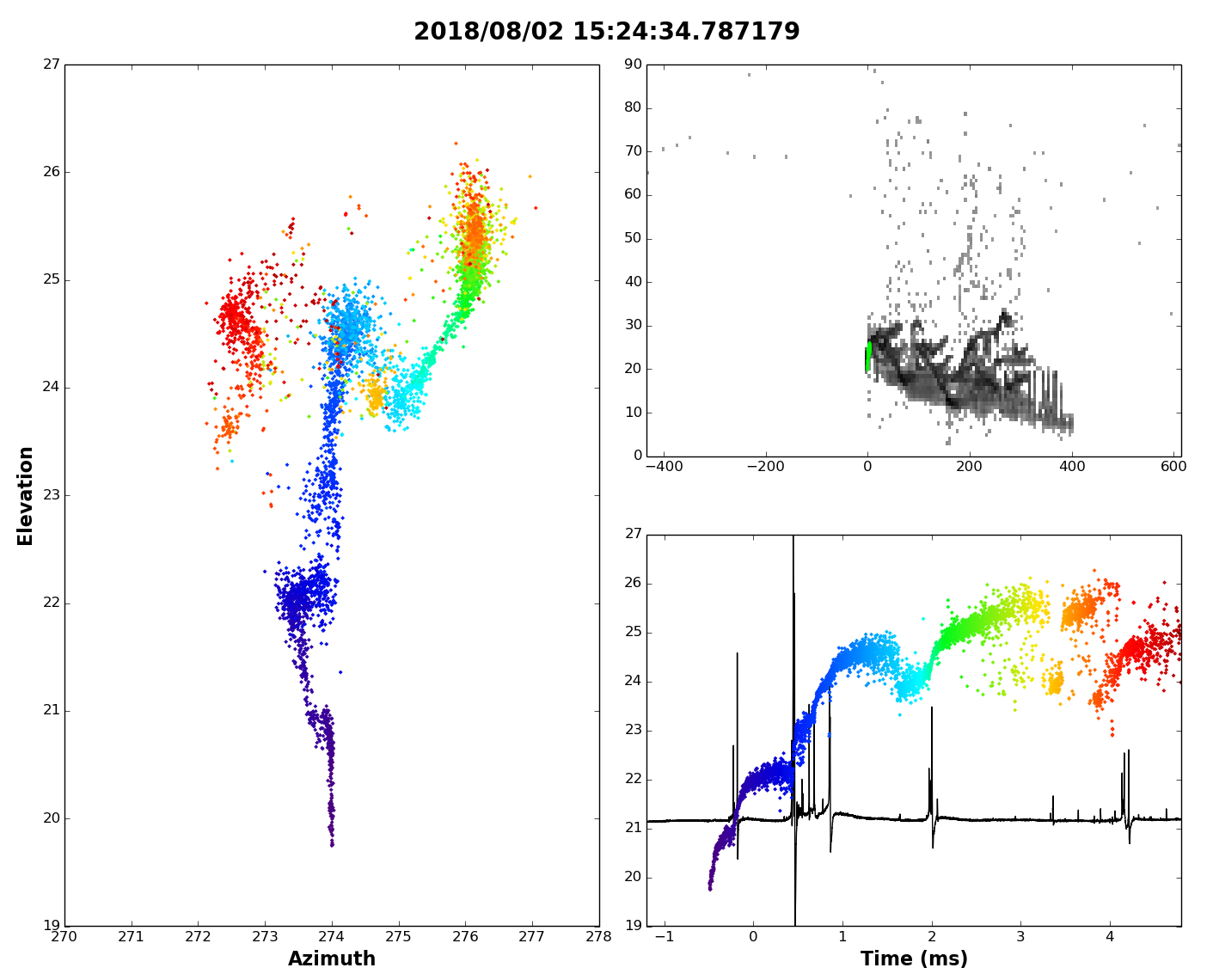} \\
\caption{
\textbf{Comparison of --CG and IC flashes.}
INTF observations of the initial
few milliseconds of the --CG flash that initiated TGF~B (top panels) and an IC flash that occurred 69~s later in the same storm (bottom panels). 
For each flash the temporal development is color-coded from blue to red, with the time scale being approximately the same for both flashes.  The upward stepping is well-delineated in the elevation vs.\ azimuth and elevation vs.\ time plots for the IC flash, and more continuous for the --CG flash.  The stepping lengths were $\simeq$350, 350, and 850~m for the IC flash before branching, and averaged $\simeq$180~m for the --CG flash (900~m in 5 steps). Upper right panels show overviews of the entire flashes.
}
\label{fig:tgfb_IC_azel} 
\end{figure}

\begin{figure}
\centering
\includegraphics[width=0.8\textwidth]{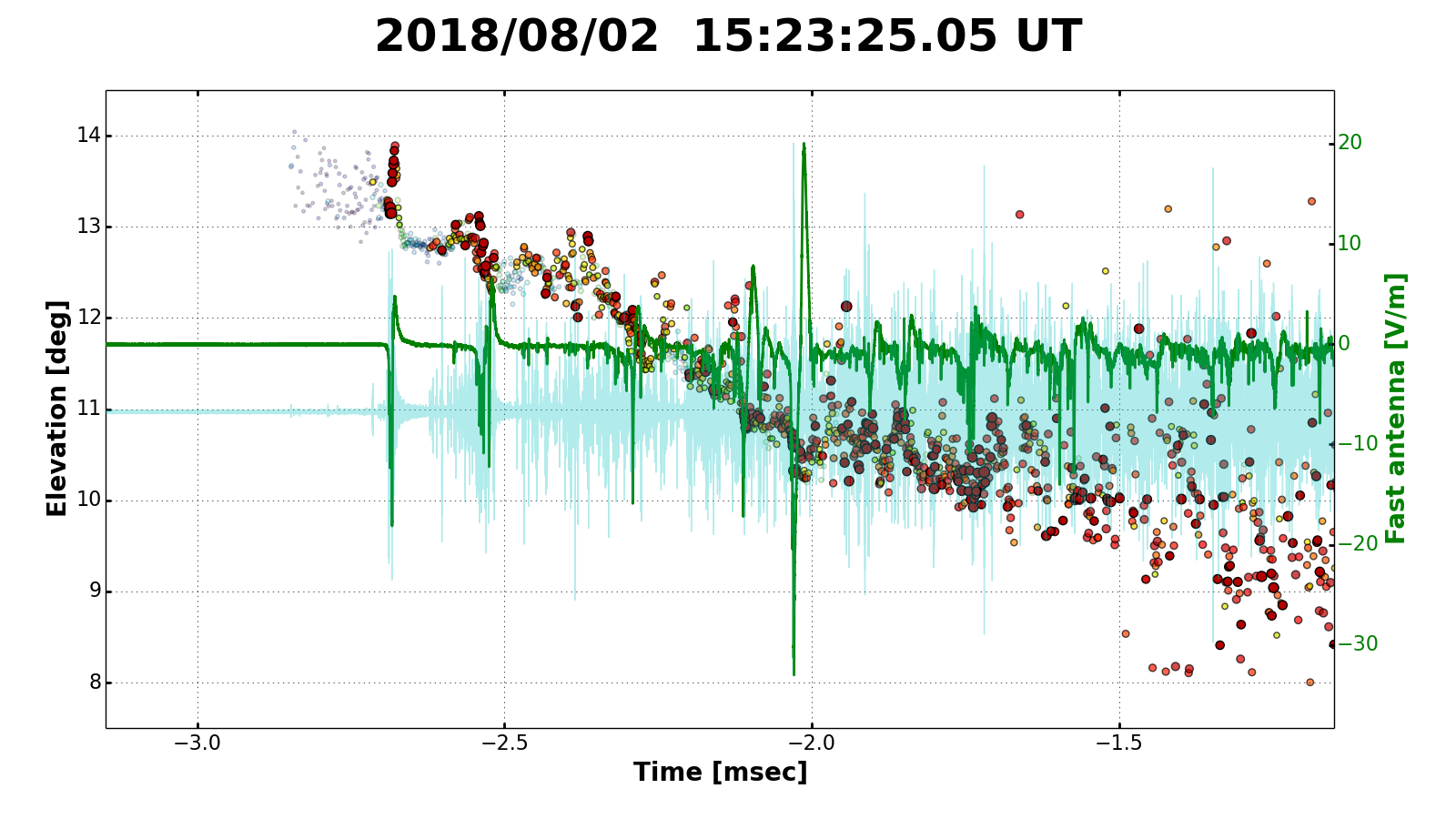} \\
\includegraphics[width=0.8\textwidth]{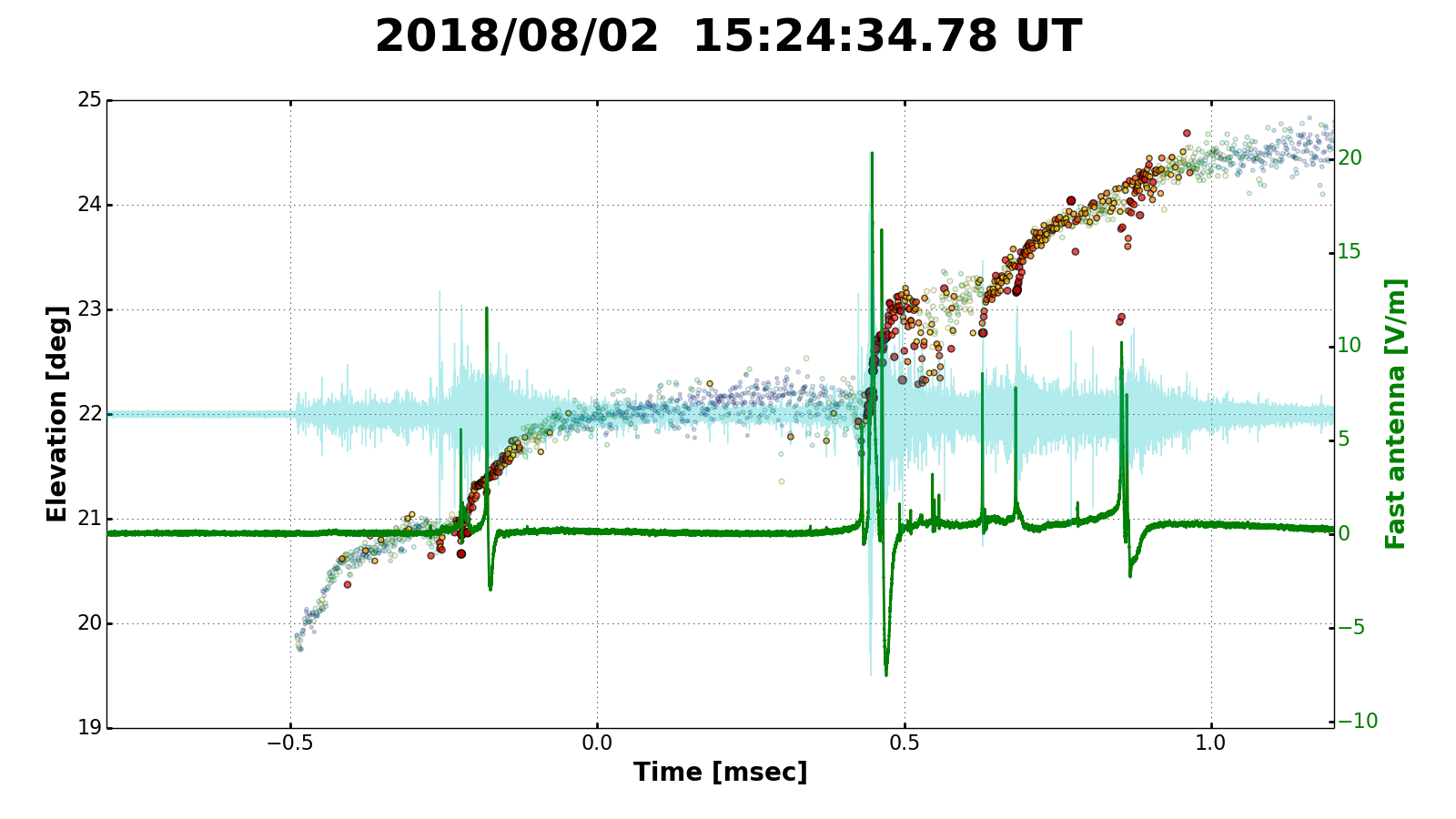} \\
\caption{
\textbf{Two millisecond zoom plots}.
Same as the lower right panels of Fig.\ S24,
but showing the first $\simeq$1.5~ms of each flash, illustrating how --CG
flashes develop rapidly and continuously, while IC flashes develop more
intermittently, with longer-duration steps and complex sequences of IBPs and
sub-pulses.  Each of the two IC steps was initiated by an IBP produced by fast
negative breakdown having speeds of $\simeq$1--2~$\times 10^7$~m/s, with the
second, complex IBP and sub-pulse sequence lasting $\simeq$500~$\mu$s.
}
\label{fig:tgfb_IC_2ms} 
\end{figure}

\begin{figure}
\centering
\includegraphics[width=0.8\textwidth]{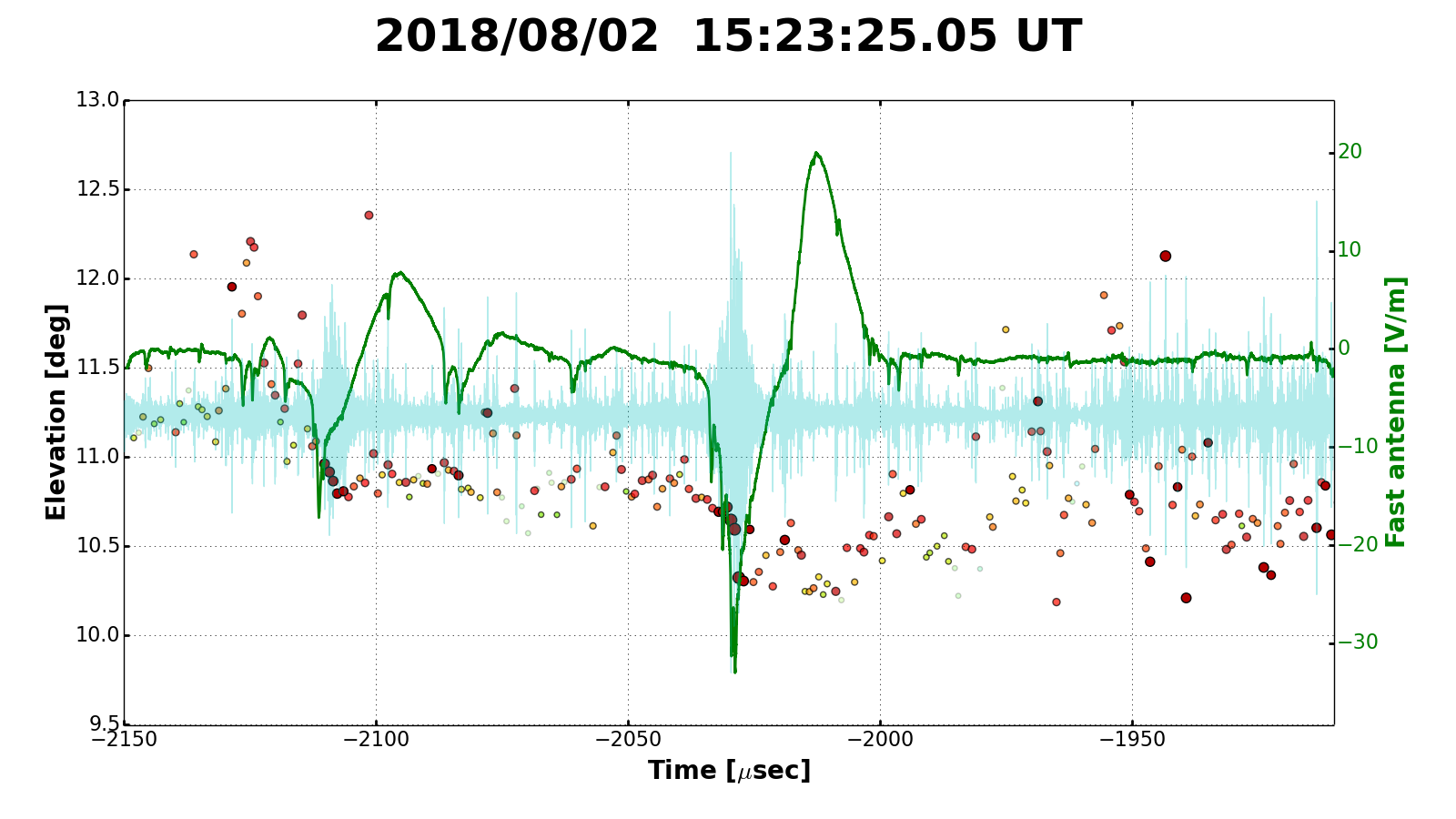} \\
\includegraphics[width=0.8\textwidth]{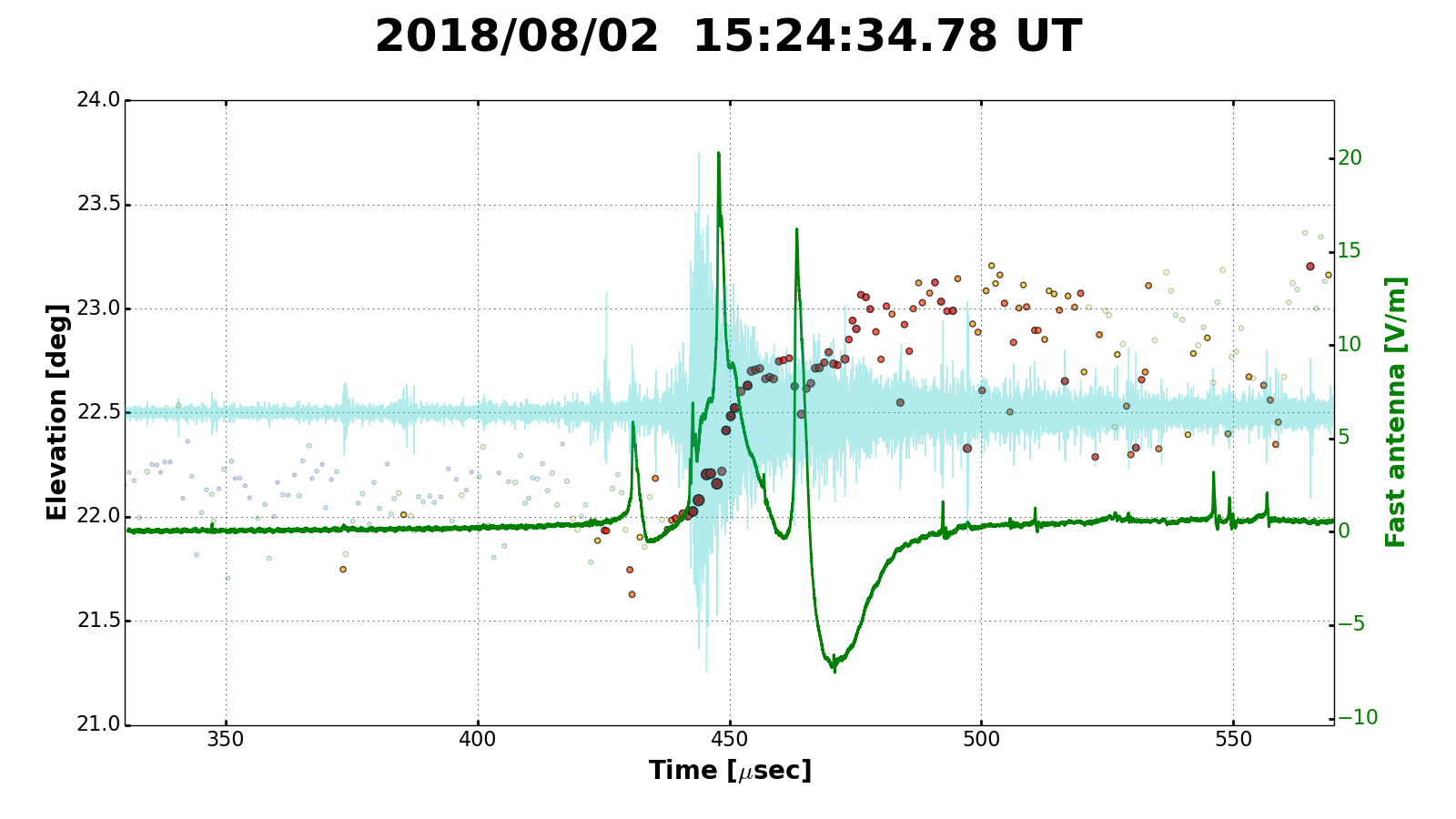} \\
\caption{
\textbf{240 microsecond zoom plots.}
Comparison of the largest IBPs of the --CG and
IC flashes in Fig.\ S25.  In both cases the IBP was produced by fast negative
breakdown, having speeds of 1--2~$\time 10^7$~m/s.  The sub-pulses of the IC
IBP were noticeably stronger and more impulsive than those of the --CG flash,
which initiated TGF~B. Note the onset of strong VHF radiation at the beginning
of the FNB, and sub-pulses occurring both before and during the
opposite-polarity field change of the IBP.  The overall duration of the IC IBP
sferic was somewhat longer than that of the --CG flash, being
$\simeq$60 and 40~$\mu$s, respectively.
}
\label{fig:tgfb_IC_240us} 
\end{figure}

\begin{figure}
\centering
\includegraphics[width=0.7\textwidth]{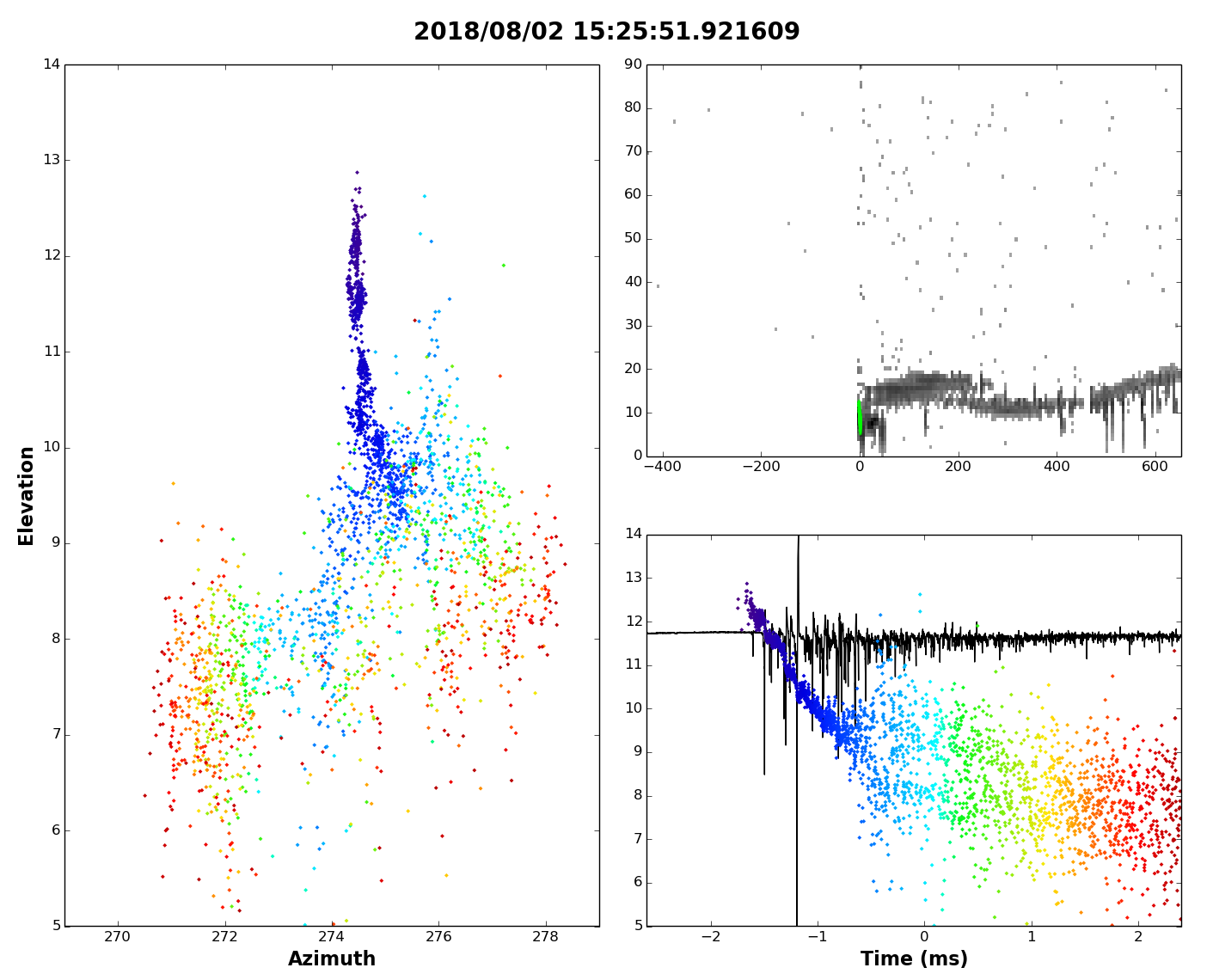} \\
\includegraphics[width=0.7\textwidth]{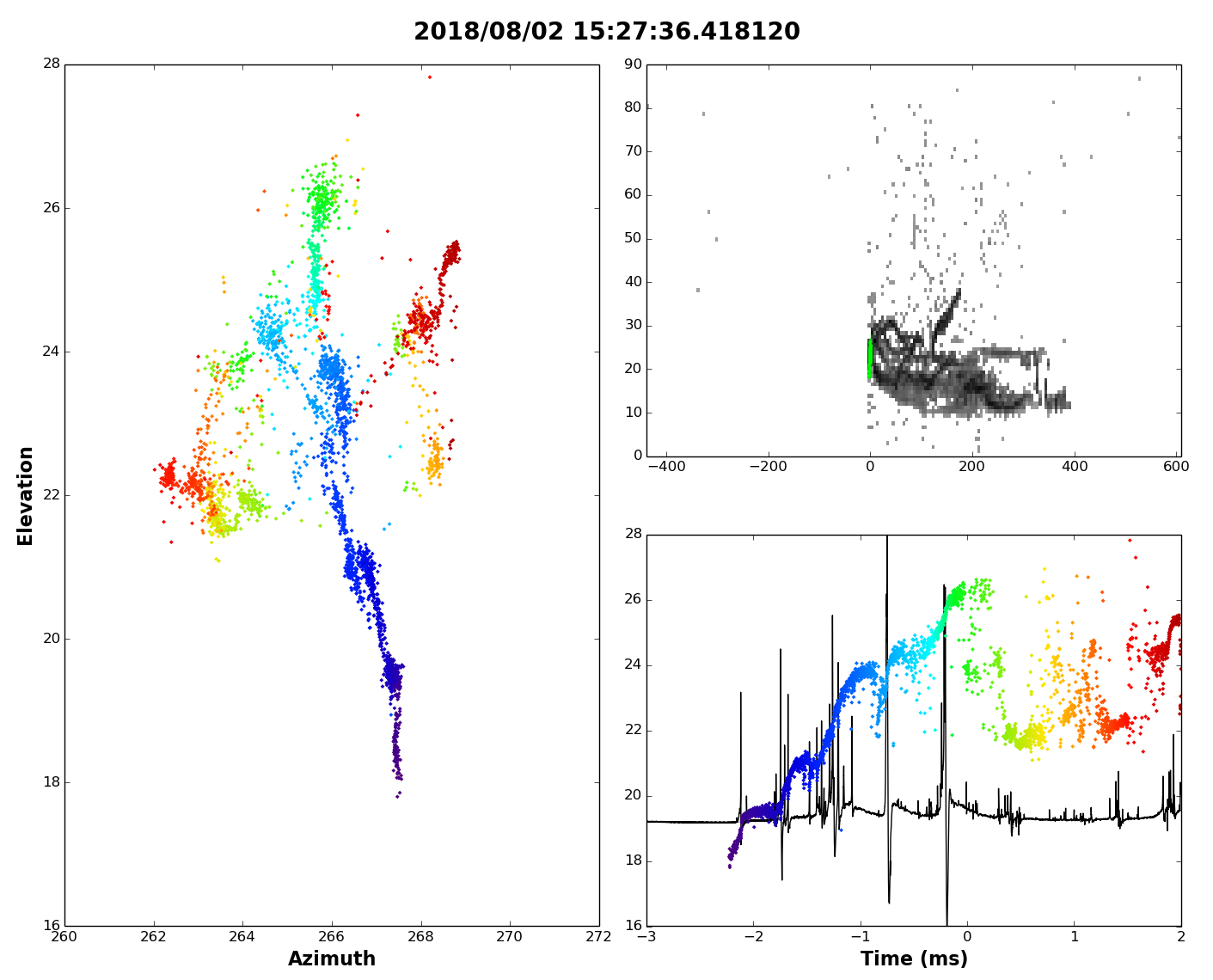} \\
\caption{
\textbf{Primary IC/CG flash comparison.}
Same as Fig.\ S24, except providing
overviews of the --CG flash that initiated TGF C (top panels) and the
following IC flash in the storm (bottom panels), corresponding to the examples
of Figures~6 and 7 of the main text.  Of particular note in both this IC and
that of Figure~S24 is that the strongest IBP occurred as the upward negative
breakdown started to branch out, at which point the strong IBPs abruptly
started dying out (lower right panel of the IC). The same observation has been
noted in connection with the IBPs of the main TGFS, as noted for Figures~S19--S22.
}
\label{fig:tgfc_IC_azel} 
\end{figure}

\begin{figure}
\centering
\includegraphics[width=0.8\textwidth]{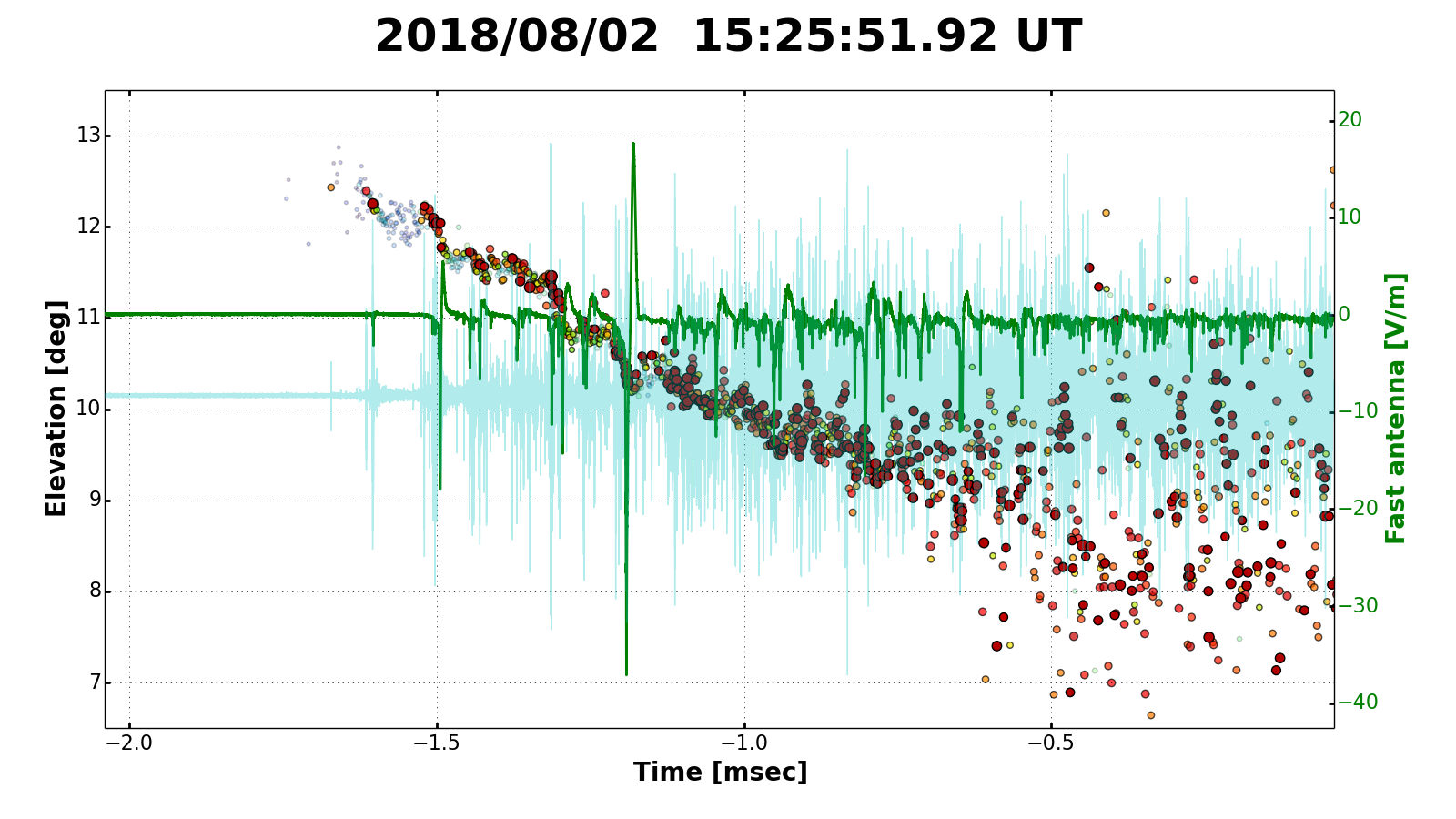} \\
\includegraphics[width=0.8\textwidth]{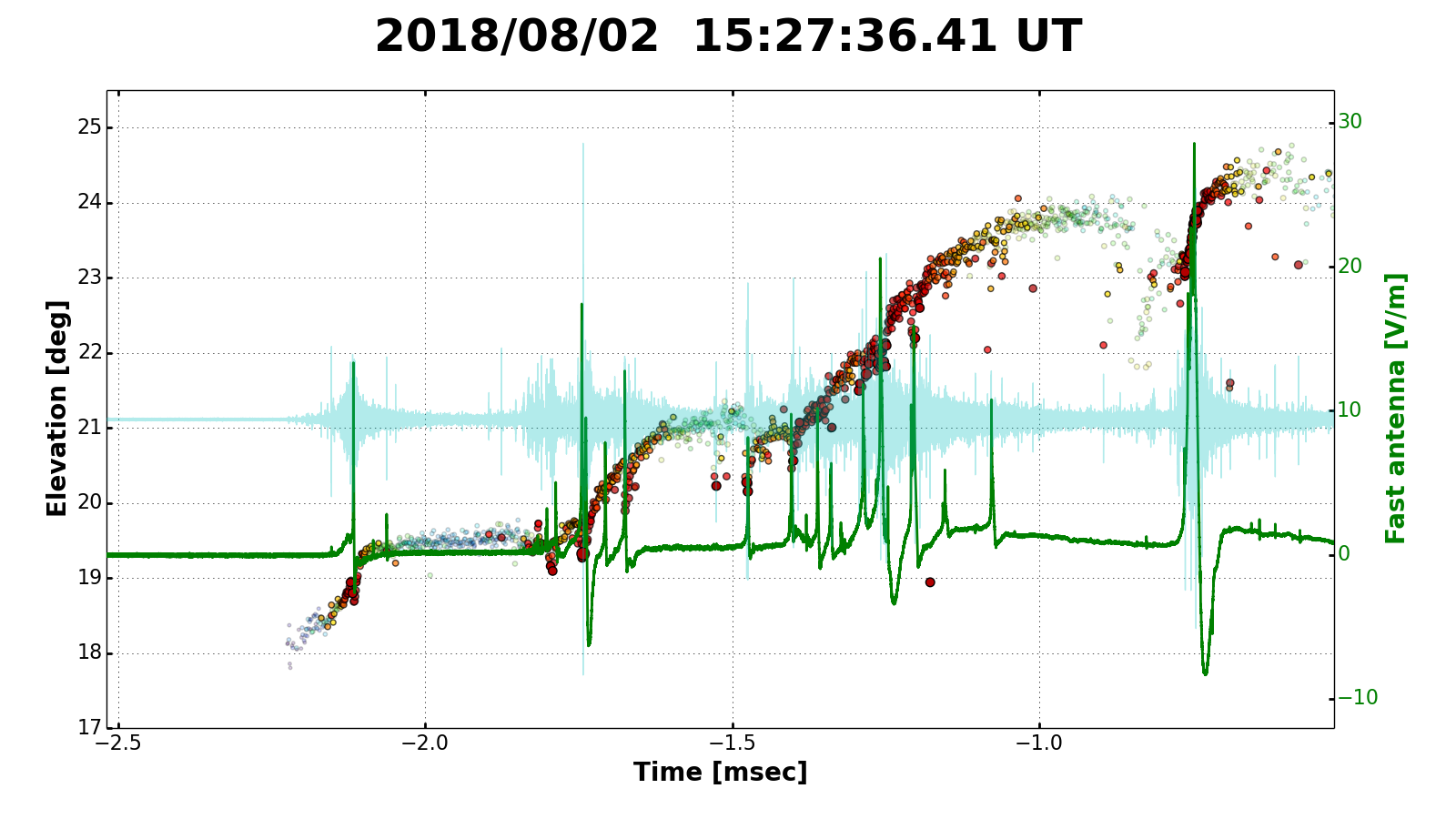} \\
\caption{
\textbf{Two millisecond IC/CG comparisons.} Same as Fig.\ S25 and enlarged
versions of the top two panels of Figure~6 of the main text, illustrating
the complex sequences of the initial breakdown during the post-TGF~C IC
flash and the role of high-power negative streamer breakdown and sub-pulses
in the upward negative breakdown of the IC (see Section~3.2 of the main text).  As in the IC/CG
comparison of Figure~S24, the stepping lengths of the IC were longer than
those of TGF~C, being $\simeq$400, 570, and 1000~m for the first three IC
steps, vs.\ $\simeq$140 and 180~m for the two largest steps of TGF~C's --CG.
}
\label{fig:tgfc_IC_2ms} 
\end{figure}

\begin{figure}
\centering
\includegraphics[width=0.8\textwidth]{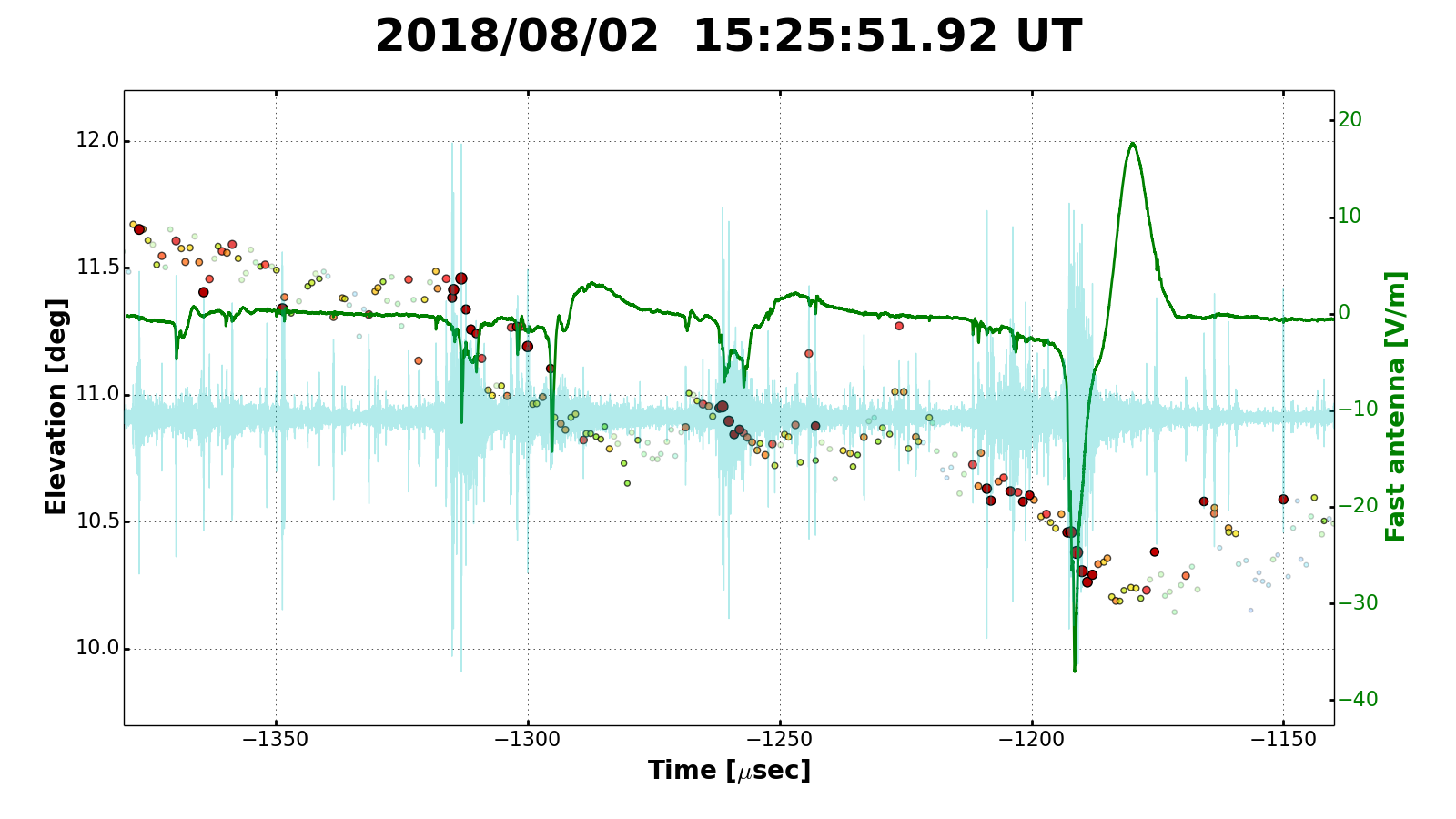} \\
\includegraphics[width=0.8\textwidth]{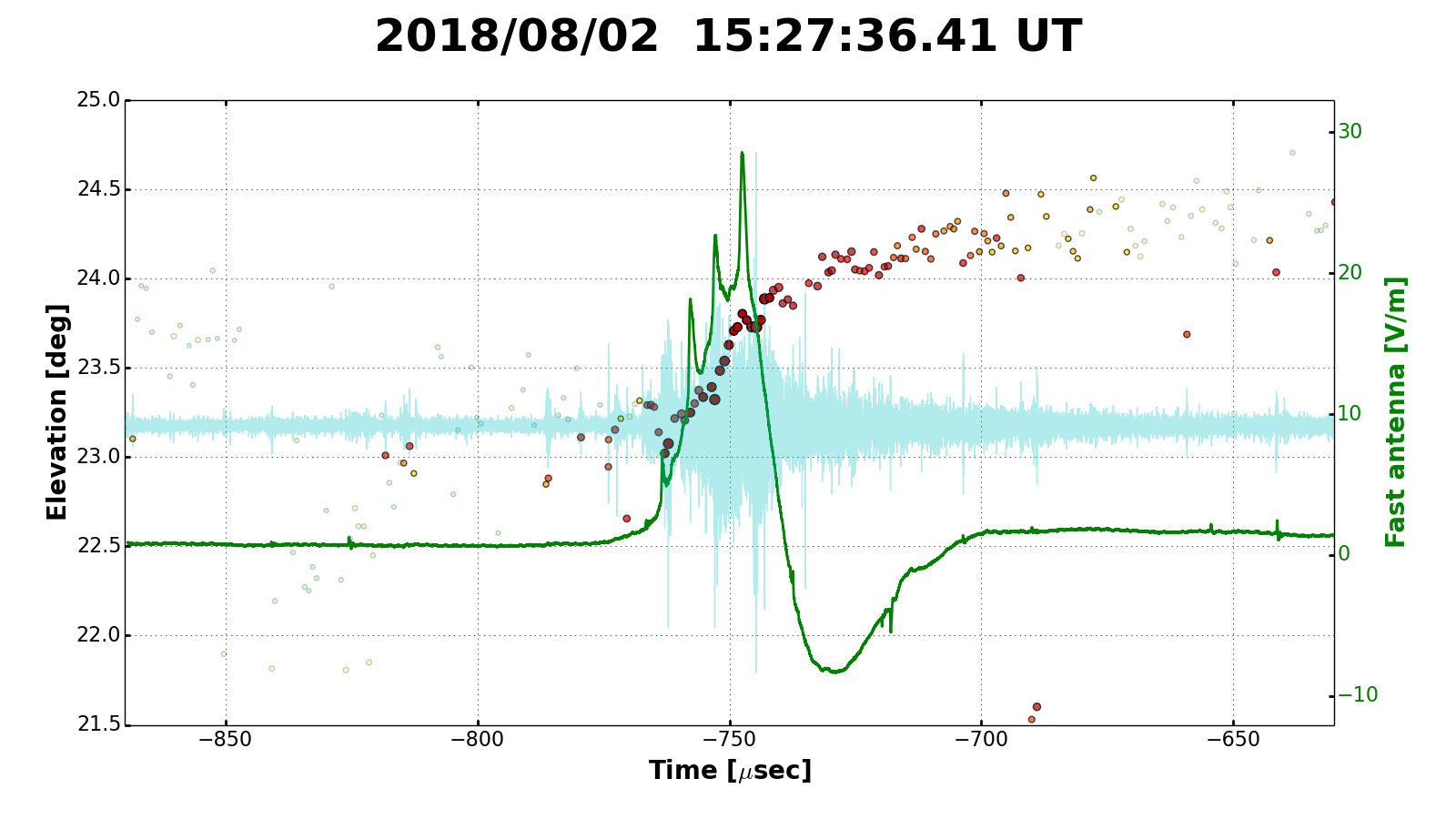} \\
\caption{
\textbf{240 microsecond plots.} 
Comparison of the fourth IBP of the IC flash
of Fig.\ S28 (bottom panel) with the IBP of TGF~C (right-hand side of top
panel), showing how the IC IBP is substantially stronger in amplitude,
duration and in the strength and impulsiveness of the sub-pulses, but
is otherwise produced by the same basic process of FNB having embedded
sub-pulses.  Like TGF~C's IBP, the IC's FNB is similarly fast (1.5~$\times
10^7$~m/s) and is initiated with brief fast positive breakdown (in this case
downward).  Note that the earlier gamma event of TGF~C (Figs.\ S8 and S12c,f)
was produced by the relatively weak sferic and FNB event at about
--1310~$\mu$s in the top panel, indicating how even weak FNB can produce
gamma-producing avalanching.
}
\label{fig:tgfc_IC_240us} 
\end{figure}

\begin{figure}
\centering
\includegraphics[width=0.8\textwidth]{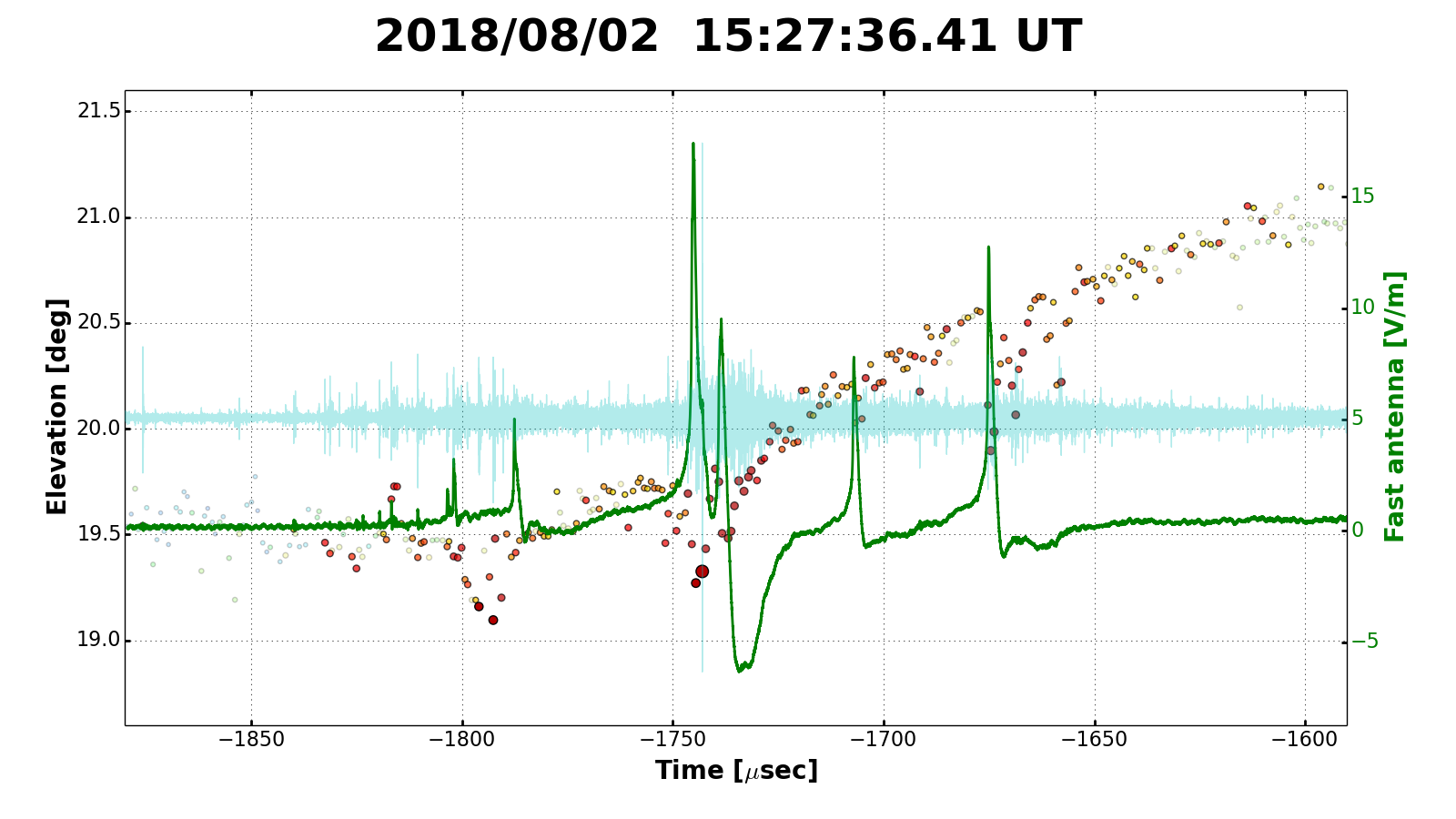} \\
\includegraphics[width=0.8\textwidth]{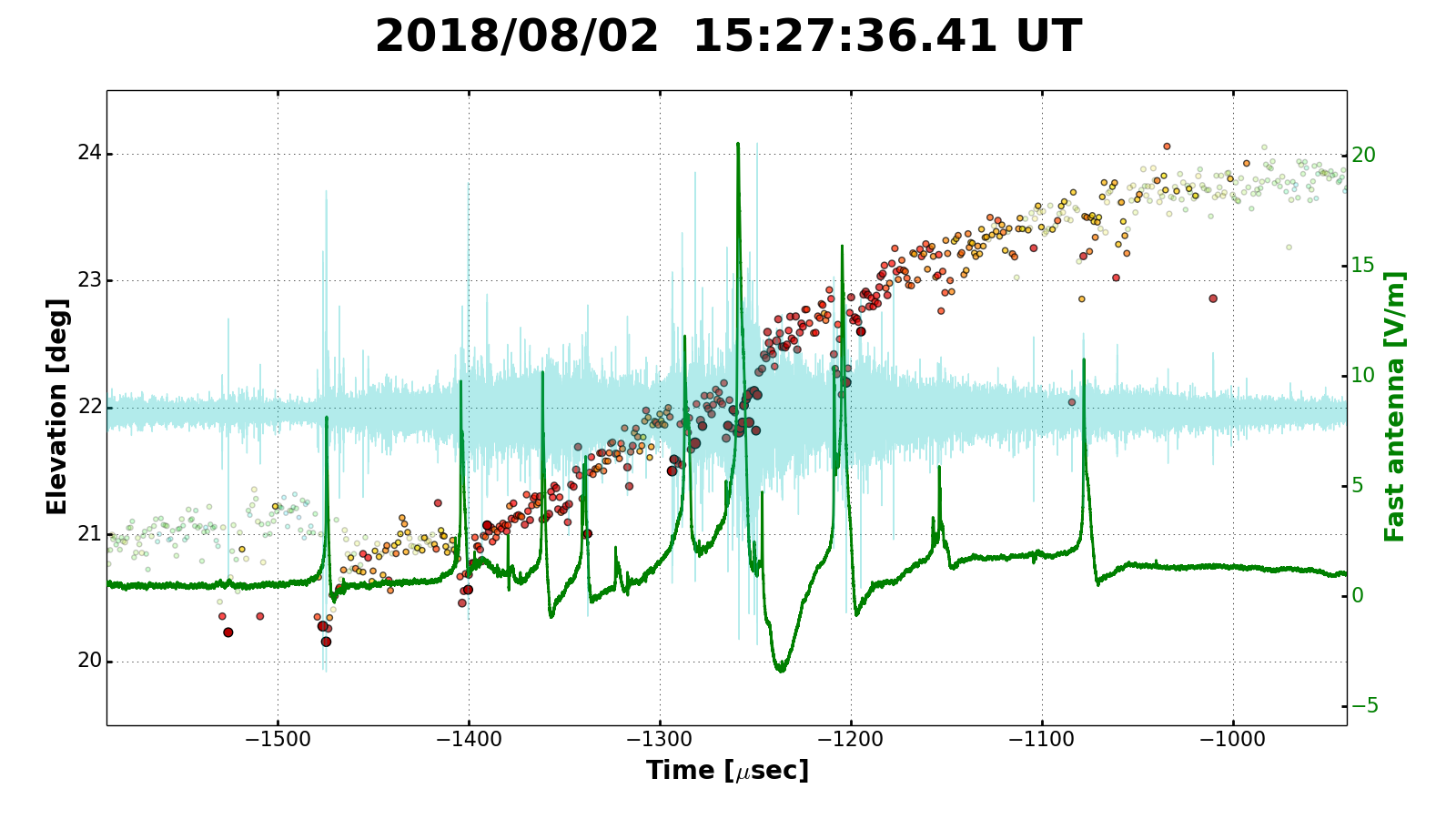} \\
\caption{
\textbf{Complex IBP/sub-pulse events.} 
Expanded views of the two complex
IBP/sub-pulse sequences of the post-TGF~C intracloud flash of Fig.\ S28 and
Figure~7 of the main text. The FNB breakdown of the IBPs and the sub-pulses
are each embedded in continuous upward negative streamer breakdown having a
propagation speed of $\simeq$2--3~$\times 10^6$~m/s, showing that
negative streamer breakdown doesn't have to travel at speeds of $10^7$~m/s to
produce the sub-pulse sparks. The overall durations of the sferics are
$\simeq$130 and 400~$\mu$s, respectively, with the first complex event being
very similar to that of a Florida IC flash that produced a satellite-detected
TGF of 50~$\mu$s duration (Sections~3.2 and 3.3 of main text).
}
\label{fig:tgfc_IC_long} 
\end{figure}

\begin{figure}
\centering
\includegraphics[width=0.45\textwidth]{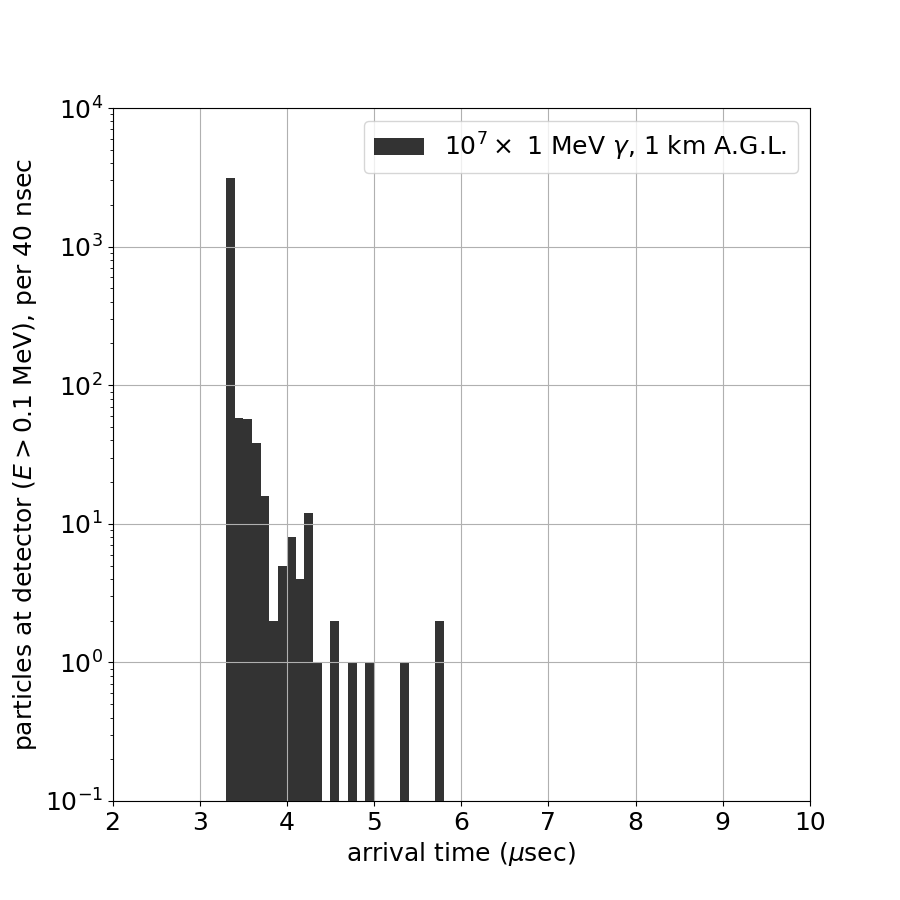} \\
\includegraphics[width=0.45\textwidth]{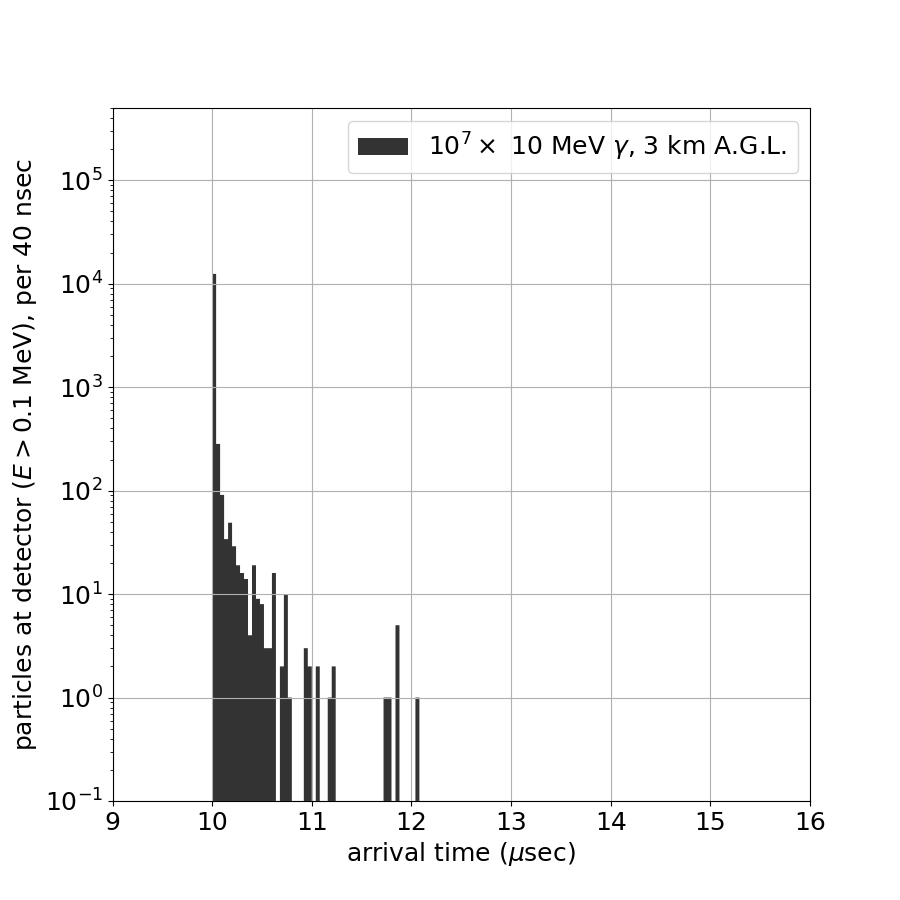}
\includegraphics[width=0.45\textwidth]{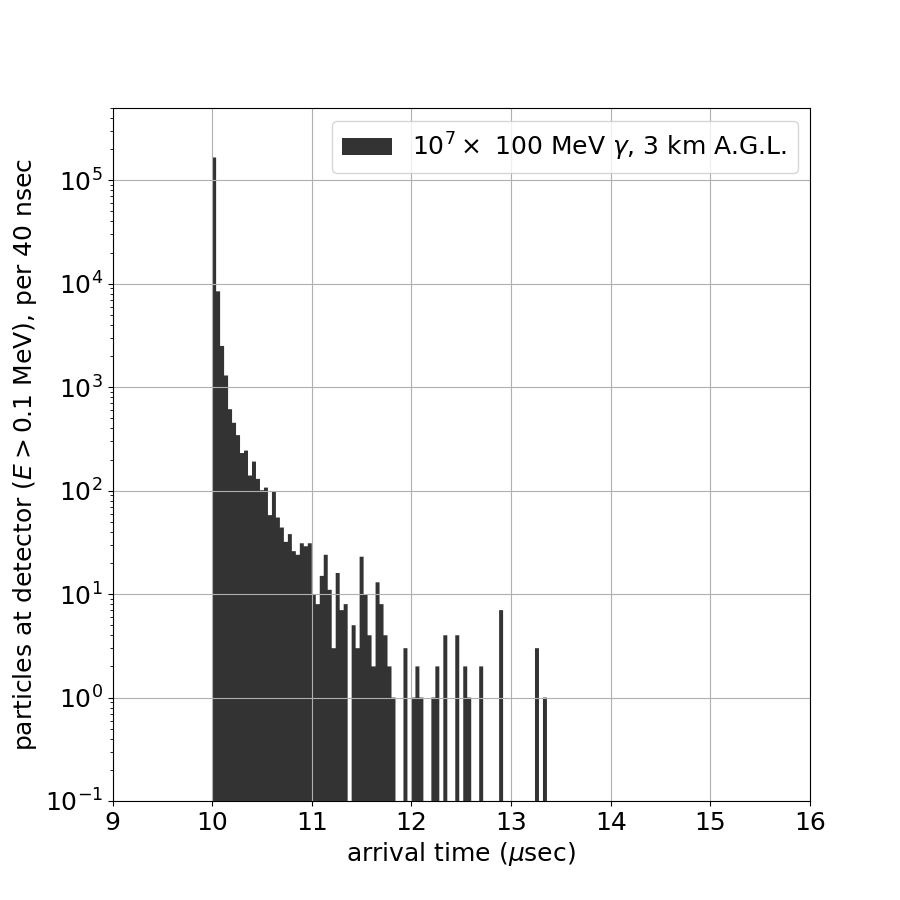} \\
\caption{
\textbf{GEANT4 simulation of the effects of Compton scattering},
pair production, and bremsstrahlung on the time-of-arrival of gamma ray showers. The lower panels with 10 and 100~MeV photons were generated 3~km above ground level; 1~MeV photons (upper panel) will not produce detectable energy deposit from 3~km, and so were generated at 1~km. In each simulation, $10^7$ monoenergetic photons were generated in air at the center of a spherical (in order to remove geometric effects) ``detector'' of radius 3~km (1~km for 1~MeV case). Particles were scored at the detector if their energies exceeded 0.1~MeV, corresponding to the minimum detectable energy deposit above background (Abbasi et al., 2018). 95\% of particles arrive within 20~ns (60~ns) for 10~MeV (100~MeV) primary photons, and within 40~ns for 1~MeV. We conclude that temporal structure in the TASD waveforms on longer time scales is indicative of the intrinsic duration of the TGF source.
}
\label{fig:compton_sim} 
\end{figure}

\clearpage

%
%
%
%
%

\begin{table}  
\settablenum{S1}
\centering
\resizebox{0.8\textwidth}{!}{ \begin{minipage}{\textwidth}
\begin{tabular}{ccrrrrr}
          &        &         & LMA & NLDN$\quad$ & TASD
                                                      
                                                      energy sum& Number of\\
   Date   &  Time  & $\mu$sec$\;$ & dBW & I$_{pk}\qquad$ &  VEM/MeV & TASDs\\
\hline
2018/08/02&14:17:20& 616655 & 20.3 &  \\              
          &        &    848 &       &  -10.0 kA C&  \\
          &        &    851 & 21.5&   \\                  
          &        &    981 &       & -36.7  kA C & \\
          &        &    987 & 27.7&   \\                  
          &        &    994 &       &          & 561/1150 & 9 \\ 
          &        & 617094 & 27.2  &  \\                
          &        &    104 &       &          & 192/393 & 8 \\ 
          &        &    191 & 20.3  &  \\                 
          &        &    288 &       & -12.0  kA C & \\
          &        &    294 & 27.6  \\ 
          &        &    444 & 21.8  \\ 
\hline
2018/08/02&15:23:25& 042253 & 25.4 & \\            
          &        &    255 &              & -10.9 kA C & \\
          &        &    259 &              & +12.0 kA C & \\
          &        &    279 &              & -3.1 kA C & \\
          &        &    329 &              & -26.9 kA C & \\
          &        &    332 & 27.2  & \\     
          &        &    341 &              & -30.1 kA C & \\
          &        &    341 &     & &     112/229  &  12\\
          &        &    416 & 18.1 & \\ 
          &        &    459 &              & -7.6 kA C & \\
\hline
2018/08/02&15:25:51& 913524 & 4.9  & \\
          &        &    615 & &-6.4 kA C  & \\     
          &        &    633 & 15.3  & \\  
          &        &    816 & 20.1 & \\        
          &        &    825 &     & &     35/72  &  5\\
          &        &    832 & &-5.0 kA C  & \\     
          &        &    937 & &-21.7 kA C  & \\     
          &        &    938 & 26.7  & \\ 
          &        &    942 &     & &     212/434  &  8\\
          &        & 914016 & 14.4  & \\ 
          &        &    076 & &-5.9 kA C  & \\    
\hline
2018/10/03&04:03:48& 687336 & 22.2  & \\
          &        &    696      & 23.2 & \\     
          &        & 688177       & 26.1 & \\  
          &        &    200       & 22.0 & \\                  
          &        &    346       & 24.6 & \\                  
          &        &    436 &     & &     100/205  &  9\\
          &        &    508 & 31.8 &  \\  
          &        &    584 &     & &     440/902  &  12\\                       
          &        &    599 &  30.5 &  \\  
\hline
 
\end{tabular}
\end{minipage}}
\caption{\textbf{Quantitative event values.}
  Quantitative values of the LMA, NLDN, and TASD observations for the flashes
  of TGFs A,B,C,D, during the initial 1--2~ms of the flashes.  Shown are the
  times of the LMA, NLDN, and TASD events [$\mu$s], the VHF source powers of
  the LMA sources [dBW], the peak currents $I_{pk}$ of the NLDN detected
  events [kA], the sum total energy deposited in all adjacent surface
  detectors triggered by the gamma bursts, in [VEM] and [MeV], and the number of TASDs contributing to the total. TASD times reported here are the time of detection at ground level, delayed by propagation from the source. LMA and NLDN times correspond to the time of the source itself.}
\label{tab:events1} 
\end{table}

\begin{table}[h!]
	\begin{center}
		\label{tab:table1}
		\centering
		\resizebox{0.85\textwidth}{!}{ \begin{minipage}{\textwidth}
		\begin{tabular}{l|r|r|r|r|r|r}

			\textbf{Event} & \textbf{SD} & \textbf{$\theta_c$ (deg)} & \textbf{z$_a$ (km)} & \textbf{$t_c$ ($\mu$s)} & \textbf{$\Delta t_b$ ($\mu$s)} & \textbf{U\slash L ratio} \\
			\hline
			TGF A & 2208 & \textbf{10.73 $\pm$ 0.03} & \textbf{3.21 $\pm$ 0.03} & \textbf{617,039.2 $\pm$ 0.7} & 45.2 $\pm$ 0.4 & 1.5 $\pm$ 1.0 \\
			& 2206 & 10.73 $\pm$ 0.03 & 3.21 $\pm$ 0.03 & 617,039.2 $\pm$ 0.9 & 45.1 $\pm$ 0.7 & 0.6 $\pm$ 0.3 \\
			& 2308 & 10.73 $\pm$ 0.03 & 3.21 $\pm$ 0.03 & 617,039.1 $\pm$ 0.6 & 45.9 $\pm$ 0.3 & \textbf{1.0 $\pm$ 0.6} \\
			& 2209 & 10.73 $\pm$ 0.03 & 3.21 $\pm$ 0.03 & 617,039.3 $\pm$ 0.7 & 42.6 $\pm$ 0.3 & 0.7 $\pm$ 0.4 \\
			& 2307 & 10.79 $\pm$ 0.03 & 3.23 $\pm$ 0.03 & 617,038.3 $\pm$ 0.6 & 46.9 $\pm$ 0.4 & 1.0 $\pm$ 0.6 \\
			& 2409 & 10.71 $\pm$ 0.02 & 3.21 $\pm$ 0.03 & 617,041.1 $\pm$ 0.6 & 43.6 $\pm$ 0.1 & 1.4 $\pm$ 0.9 \\
			& 2207 & 10.73 $\pm$ 0.03 & 3.21 $\pm$ 0.03 & 617,039.8 $\pm$ 0.8 & 45.9 $\pm$ 0.5 & 1.6 $\pm$ 0.9 \\
			& 2408 & 10.76 $\pm$ 0.03 & 3.22 $\pm$ 0.03 & 617,038.7 $\pm$ 0.6 & 45.7 $\pm$ 0.2 & 0.9 $\pm$ 0.6 \\
			& 2309 & 10.78 $\pm$ 0.03 & 3.23 $\pm$ 0.03 & 617,038.6 $\pm$ 0.7 & 43.5 $\pm$ 0.2 & 0.8 $\pm$ 0.4 

\\			\hline
			TGF B & 1419 & 9.93 $\pm$ 0.03 & 2.91 $\pm$ 0.02 & 42,388.4 $\pm$ 0.4 & 45.0 $\pm$ 0.2 & 1.3 $\pm$ 0.8 \\
			& 1421 & 10.06 $\pm$ 0.02 & 2.95 $\pm$ 0.01 & 42,384.0 $\pm$ 0.3 & 45.7 $\pm$ 0.2 & 1.0 $\pm$ 0.6 \\
			& 1520 & \textbf{9.96 $\pm$ 0.03} & \textbf{2.92 $\pm$ 0.02} & 42,387.2 $\pm$ 0.3 & 44.35 $\pm$ 0.09 & \textbf{0.9 $\pm$ 0.5} \\
			& 1320 & 9.87 $\pm$ 0.02 & 2.90 $\pm$ 0.02 & 42,391.0 $\pm$ 0.4 & 46.5 $\pm$ 0.3 & 0.3 $\pm$ 0.2 \\
			& 1519 & 9.92 $\pm$ 0.02 & 2.91 $\pm$ 0.02 & 42,389.3 $\pm$ 0.4 & 43.8 $\pm$ 0.2 & 0.7 $\pm$ 0.5 \\
			& 1620 & 9.96 $\pm$ 0.03 & 2.92 $\pm$ 0.02 & \textbf{42,387.4 $\pm$ 0.4} & 41.72 $\pm$ 0.05 & 3.0 $\pm$ 2.0 \\
			& 1521 & 9.98 $\pm$ 0.02 & 2.93 $\pm$ 0.02 & 42,386.4 $\pm$ 0.3 & 44.1 $\pm$ 0.1 & 0.8 $\pm$ 0.6 \\
			& 1621 & 9.96 $\pm$ 0.03 & 2.92 $\pm$ 0.02 & 42,387.2 $\pm$ 0.4 & 41.48 $\pm$ 0.07 & 0.4 $\pm$ 0.3 \\
			& 1720 & 9.87 $\pm$ 0.02 & 2.89 $\pm$ 0.02 & 42,391.4 $\pm$ 0.4 & 38.58 $\pm$ 0.03 & 1.6 $\pm$ 1.0 

\\			\hline
			TGF C & 1424 & 9.94 $\pm$ 0.01 & 2.80 $\pm$ 0.01 & 913,983.5 $\pm$ 0.3 & 39.9 $\pm$ 0.2 & 2.9 $\pm$ 2.3 \\
			& 1521 & \textbf{9.83 $\pm$ 0.03} & \textbf{2.77 $\pm$ 0.01} & 913,989.4 $\pm$ 0.2 & 43.70 $\pm$ 0.05 & 1.4 $\pm$ 1.0 \\
			& 1522 & 9.83 $\pm$ 0.03 & 2.77 $\pm$ 0.01 & 913,989.4 $\pm$ 0.2 & 43.45 $\pm$ 0.07 & \textbf{1.0 $\pm$ 0.7} \\
			& 1421 & 9.83 $\pm$ 0.03 & 2.77 $\pm$ 0.01 & \textbf{913,989.6 $\pm$ 0.2} & 44.7 $\pm$ 0.1 & 0.9 $\pm$ 0.5 \\
			& 1423 & 9.85 $\pm$ 0.02 & 2.78 $\pm$ 0.01 & 913,988.9 $\pm$ 0.3 & 42.6 $\pm$ 0.2 & 2.3 $\pm$ 1.7 \\
			& 1422 & 9.83 $\pm$ 0.03 & 2.77 $\pm$ 0.01 & 913,989.6 $\pm$ 0.2 & 44.4 $\pm$ 0.1 & 0.9 $\pm$ 0.5 \\
			& 1323 & 9.77 $\pm$ 0.03 & 2.75 $\pm$ 0.01 & 913,991.3 $\pm$ 0.3 & 42.2 $\pm$ 0.2 & 0.8 $\pm$ 0.5 \\
			& 1322 & 9.81 $\pm$ 0.03 & 2.76 $\pm$ 0.01 & 913,990.0 $\pm$ 0.2 & 43.9 $\pm$ 0.2 & 1.0 $\pm$ 0.5 \\
			& 1222 & 9.78 $\pm$ 0.03 & 2.76 $\pm$ 0.01 & 913,991.0 $\pm$ 0.3 & 42.0 $\pm$ 0.2 & 10.6 $\pm$ 6.6 

\\			\hline
			TGF D & 1704 & 7.20 $\pm$ 0.02 & 3.02 $\pm$ 0.04 & 688,680.1 $\pm$ 1.4 & 69.3 $\pm$ 0.8 & 1.0 $\pm$ 0.6 \\
			& 1603 & 7.20 $\pm$ 0.02 & 3.02 $\pm$ 0.04 & 688,680.1 $\pm$ 1.5 & 68.7 $\pm$ 1.1 & 0.7 $\pm$ 0.5 \\
			& 1705 & \textbf{7.19 $\pm$ 0.02} & \textbf{3.02 $\pm$ 0.04} & \textbf{688,680.2 $\pm$ 1.4} & 67.3 $\pm$ 0.6 & 1.0 $\pm$ 0.6 \\
			& 1702 & 7.21 $\pm$ 0.02 & 3.03 $\pm$ 0.04 & 688,679.9 $\pm$ 1.7 & 69.5 $\pm$ 1.3 & 4.0 $\pm$ 3.0 \\
			& 1805 & 7.16 $\pm$ 0.03 & 3.01 $\pm$ 0.04 & 688,681.0 $\pm$ 1.3 & 67.3 $\pm$ 0.3 & 6.8 $\pm$ 6.4 \\
			& 1703 & 7.20 $\pm$ 0.02 & 3.02 $\pm$ 0.04 & 688,680.1 $\pm$ 1.4 & 70.2 $\pm$ 1.0 & 1.4 $\pm$ 1.0 \\
			& 1803 & 7.17 $\pm$ 0.02 & 3.01 $\pm$ 0.04 & 688,681.0 $\pm$ 1.3 & 70.2 $\pm$ 0.8 & 1.0 $\pm$ 0.7 \\
			& 1504 & 7.17 $\pm$ 0.02 & 3.01 $\pm$ 0.04 & 688,680.9 $\pm$ 1.6 & 65.7 $\pm$ 1.1 & \textbf{1.1 $\pm$ 0.7} \\
			& 1804 & 7.14 $\pm$ 0.03 & 3.00 $\pm$ 0.04 & 688,681.7 $\pm$ 1.3 & 69.4 $\pm$ 0.5 & 1.8 $\pm$ 1.2 \\
			& 1602 & 7.19 $\pm$ 0.02 & 3.02 $\pm$ 0.04 & 688,680.1 $\pm$ 1.7 & 68.1 $\pm$ 1.4 & 0.9 $\pm$ 0.6 \\
			& 1604 & 7.21 $\pm$ 0.02 & 3.03 $\pm$ 0.04 & 688,679.9 $\pm$ 1.5 & 67.9 $\pm$ 1.0 & 1.4 $\pm$ 1.0 \\
			& 1505 & 7.16 $\pm$ 0.03 & 3.01 $\pm$ 0.04 & 688,681.1 $\pm$ 1.6 & 64.1 $\pm$ 1.1 & 0.6 $\pm$ 0.4 

		\end{tabular}
		\end{minipage}}
	\caption{\textbf{Calculated source values.} Converged iteration values and associated uncertainties for each TGF, calculated independently for each surface detector. SD is the surface detector's identifying number XXYY identifying their easterly (XX) and northerly (YY) locations within the array in 1.2~km grid spacing units. $\theta_c$ is the elevation angle corresponding to z$_a$, which is the source altitude (above a reference plane of 1.4~km). $t_c$ is the determined microsecond of TGF signal arrival at the INTF and $\Delta t$ is the relative timing difference between INTF and SD signals, and U\slash L Ratio is the ratio of energy deposit between upper and lower levels of scintillator. Values in bold are the medians for that column and indicate the burst's median onset time/elevation.}
	\end{center}

\end{table}

\begin{table}[h!]
	\begin{center}
		\label{tab:table3}
		\resizebox{0.8\textwidth}{!}{ \begin{minipage}{\textwidth}
		\begin{adjustbox}{center}
		\begin{tabular}{l|r|r|r|r|r|r|r}

			\textbf{Event} & \textbf{D (km)} & \textbf{z$_a$ (km)} & \textbf{t$_a$ ($\mu$s)} & \textbf{$\Delta \theta _{FNB}$ (deg)} & \textbf{$\Delta$z$_{FNB}$ (m)} & \textbf{$\Delta$t$_{FNB}$ ($\mu$s)} & \textbf{v$_{FNB}$ (m/s)} \\
			\hline
			TGF A & 16.96 $\pm$ 0.15 & 3.21 $\pm$ 0.03 & 616,981.7 $\pm$ 0.6 & 0.55 & 150 & 10.0 & 1.5$\times 10^7$

\\			\hline
			TGF B & 16.64 $\pm$ 0.08 & 2.92 $\pm$ 0.02 & 42,331.7 $\pm$ 0.3 & 0.32 & 100 & 3.7 & 2.7$\times 10^7$

\\			\hline
			TGF C & 15.98 $\pm$ 0.04 & 2.77 $\pm$ 0.01 & 913,935.1 $\pm$ 0.2 & 0.40 & 120 & 4.7 & 2.6$\times 10^7$

\\			\hline
			TGF D & 23.9 $\pm$ 0.3 & 3.02 $\pm$ 0.04 & 688,600.1 $\pm$ 1.4 & 0.56 & 240 & 13.4 & 1.8$\times 10^7$

		\end{tabular}
		\end{adjustbox}
		\end{minipage}}
	\vspace{1cm}
	\caption{\textbf{Observed fast breakdown characteristics.} Extent and duration of fast breakdown occurring during the brightest event for each of the four TGFs, specified by the first column. Second column gives D, the plan distance between each TGF and the INTF. The third column $z_a$ is the median altitude result of the iteration process (Table~S2). The fourth column, $t_a$, is the reconstructed source time. $\Delta \theta_{FNB}$ is the angular extent of downward breakdown which, combined with D, gives the propagation distance, $\Delta$z$_{FNB}$. The fifth column, $\Delta$t$_{FNB}$, is the breakdown's duration in time, which allows for an estimation of the fast breakdown speed shown in the final column v$_{FNB}$. Note that the final four columns do not include uncertainties, as their values are estimated by simply assuming a linear descent of FNB based on data shown in Figure~4 (and Figures~S15-S18).}
	\end{center}

\end{table}

%